\documentclass[10pt,journal,compsoc]{IEEEtran}
\ifCLASSOPTIONcompsoc
  \usepackage[nocompress]{cite}
\else
  \usepackage{cite}
\fi

\ifCLASSINFOpdf
\else
\fi

\usepackage[numbers,sort&compress]{natbib}
\usepackage{hyperref}
\hypersetup{
    colorlinks=true, 
    linkcolor=black,
    citecolor=black,
    urlcolor=black, 
}

\usepackage{threeparttable}
\usepackage{graphicx}
\usepackage{amsmath}
\usepackage{amssymb}
\usepackage{multirow}
\usepackage{hhline}
\usepackage{makecell}
\usepackage{pifont}
\usepackage{enumitem}
\usepackage{booktabs}

\renewcommand{\citep}[1]{\citeauthor{#1}~(\citeyear{#1})}
\usepackage{tikz}

\usepackage[edges]{forest}
\definecolor{hidden-draw}{RGB}{0,0,0}
\definecolor{hidden-pink}{rgb}{0.98, 0.94, 0.75}
\definecolor{level0}{rgb}{0.98, 0.68, 0.4}
\definecolor{level1}{rgb}{0.98, 0.92, 0.84}
\definecolor{level2}{rgb}{0.8, 0.8, 1.0}
\definecolor{level3}{rgb}{1.0, 0.71, 0.76}
\definecolor{level4}{rgb}{0.49, 0.99, 0.0}

\definecolor{lawngreen}{rgb}{0.49, 0.99, 0.0}
\definecolor{pink}{rgb}{1, 0, 0.5}
\definecolor{airforce}{rgb}{0.36, 0.54, 0.66}

\hypersetup{
    linkbordercolor=airforce,
    urlbordercolor=pink,
    citebordercolor=green,
}
\begin{document}

\title{A Survey of Large Language Models for Financial Applications:
Progress, \\ Prospects and Challenges}

\author{Yuqi Nie\IEEEauthorrefmark{1}, Yaxuan Kong\IEEEauthorrefmark{1}, Xiaowen Dong, John M. Mulvey\IEEEauthorrefmark{2}\IEEEauthorrefmark{3}, H. Vincent Poor, \\ Qingsong Wen, Stefan Zohren\IEEEauthorrefmark{2}% <-this % stops a space

\IEEEcompsocitemizethanks{\IEEEcompsocthanksitem Yuqi Nie and H.Vincent Poor are with the Department of Electrical and Computer Engineering, Princeton University, Princeton, NJ 08540 USA (e-mail: ynie@princeton.edu; poor@princeton.edu).
\IEEEcompsocthanksitem John M. Mulvey is with the Department of Operation Research and Financial Engineering, Princeton University, Princeton, NJ 08540 USA (e-mail: mulvey@princeton.edu).
\IEEEcompsocthanksitem Yaxuan Kong, Stefan Zohren and Xiaowen Dong are with the Department of Engineering Science, University of Oxford, Oxford, OX1 2JD, UK (e-mail: yaxuan.kong@eng.ox.ac.uk; stefan.zohren@eng.ox.ac.uk; xiaowen.dong@eng.ox.ac.uk).
\IEEEcompsocthanksitem Qingsong Wen is with Squirrel AI Learning, WA 98004 USA (email: qingsongedu@gmail.com)}

% \IEEEcompsocitemizethanks{\IEEEcompsocthanksitem Yuqi Nie, John M. Mulvey and H.Vincent Poor are with Princeton University, Princeton, NJ 08540 USA (e-mail: ynie@princeton.edu; poor@princeton.edu).
% \IEEEcompsocthanksitem Yaxuan Kong, Stefan Zohren and Xiaowen Dong are with University of Oxford, Oxford, OX1 2JD, UK (e-mail: yaxuan.kong@eng.ox.ac.uk; stefan.zohren@eng.ox.ac.uk; xiaowen.dong@eng.ox.ac.uk).
% \IEEEcompsocthanksitem Qingsong Wen is with Squirrel AI Learning by Yixuan Education Inc. (email: qingsongedu@gmail.com)}

% note need leading \protect in front of \\ to get a newline within \thanks as
% \\ is fragile and will error, could use \hfil\break instead.
% stops a space
\thanks{\IEEEauthorrefmark{1} Equal Contribution. Random Ordering.}
\thanks{\IEEEauthorrefmark{2} Main Advisors.}
\thanks{\IEEEauthorrefmark{3} Corresponding Author.}
\thanks{Note: After the first two primary contributors, the remaining authors are listed in alphabetical order by surname.}}

% The paper headers
%\markboth{Journal of \LaTeX\ Class Files,~Vol.~14, No.~8, August~2015}%
%{Shell \MakeLowercase{\textit{et al.}}: Bare Advanced Demo of IEEEtran.cls for IEEE Computer Society Journals}
\IEEEtitleabstractindextext{%
\begin{abstract}
Recent advances in large language models (LLMs) have unlocked novel opportunities for machine learning applications in the financial domain. These models have demonstrated remarkable capabilities in understanding context, processing vast amounts of data, and generating human-preferred contents. 
In this survey, we explore the application of LLMs on various financial tasks, focusing on their potential to transform traditional practices and drive innovation. 
We provide a discussion of the progress and advantages of LLMs in financial contexts, analyzing their advanced technologies as well as prospective capabilities in contextual understanding, transfer learning flexibility, complex emotion detection, etc. 
We then highlight this survey for categorizing the existing literature into key application areas, including linguistic tasks, sentiment analysis, financial time series, financial reasoning, agent-based modeling, and other applications. For each application area, we delve into specific methodologies, such as textual analysis, knowledge-based analysis, forecasting, data augmentation, planning, decision support, and simulations. 
Furthermore, a comprehensive collection of datasets, model assets, and useful codes associated with mainstream applications are presented as resources for the researchers and practitioners.
Finally, we outline the challenges and opportunities for future research, particularly emphasizing a number of distinctive aspects in this field.
We hope our work can help facilitate the adoption and further development of LLMs in the financial sector.
\end{abstract}

% Note that keywords are not normally used for peerreview papers.
\begin{IEEEkeywords}
Large Language Models (LLMs), Financial Applications, Deep Learning, Foundation Models, Linguistic Tasks, Sentiment Analysis, Time Series, Reasoning, Agent-based Modeling.
\end{IEEEkeywords}}

% make the title area
\maketitle

% \setcounter{tocdepth}{2}
%\newpage
\tableofcontents
%\newpage

% To allow for easy dual compilation without having to reenter the
% abstract/keywords data, the \IEEEtitleabstractindextext text will
% not be used in maketitle, but will appear (i.e., to be "transported")
% here as \IEEEdisplaynontitleabstractindextext when compsoc mode
% is not selected <OR> if conference mode is selected - because compsoc
% conference papers position the abstract like regular (non-compsoc)
% papers do!
\IEEEdisplaynontitleabstractindextext
% \IEEEdisplaynontitleabstractindextext has no effect when using
% compsoc under a non-conference mode.

% For peer review papers, you can put extra information on the cover
% page as needed:
% \ifCLASSOPTIONpeerreview
% \begin{center} \bfseries EDICS Category: 3-BBND \end{center}
% \fi
%
% For peerreview papers, this IEEEtran command inserts a page break and
% creates the second title. It will be ignored for other modes.
\newpage

\IEEEpeerreviewmaketitle

\begin{figure*}[htbp!]
    \centering
    \includegraphics[width=.85\linewidth]{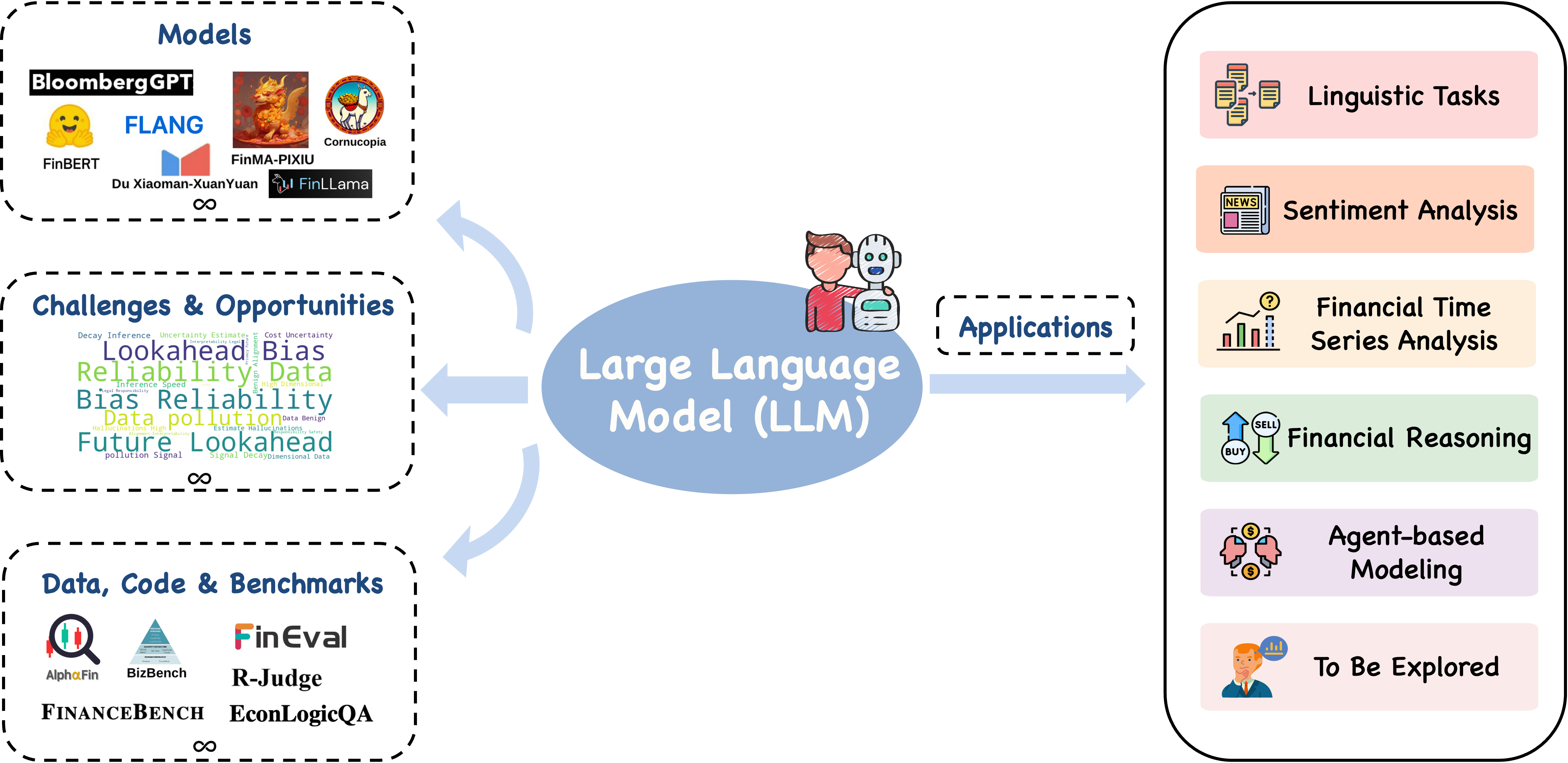} % Uncomment and replace with actual image
    % \fbox{\rule{0pt}{2in} \rule{0.9\linewidth}{0pt}} % Placeholder box
    \caption{An overview of our paper structure, focusing on models, applications, data, code and benchmarks, and challenges and opportunities.}
    \vspace{-1em}
    \label{fig:overview}
\end{figure*}

\section{Introduction}
\label{sec:introduction}

The financial domain has always been characterized by complexity, uncertainty, and rapid evolution. With the advent of technology, the integration of advanced computational models in finance has gained significant momentum \cite{mulvey2022applications}. Among these advancements, large language models (LLMs) have emerged as a powerful tool, demonstrating remarkable capabilities in understanding context, processing vast amounts of data, and generating human-like text. The application of LLMs in finance promises to transform traditional practices, drive innovation, and unlock novel opportunities across various financial tasks.

LLMs, such as GPT-series, BERT, and their financial-specific variants like FinBERT, have shown impressive performance in natural language processing (NLP) tasks. These models leverage sophisticated algorithms and extensive pre-training on vast datasets to achieve advanced contextual understanding, customization capabilities, and scalability for real-time analysis. Their ability to detect complex emotional states and provide accurate interpretations makes them particularly valuable in the financial sector, where understanding market sentiment and making informed decisions are crucial.

In recent years, the financial domain has witnessed a growing interest in applying LLMs across various applications. These applications are not only reshaping the landscape of financial analysis but also offering new perspectives on market behavior and economic activities. For instance, in \textit{Linguistic Tasks}, LLMs excel in summarizing and extracting key information from extensive financial documents, thereby streamlining complex financial narratives into concise summaries and enabling more efficient information processing. \textit{Sentiment Analysis}, as one of the most crucial applications in finance, has been widely explored for decades. The advancement of LLMs has made them pivotal in quantifying market sentiment from financial news, social media, and corporate disclosures, thereby providing critical insights that influence market movements and investment decisions. Additionally, LLMs have shown potential capabilities in \textit{Financial Time Series Analysis}, including forecasting market trends, detecting anomalies, and classifying financial data, although their efficacy remains under debate. These models aim to enhance prediction accuracy and robustness by leveraging their deep learning architecture to capture complex temporal dependencies and patterns within financial datasets. One of the most promising areas of research where LLMs distinctly surpass previous deep learning methods is their capability of reasoning, which enables them not only to fit the data but also to emulate reasoning processes similar to human cognition. In \textit{Financial Reasoning}, LLMs support financial planning, generate investment recommendations, and assist in decision-making by processing and synthesizing vast amounts of financial data from diverse sources. Leveraging their ability to imitate human decision-making processes, LLMs are further applied in \textit{Agent-based Modeling}. This application extends the reasoning capabilities of LLMs to interactions between agents and their environments, markets, and humans, enabling the simulation of market behaviors, economic activities, and the dynamics of financial ecosystems.

\begin{table}[!t]
            \centering
	    \caption{Comparison between our survey and related surveys. Circles indicate areas covered but lacking extensive detail.}
            \vspace{-1em}
           \setlength\tabcolsep{3 pt}
    	\resizebox{\linewidth}{!}{
    		\begin{tabular}{cccccc}
    			\toprule
    			Survey & Financial LLMs & Benchmarks & Applications & Challenges\\
    			\midrule
    			  \citet{lee2024survey} & \ding{52} & \ding{52} & \ding{109} & \ding{109} \\
    			  \citet{li2023large} & \ding{52} & \ding{56} & \ding{109} & \ding{109} \\
                    \citet{dong2023scoping} & \ding{56} & \ding{56} & \ding{52} & \ding{109}
                   \\
                    \citet{zhao2024revolutionizing}  & \ding{56} & \ding{52} & \ding{52} & \ding{109}\\
    			\midrule
    			\midrule
    			  Ours  & \textbf{\ding{52}} & \textbf{\ding{52}}  & \textbf{\ding{52}} & \textbf{\ding{52}} \\
    			\bottomrule
    		\end{tabular}
    	}
             \vspace{-1.5em}
\end{table} 

Despite the promising advancements, the application of LLMs in finance also presents several challenges, such as lookahead bias in backtesting, legal concerns surrounding robot-generated content, data pollution, signal decay, inference speed, cost, uncertainty estimation, dimensionality considerations, interpretability, legal responsibility, safety, and privacy. Addressing these challenges is essential to ensure the ethical and effective deployment of LLMs in financial applications.
\\~\\
\textbf{Related Work:} Recently, several surveys have explored the applications of LLMs in the financial domain. For instance, \citet{lee2024survey} present an overview of financial LLMs from the model perspective. \citet{li2023large} review the current approaches employing LLMs in finance and propose a decision framework to guide their adoption. \citet{dong2023scoping} provide a scoping review on ChatGPT and related LLMs in the fields of accounting and finance. \citet{zhao2024revolutionizing} focus on the integration of LLMs into a variety of financial tasks. 

Despite these contributions, existing surveys often lack a deep dive into the practical challenges and opportunities specific to finance, or they focus primarily on the technical aspects without addressing the broader implications for financial decision-making and industry practices. This survey aims to fill these gaps by not only reviewing the state-of-the-art but also presenting a detailed analysis of specialized models, useful benchmarks, innovative applications, and fundamental challenges. Our work uniquely positions itself by providing a holistic view that is driven by \textbf{real-world applications in finance}, thus offering valuable insights for both researchers and practitioners.
\\~\\
\textbf{Contributions:} Our main contributions include:
\begin{itemize}[leftmargin=15pt]
    \item \textit{Holistic View of Financial Applications and Practical Insights.} Our survey bridges the gap between academic research and practical implementation by providing a thorough examination of LLM applications in finance. This holistic view ensures relevance to both researchers and practitioners, highlighting the transformative potential of LLMs in diverse financial tasks.
    \item \textit{Comprehensive Coverage of Models, Data, and Benchmarks.} We examine specific LLMs for financial applications, analyzing their architecture, pre-training methods, and customization. We also analyze the datasets and benchmarks, providing a valuable collection of resources.
    \item \textit{Novel Challenges and Opportunities.} Our survey addresses unique challenges in applying LLMs to finance, such as lookahead bias, legal concerns, data pollution, and interpretability. We explore potential solutions and future research directions, providing a foundation for further development in the financial sector. 
\end{itemize} 
\vspace{1cm}
\textbf{Paper Organization:} The paper is structured as follows: In Section \ref{sec: models}, we discuss the various LLMs that are specifically designed or fine-tuned for financial applications. Section \ref{sec: app} provides a comprehensive survey on various application areas, including linguistic tasks, sentiment analysis, financial time series analysis, financial reasoning, and agent-based modeling. Section \ref{sec: data} delves into the data, code, and benchmarks available for financial LLM research. Finally, Section \ref{sec: challenges} explores the challenges and opportunities associated with the deployment of LLMs in finance. This survey aims to provide a comprehensive overview of the current state of LLM applications in finance, highlighting the progress, prospects, and challenges. By presenting a detailed survey of the current landscape, we hope it to facilitate the adoption and further development of LLMs in the financial sector, paving the way for innovative solutions and enhanced decision-making processes.

\section{Models}
\label{sec: models}
\subsection{Collections of Models}
LLMs have demonstrated remarkable capabilities across a wide range of domains \cite{wu2023bloomberggpt}, \cite{liu2023medical}, \cite{wang2024large}. While general-domain LLMs such as GPT-series, Llama-series, and BERT have shown impressive performance on various NLP tasks, there has also been growing interest in developing financial domain-specific LLMs. These specialized models are trained on vast amounts of financial data, allowing them to better understand and generate content related to finance, economics, and business. In this section, we will introduce several prominent financial domain-specific LLMs, discussing their strengths, limitations, and potential applications in downstream financial tasks.
\\~\\
\textbf{GPT-series:} 
 One of the most well-known general-domain LLMs is the GPT (Generative pre-trained transformers) series, developed by OpenAI \cite{radford2018improving}, \cite{radford2019language}, \cite{brown2020language}, \cite{achiam2023gpt}. GPT models, based on the transformer architecture, leverage self-attention mechanisms and positional embeddings to capture long-range dependencies in text. Recently, \textbf{Ploutos} \cite{tong2024ploutos}, a novel financial LLM framework derived from GPT-4, has been proposed for interpretable stock movement prediction. Ploutos consists of two main components: PloutosGen and PloutosGPT. PloutosGen addresses the challenge of fusing textual and numerical information by integrating multimodal data through a diverse expert pool, including sentiment, technical, and human analysis experts, which generate quantitative strategies from different perspectives. On the other hand, PloutosGPT tackles the lack of clarity in traditional methods by using rearview-mirror prompting, which leverages historical stock data and expert analysis to guide the model, and dynamic token weighting to generate accurate and interpretable rationales for stock predictions. While Ploutos demonstrates enhanced prediction accuracy and interpretability, it is constrained by potential expert selection bias, computational complexity, and limited data types. Future research could potentially focus on optimizing efficiency, expanding data variety, and mitigating biases to further improve the framework’s performance.
\begin{figure*}[htbp!]
    \centering
    \includegraphics[width=.9\linewidth]{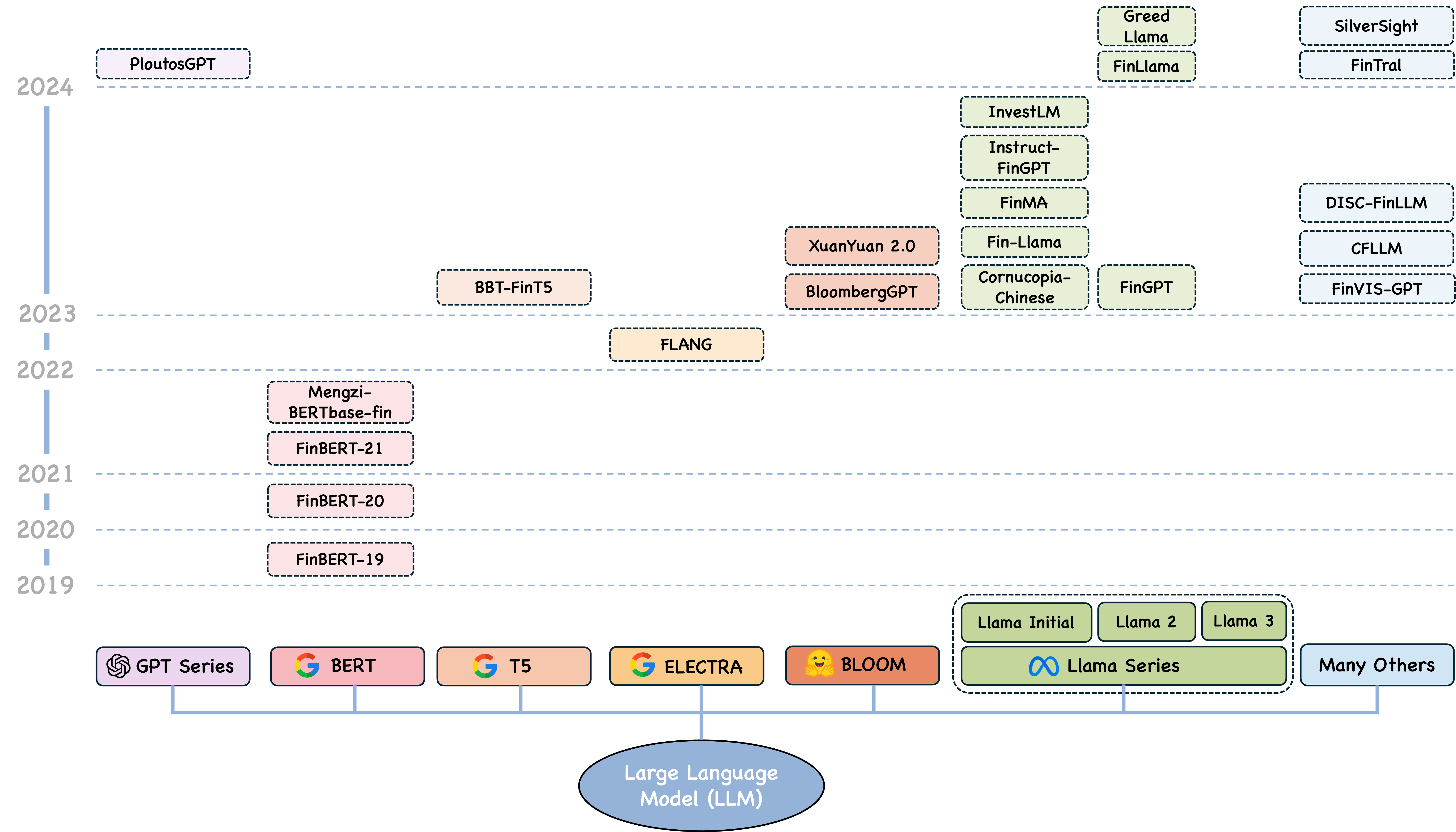} % Uncomment and replace with actual image
    % \fbox{\rule{0pt}{2in} \rule{0.9\linewidth}{0pt}} % Placeholder box
    \caption{Overview of financially specialized large language models (LLMs) from 2019, categorized by their foundational model types and many others.}
    \vspace{-1em}
    \label{fig:model}
\end{figure*}
\\~\\
\textbf{BERT:} 
In 2018, BERT (Bidirectional Encoder Representations from Transformers) \cite{devlin2018bert} revolutionized the field of NLP with its deep bidirectional architecture that could learn contextual representations. This breakthrough led to the development of several domain-specific variants, particularly in the financial sector. Building upon BERT’s foundation, \textbf{FinBERT-19} \cite{araci2019finbert} was developed by continually pre-training BERT on financial text to enhance its sentiment analysis capabilities. The following year, \textbf{FinBERT-20} \cite{yang2020finbert} further evolved this approach by conducting domain-specific pre-training from scratch, focusing solely on financial communications and utilizing a large-scale financial corpus. In 2021, \textbf{FinBERT-21} \cite{liu2021finbert} introduced a mixed-domain pre-training strategy, leveraging both general corpora (Wikipedia and BooksCorpus) and financial domain corpora (FinancialWeb, YahooFinance, and RedditFinanceQA). By simultaneously training on general and financial domain corpora, FinBERT-21 aims to capture a broader range of language knowledge and semantic information relevant to financial text mining. These FinBERT models have demonstrated their effectiveness in various financial downstream tasks, such as sentiment analysis, named entity recognition, question answering, and text classification within the financial domain. In addition to the Fin-BERT models mentioned above, RoBERTa \cite{liu2019roberta}, introduced in 2019, is another variant of BERT. \textbf{Mengzi-BERTbase-fin} \cite{zhang2021mengzi}, trained with 20GB of financial news and research reports, is a specialized version of RoBERTa designed for financial applications.
\\~\\
\textbf{T5:} 
In 2019, Google introduced the Text-to-Text Transfer Transformer (T5) \cite{raffel2020exploring}, a unified framework that treats every text processing task as a “text-to-text” problem. This model utilizes an encoder-decoder architecture and is pre-trained using a self-supervised learning objective called “span corruption”. This involves randomly masking contiguous spans of text in the input sequence and training the model to reconstruct the original text. Building on this, the \textbf{BBT (Big Bang Transformer)-FinT5} \cite{lu2023bbt} was developed specifically for the Chinese financial sector. This model incorporates knowledge-enhanced pre-training methods and is built on the BBT-FinCorpurs – a large-scale financial corpus comprising diverse sources, including corporate reports, analyst reports, social media, and financial news. BBT-FinT5 benefits from the text-to-text framework of T5, allowing it to tackle both language understanding and generation tasks within the financial domain. However, being a domain-specific model, its performance on general NLP tasks outside of finance might be limited. BBT-FinT5 can be fine-tuned for various financial applications including news classification, summarization, relation extraction, sentiment analysis, and event-based question answering.
\\~\\
\textbf{ELECTRA:} 
In 2020, ELECTRA \cite{clark2020electra} introduced an innovative generator-discriminator framework for pre-training language models. The model improves efficiency by training the discriminator to distinguish between real and synthetically generated tokens. Building upon this, researchers developed \textbf{FLANG} \cite{shah2022flue}, a specialized variant of ELECTRA tailored for the financial domain. FLANG integrates specific adaptations such as selective token masking and span boundary objectives to effectively handle the complexities of financial language. While FLANG excels in handling financial terminology and delivers enhanced performance on tasks such as sentiment analysis and entity recognition within financial documents, its specialization may limit its effectiveness in non-financial contexts without further fine-tuning. Despite this limitation, FLANG has demonstrated its value in various downstream financial tasks. It enables precise analysis of market reports, accurate classification of financial headlines, and reliable identification of key financial entities.
\\~\\
\textbf{BLOOM:} 
In 2022, BLOOM \cite{le2023bloom} was released as a fundamental multilingual LLM with 176 billion parameters. It was pre-trained on a vast corpus of text that included 46 natural languages and 13 programming languages. BLOOM is notable for its diversity and accessibility as an open-source model that supports a variety of languages. From BLOOM, specialized versions focused on financial applications have been created, including \textbf{BloombergGPT} \cite{wu2023bloomberggpt} and \textbf{XuanYuan 2.0} \cite{zhang2023xuanyuan}. BloombergGPT, with its 50 billion parameters, was designed for the financial sector by training on Bloomberg’s financial data sources. This model demonstrates enhanced performance on specific financial tasks while maintaining competitive overall competence. XuanYuan 2.0, created for the Chinese financial market, is a large open-source Chinese financial chat model. It proposes a novel hybrid-tuning strategy that combines general and financial-specific data, allowing the model to retain general language capabilities while excelling at domain-specific tasks such as financial advisory and market analysis. This strategy lowers the likelihood of catastrophically forgetting previous knowledge and enhances accuracy on finance-related tasks.
\\~\\
\textbf{Llama-series:} 
Llama \cite{touvron2023llama}, an LLM introduced in 2023, offers flexibility with model sizes ranging from 7B to 65B parameters. Trained on publicly available datasets for transparency, Llama outperforms larger models, including GPT-3, on most benchmarks despite its smaller size. Its financial variants, which include \textbf{FinMA} \cite{xie2023pixiu}, \textbf{Fin-Llama} \cite{Fin-LLAMA}, \textbf{Cornucopia – Chinese} \cite{Cornucopia-LLaMA-Fin-Chinese}, \textbf{Instruct-FinGPT} \cite{zhang2023instruct} and \textbf{InvestLM} \cite{yang2023investlm}, provide specialized capabilities for various financial tasks. Among them, InvestLM, based on LLaMA-65B and a diverse investment-related dataset, offers investment recommendations comparable to cutting-edge commercial models. Llama 2 \cite{touvron2023llama2}, which was released later, included various enhancements over Llama, including a 40\% larger pretraining corpus, a doubled context length, and the adoption of grouped-query attention for improved inference scalability. It has financial variants such as \textbf{FinGPT} \cite{yang2023fingpt}, \textbf{FinLlama} \cite{konstantinidis2024finllama}, and \textbf{GreedLlama} \cite{yu2024greedllama}. Particularly, FinGPT is an open-source model that focuses on providing accessible and transparent resources for developing financial LLMs. Despite having relatively small training data compared to BloombergGPT, FinGPT claims to offer a more accessible, flexible, and cost-effective solution for financial language modeling. In April 2024, Meta introduced Llama 3 \cite{meta2024llama3}, featuring 8B and 70B parameter models that showcase state-of-the-art performance and improved reasoning capabilities, marking them as the most capable openly available LLMs to date. The LLM community is visibly excited, and we expect more Llama 3 variants for financial LLM models to emerge soon.
\\~\\
In addition to the models mentioned above, there are also other financial domain-specific LLMs such as FinTral \cite{bhatia2024fintral}, driven from Mistral 7B \cite{jiang2023mistral}; SilverSight \cite{zhou2024silversight}, based on the Qwen 1.5-7B chat model \cite{bai2023qwen}; DISC-FinLLM \cite{chen2023disc}, used Baichuan-13B \cite{baichuan2023} as the backbone; CFLLM \cite{li2023cfgpt}, based on InternLM-7B \cite{internlm}; FinVIS-GPT \cite{wang2023finvis} which is a multimodal LLM for financial chart analysis, based on LLaVA \cite{liu2023llava}. These domain-specific LLMs utilize vast financial datasets and advanced training techniques to provide more accurate and context-aware financial analysis than general-domain models. As research in this area continues to progress, we expect the development of even more sophisticated financial LLMs that could transform various sectors of the financial industry, including investment strategies, risk management, forecasting, and customer service. However, it’s crucial to acknowledge the limitations and potential biases of these models and to employ them thoughtfully alongside human expertise and judgment.

% Discuss the pros and cons of them.

% Categorize LLMs.

% FinBERT-19 - A specialized version of BERT fine-tuned for financial sentiment analysis to understand the nuances of language used in the financial domain. BERT variants [2019.09]

% FinBERT-20 - An adaptation of BERT tailored for financial sentiment analysis, updated to better capture the nuances of financial communication. [2020.07] BERT variants

% FinBERT-21 - Further refinement of FinBERT for financial text, incorporating more recent data and advancements in NLP to improve performance. [2021.01] BERT variants

% FLANG - A variant focusing on fine-tuning and optimizing ELECTRA models for specific tasks or languages. [2022.10] ELECTRA variants

% BloombergGPT - [2023.03] BLOOM variants

% XuanYuan2.0 - BLOOM variants

% Fin-T5 - T5 Variants

% FinMA - [2023.06] LLaMA variants

% Instruct-FinGPT - LLaMA variants

% Fin-LLaMA - LLaMA variants

% Cornucopia (Chinese) - LLaMA variants

% InvestLM - [2023.09] LLaMA variants

% FinGPT - [2023.10] LLaMA2 variants

\subsection{Zero-shot vs Fine-tuning}

Zero-shot and fine-tuning are two distinct adaptation methods in the applications of LLMs. Zero-shot (or few-shot) learning refers to the ability of a model to correctly predict or perform tasks it has not explicitly been trained to handle, based on its pre-existing knowledge and generalization capabilities. Fine-tuning, on the other hand, involves adjusting a pre-trained model on a specific dataset or for a particular task to improve its accuracy and performance on that task \cite{li2023large}. 

%Fine-tuning has emerged as a prevalent method for adapting LLMs to specific downstream tasks and domains, which involves adjusting the parameters of an already pretrained model on a targeted dataset. 

Fine-tuning is favored when domain-specific accuracy is essential, adaptability to real-time changes is required, or when customization and privacy are critical considerations. In practice, the integration of financial-related text data is a common approach in fine-tuning LLMs. \citet{araci2019finbert} develops FinBERT, a tailored version of the BERT language model, achieved through extended pre-training on a comprehensive financial dataset, including news, articles, and tweets, alongside strategic fine-tuning methods. FinBERT sets a new benchmark in financial-related text analysis, eclipsing earlier deep learning methodologies in the field. 

Several technologies have been proposed to make fine-tuning more efficient. Instruction tuning \cite{wei2021finetuned} is a fine-tuning method for language models where the model is trained to follow specific instructions, not only improves performance on the target tasks but also enhances the model's zero-shot and few-shot learning capabilities, making it popular among various financial applications and models. \citet{zhang2023instruct} propose an instruction-tuned FinGPT model that enhances the financial sentiment analysis capabilities of LLMs by employing instruction tuning, which transforms a small portion of supervised financial sentiment data into instruction data, thereby improving the model's numerical sensitivity and contextual understanding. Furthermore, \citet{zhang2023enhancing} integrate instruction-tuned LLMs with a retrieval-augmentation module, which is a technique that enhances language models by supplementing their input with relevant information retrieved from external sources, to enhance the models' predictive performance by providing a richer context. Besides instruction tuning, people have also applied low-rank adaptation (LoRA) \cite{hu2021lora} or quantized LLMs \cite{ma2024era}, \cite{dettmers2024qlora} for more efficient adaptation on financial tasks, such as FinGPT \cite{zhang2023instruct}, FinGPT-HPC \cite{liu2024fingpt} and Llama-based models \cite{pavlyshenko2023financial}.

Another prevalent approach involves the consideration of smaller models, as energy efficiency and the lightweight nature of models are crucial in today's machine learning landscape \cite{yakar2019energy}, \cite{nie2021neural}, \cite{wang2022novel2}. \citet{rodriguez2024large} demonstrate that smaller LLMs can be effectively fine-tuned on financial documents and instructions to achieve comparable or superior performance to larger models. \citet{deng2023llms} presents a case study on utilizing an LLM for semi-supervised financial sentiment analysis on Reddit data, where the LLM generates weak sentiment labels through in-context learning and chain of thought reasoning, which are then used to train a smaller model for production use, achieving competitive performance with minimal human annotation. 

While pre-training and fine-tuning allow these models to adapt to the specific linguistic characteristics and styles of various applications, zero-shot learning is preferred when labeled data is limited, rapid deployment is crucial, or when modular development and interpretability are prioritized. 
The zero-shot and few-shot capabilities of LLMs underscore their efficiency by allowing for direct application without the need for extensive dataset-specific training. This efficiency is due to the transfer learning from the vast datasets on which LLMs are trained, as well as their emergent abilities to generate new insights or address unexpected problems during information processing \cite{wei2022emergent}. These features significantly broaden their usefulness across various fields without the need for further training. 
For instance, \citet{steinert2023linking} explore the zero-shot capability of GPT-4 to predict same-day stock price movements of Apple and Tesla in 2017 with microblogging messages, and by comparing its performance to BERT, they highlight the importance of prompt engineering in extracting sophisticated sentiments from GPT-4 for financial applications.

\subsection{Why Applying LLMs in Finance?}
\label{sec: why llm}
% Some general points:
% \begin{itemize}
%     \item Advanced Contextual Understanding: LLMs' ability to grasp context, including financial jargon and nuanced expressions, greatly enhances sentiment analysis accuracy. 
%     \item Customization for Financial Analysis: LLMs can be customized to detect sentiments specific to various financial instruments or market conditions, offering tailored analysis capabilities.
%     \item Transfer Learning Flexibility: Pre-trained on extensive internet text, LLMs bring a broad understanding of language that can be fine-tuned for specific financial sentiments, reducing the need for large domain-specific datasets. LLMs can apply knowledge learned from one domain to another, enabling efficient adaptation to financial sentiment analysis with minimal domain-specific training data.
% \end{itemize}

% Some specific points:
% \begin{itemize}
%     \item Scalability for Real-time Analysis: Their ability to quickly process large volumes of text allows LLMs to offer timely sentiment insights essential for fast-paced financial decision-making.
%     \item Advanced Emotional Detection. Beyond basic sentiments, LLMs can identify complex emotional states (e.g. sarcasm, subtle sentiments) in financial texts and social media, providing deeper insights into market sentiments. 
%     \item benefit of attention in time series (like in language)
%     \item interpretability
% \end{itemize}
The integration of LLMs in financial analysis represents a revolutionary transformation in how data-driven decisions are made within the financial sector. These models' unique capabilities are driven by advanced machine learning techniques that interpret and process natural language at an unprecedented scale and complexity. Here, we delve into the core reasons for leveraging LLMs in financial applications, emphasizing both general and specific advantages.
\\~\\
\textbf{Advanced Contextual Understanding:} LLMs are distinguished by their profound ability to understand context. This includes a comprehensive understanding of financial terminologies, jargon, and refined expressions. Such advanced contextual understanding significantly enhances the accuracy of sentiment analysis, a critical aspect when dealing with the complex and often ambiguous language found in financial documents and news articles.
\\~\\
\textbf{Transfer Learning Flexibility:} LLMs are initially pre-trained on a vast corpus of internet text, encompassing a wide range of topics and languages. This pre-training equips LLMs with a broad understanding of language, which can then be fine-tuned for specific financial tasks. Such flexibility in transfer learning reduces the reliance on large, domain-specific datasets, allowing for efficient adaptation to new tasks with minimal domain-specific training data in finance.
\\~\\
\textbf{Scalability for Real-time Analysis:} The financial market's fast-paced nature demands tools that can offer timely insights. LLMs excel in processing large volumes of text rapidly, enabling real-time reasoning and sentiment analysis. This capability ensures that financial decision-makers can receive immediate insights from news articles, market information, reports, and social media, facilitating more informed and timely decisions.
\\~\\
\textbf{Multimodality:} Multimodal capabilities of LLMs extend their application beyond text to include other data forms such as images, audio, and structured data \cite{jiang2023re}, \cite{wang2023sst}. In finance, this is particularly useful for integrating various data sources, such as text from news articles, numerical data from financial statements, and visual data from market charts. For instance, combining textual analysis of news with visual analysis of stock price movements can provide a more comprehensive understanding of market trends and investor sentiment. This integration of different data types enhances the robustness and depth of financial analysis.
\\~\\
\textbf{Interpretability:} While deep learning models are often seen as 'black boxes', LLMs' ability to generate human-like outputs opens doors to explainability. This characteristic facilitates the provision of both results and their underlying explanations, thereby enhancing the comprehensibility of the reasoning processes within LLMs, and increasing trust and transparency in their financial applications.
\\~\\
\textbf{Customization:} LLMs exhibit a significant degree of adaptability, enabling customization to accommodate specific financial instruments or market conditions. By integrating domain-specific data and parameters, LLMs can be trained to focus on particular aspects of financial markets, such as risk assessment for bonds or trend prediction in stock markets. This approach enhances the analytical capabilities of LLMs, allowing them to generate insights that are finely tuned to the complexities of different financial environments.

\section{Applications}
\label{sec: app}

\subsection{Linguistic Tasks}
\label{sec:linguistic}

\subsubsection{Textual Work}
Many earlier models, such as those based on Recurrent Neural Networks (RNNs), specifically Long Short-Term Memory (LSTM), have demonstrated a capacity for achieving a degree of language understanding over text sequences and performing textual work \cite{lipton2015critical}. However, due to these models’ architectural constraints, they struggled with long-term dependencies. Specifically, they encountered challenges in maintaining context over long text sequences, understanding complex expressions, dealing with large datasets and handling unstructured data efficiently \cite{lipton2015critical}, \cite{staudemeyer2019understanding}. This limitation is particularly evident when applying in financial sector, where the volume of documentation is vast, and the need for accurate and concise summaries is critical \cite{zmandar2021joint}. 
\\~\\
LLMs, which leverage the transformer model architecture, on another hand, have significantly advanced the field’s capabilities. The transformer architecture, characterized by its innovative self-attention mechanism, allows LLMs to process, understand and generate text based on massive datasets they have been trained on \cite{hadi2023large}, \cite{raiaan2024review}. This breakthrough is instrumental in overcoming the challenges faced by earlier models. By efficiently managing long-term dependencies and contextual information over large volumes of text, LLMs can streamline complex financial narratives into concise summaries and extract relevant information \cite{hadi2023large}, \cite{raiaan2024review}. This process retains essential insights and enables more efficient information processing. 
\\~\\
\textbf{Summarization and Extraction:}
Recent research has effectively utilized LLMs to summarize and extract financial document information \cite{abdaljalil2021exploration}, \cite{la2020end}, \cite{ni2023unified}. Given that these financial documents are often lengthy, which can exceed the token limits of many LLMs, various studies have introduced frameworks by dividing long documents into shorter segments or utilized specific models to address the challenges of processing extensive financial texts \cite{xia2022fetilda}, \cite{vanetik2023summarizing}. Recently, \citet{yepes2024financial} propose an expanded approach to chunking documents for Retrieval Augmented Generation (RAG) by using structural elements rather than just paragraphs, which improves chunk size determination without tuning. Furthermore, some papers propose segmenting long reports into ten distinct sections, such as management’s discussion and analysis, financial highlights, and business overview, to streamline the summarization process \cite{shukla2022dimsum}, \cite{shukla2023generative}. Similarly, \citet{khanna2022transformer} utilize the Longformer-Encoder-Decoder (LED) model, a transformer architecture first introduced by \citet{beltagy2020longformer}, which employs a self-attention mechanism scalable with sequence length, making it suitable for analyzing long financial reports. 

\begin{figure}[t]
    \centering
    \includegraphics[width=\linewidth]{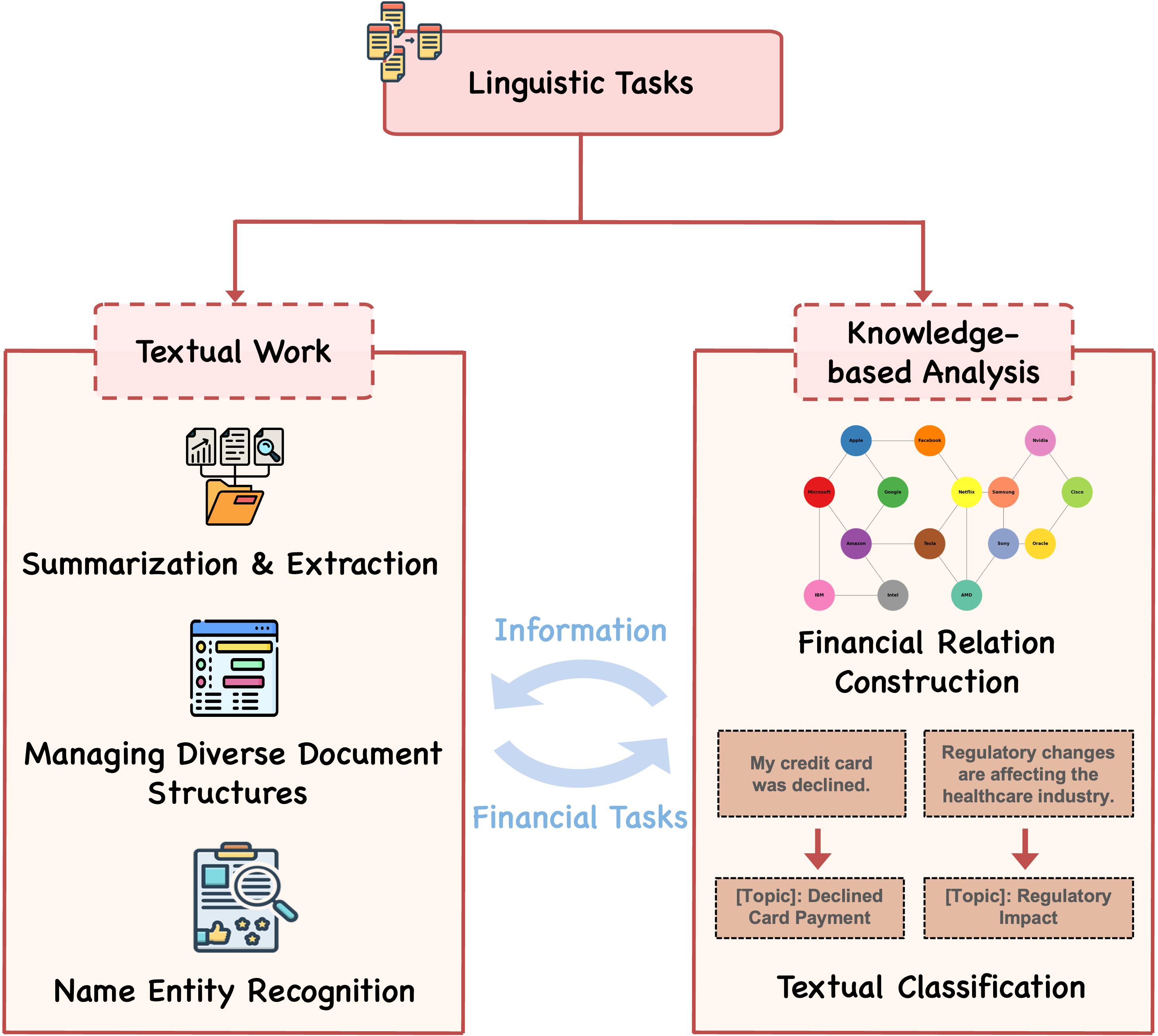} % Uncomment and replace with actual image
    % \fbox{\rule{0pt}{2in} \rule{0.9\linewidth}{0pt}} % Placeholder box
    \caption{Illustration of various linguistic tasks in finance.}
    \label{fig:lingusistic tasks}
    \vspace{-1em}
\end{figure}

Beyond handling the long size of the document, research has expanded into multilingual and domain-specific challenges. This includes summarizing financial documents across multiple languages \cite{foroutan2022multilingual}; customizing language models to tackle the adaptation challenges to Japanese financial terminology \cite{suzuki2023constructing}; automating the finetuning process for text summarization models in the cryptocurrency domain without requiring human annotation \cite{avramelou2023domain}; adopting multitask learning strategies to classify, detect and summarize financial event \cite{li2021unified}; addressing the challenge of ensuring accuracy and reducing errors in financial information extraction \cite{sarmah2023towards}; extracting information from annual report to enhance stock investment strategies \cite{gupta2023gpt}.  
\\~\\
\textbf{Managing Diverse Document Structures:}
Despite the effectiveness of LLMs in handling textual financial data, they often face challenges with PDF document formats that incorporate images, charts, and tables. This challenge may arise from their primarily text-based nature, which finds it difficult to interpret the complex spatial layouts crucial for understanding such multimodal documents \cite{li2023extracting}. One simple approach to this issue involves converting PDF files into machine-readable plain text. For instance, in the Automated Financial Information Extraction (AFIE) framework proposed by \citet{yue2023leveraging}, tables are transformed into text using PLAIN serialization. This method uses spaces and newline characters to separate cells and rows, respectively. This effectively integrated table data with regular paragraphs for LLMs to process uniformly. 

However, this conversion process may change the document’s spatial layout and potentially lead to the loss of crucial information embedded in charts or tables. To address this, the team at JP Morgan developed DocLLM \cite{wang2023docllm}, a layout-aware generative language model specifically designed for multimodal document understanding. DocLMM utilizes bounding box information to understand the spatial arrangement of elements within the documents. It enhances document understanding by modifying the attention mechanism in transformers to concentrate on the cross-alignment between textual and spatial modalities. 
\\~\\
\textbf{Name-Entity Recognition:}
Name-entity recognition (NER) is a subtask of information extraction and plays a crucial role in extracting meaningful information from various financial sources \cite{li2020survey}, \cite{ehrmann2023named}. In the financial domain, it is used to extract specific entities such as company names, financial terminology, stock symbols, financial indicators, monetary values from news articles, financial reports and market summaries \cite{swaileh2020named}. This information is crucial for financial downstream tasks, such as industry classification, sentiment analysis, credit scoring, fraud detection and regulatory compliance reporting \cite{alvarado2015domain}. 

Traditionally, NER is approached through Rule-Based Methods, Machine Learning Techniques, or Deep Learning Techniques \cite{nadeau2007survey}. Rule-based methods depend on handcrafted linguistic and grammatical rules. They offer high precision for well-defined patterns but suffer from limited scalability \cite{li2020survey}. Machine Learning Techniques include both supervised and unsupervised approaches. Supervised approaches utilize a comprehensive set of engineered features, such as word-level characteristics and list lookups, alongside machine learning algorithms, such as Hidden Markov Models \cite{EDDY1996361}, Decision Trees \cite{Induction_of_decision_trees} and Support Vector Machine \cite{support_vector_machines}, to identify and classify entities in text. Unsupervised learning approaches extract and classify named entities by employing clustering, leveraging lexical resources and patterns, and analyzing corpus statistics \cite{nadeau2007survey}. While machine learning offers flexibility and can handle diverse data types, it relies heavily on the availability of labeled data for supervised learning and can lack interpretability in unsupervised learning \cite{li2020survey}. Deep learning methods utilize advanced architectures such as Bidirectional Long Short-Term Memory (BiLSTM) networks, self-attention-based transformers, and Conditional Random Fields (CRF) for tag decoding to effectively learn and represent both word and character-level features from large datasets. These approaches significantly enhance model performance by enabling the capture of complex patterns and long-range dependencies in text \cite{li2020survey}.

With the emergence of deep learning methods, LLMs are now increasingly used in NER within the financial domain \cite{pakhale2023comprehensive}, \cite{wang2023named}. The ability of LLMs to leverage vast pre-trained knowledge and sophisticated language understanding can significantly enhance the accuracy and efficiency of entity recognition in complex financial texts \cite{pakhale2023comprehensive}. Recently, \citet{hillebrand2022kpi} propose KPI-BERT, a new system that utilizes advanced techniques NER and Relation Extraction (RE) to identify and connect key performance indicators (KPIs) such as "revenue" or "interest expenses" within German financial documents. This system relies on an end-to-end trainable architecture based on BERT. It combines an RNN with conditional label masking for sequential entity tagging, followed by relation classification. Further research has utilized LLMs for NER to improve the efficiency and accuracy of XBRL (eXtansive Business Reporting Language) tagging \cite{loukas2022finer}; identify similar peer companies \cite{covas2023named}; detect key entities of negative news information \cite{zhao2021bert}; extract relevant phrases for entities \cite{gupta2021context}. 

Despite LLMs have demonstrated exceptional generalization capabilities, they sometimes come with high training and inference costs, especially when processing long financial documents. To address these issues, \citet{zhou2023universalner} propose UniversalNER, a model that employs targeted distillation with mission-focused instruction tuning to train cost-efficient student models for open NER. This approach not only reduces the computational burden but also achieves remarkable NER accuracy without direct supervision.

\subsubsection{Knowledge-based Analysis}

In financial text analysis, summarizing and extracting key information from documents is crucial for quickly understanding and processing important data within lengthy and complex texts \cite{xue2023weaverbird}. Following the extraction of pertinent information, the next step involves utilizing this information for solving downstream financial tasks. This section will introduce two main activities central to this application: constructing financial relationships and textual classification. These efforts are vital for leveraging the extracted information to enhance decision-making and analytical processes in the finance sector. 
\\~\\
\textbf{Financial Relation Construction:}
Constructing financial relationships, particularly through the use of knowledge graphs, represents a powerful methodology for organizing and making sense of the extracted entities and their interrelations from extensive and complex financial datasets \cite{Deloitte2020KnowledgeGraphs}. Knowledge graphs consist of interconnected descriptive structures about entities (objects, events, people, etc.), the attributes of those entities, and the relationships that link them together. This framework offers a structured way of representing relationships within data and enables sophisticated analyses to be derived from them \cite{jiang2023evolution}, \cite{pan2023large}.

Upon the identification and extraction of entities (such as companies, individuals, financial instruments, events, etc.), along with the relationships between these entities (such as ownership, transactions, legal disputes, etc.), this information can be systematically organized into a graph format for further construction. Within a knowledge graph, entities are represented as nodes, and relationships are denoted as edges that connect these nodes. This structure provides a visual and programmable method to explore and comprehend the connections among different entities within the financial ecosystem. With the construction of the knowledge graph, financial analysts and systems can employ graph analytics and machine learning algorithms to discover insights, recognize patterns, and predict future events \cite{pan2024unifying}.

Recent advancements in LLMs have led researchers to explore the potential of using information extracted by LLMs to construct and analyze knowledge graphs in the financial sector \cite{trajanoska2023enhancing}, \cite{ouyang2024modal}, \cite{wang2022relation}. Notably, \citet{trajanoska2023enhancing} generate a knowledge graph by leveraging LLMs to extract structured Environmental, Social, and Governance (ESG) information from sustainability reports, using a format of triples consisting of node-edge-node, to enable deeper analysis and understanding of corporate sustainability practices. Similarly, \citet{cheng2022democratizing} develop a Semantic-Entity Interaction Module. This module combines a language model with a Conditional Random Field (CRF) layer to comprehend the interaction between entities and their semantic contexts in texts. It automatically constructs financial knowledge graphs from brokerage research reports without the need for explicit financial knowledge or extensive manual rules. 

Moreover, financial research analysts often face challenges in identifying critical documents, key entities, and important events during their research on complex financial subjects. \citet{mackie2022query} tackle these issues by developing automated methods to create detailed, query-specific knowledge graphs from documents and entities. 

As illustrated above, knowledge graphs have demonstrated their utility in information retrieval. A special case within this domain is the translation of Natural Language (NL) into Graph Query Language (GQL). This process enhanced querying experiences by leveraging the relational data within knowledge graphs, offering advantages over traditional text-to-SQL methods. However, this approach is challenged by the complexity of accurately mapping NL to GQL syntax and the lack of domain-specific examples, making it difficult to fine-tune LLMs for precise alignment with graph databases in specialized fields \cite{pan2023large}. To address this, \citet{liang2024aligning} develop a pipeline that employs LLMs to generate NL-GQL pairs from financial graph databases without labeled data. This process involved creating template pairs with ChatGPT and refining them through a self-instruction method. Subsequently, LLMs were fine-tuned with these pairs using the LoRA technique to align the models with the specific knowledge contained in graph databases.

Knowledge graphs can also be used to significantly enhance question-answering systems. \citet{wang2024knowledge} introduce an innovative Knowledge Graph Prompting (KGP) for multi-document question answering (MD-QA). Their approach constructs a knowledge graph from multiple documents, highlighting semantic or lexical relationships between passages or document structures. An LLM-based graph traversal agent then uses this knowledge graph to gather contextually relevant information, thereby enhancing the LLM’s accuracy in answering questions

Another beneficial aspect of knowledge graphs is their ability to be enriched over time through the use of LLMs. \citet{li2023findkg} presents FinDKG, a dynamic knowledge graph with LLMs used in financial domain. FinDKG incorporates a temporal layer in its structure, which allows it to reflect and adapt to changes in financial markets, economic indicators, and thematic trends. This dynamic approach provides valuable insights for thematic investing, making it possible to identify and leverage long-term industry shifts and economic trends for strategic investment decision-making.

There exist other financial relation extraction studies using LLMs, though not necessarily for knowledge graph construction \cite{ok2023fintree}, \cite{kaur2023refind}, \cite{chai2023fin}, \cite{tian2019chinese}. \citet{ghosh2023mask} propose the Mask One At a Time (MOAT) framework, which masks one entity at a time, extracts contextual embeddings using a domain-specific language model (SEC-BERT), and combines these embeddings with additional features to train a neural network for accurately classifying relationships between financial entities. Similarly, \citet{rajpoot2023gpt} employ in-context learning with GPT models, utilizing both a learning-free dense retriever (KNN with OpenAI embeddings) that relies on the similarity of embeddings to find the most relevant examples, and a learning-based retriever trained to select the most similar example in the training set for each test example by estimating the probability of the output given the input and a candidate training example as the prompt. Focusing on multi-type Chinese financial event relation extraction, \citet{wan2023cfere} propose the CFERE framework, which employs a Core Verb Chain for event identification, constructs a Syntactic Semantic Dependency Parsing graph to combine events into pairs, and enhances BERT with an Event Core Embeddings layer to capture semantic meanings. These studies demonstrate the potential of LLMs and innovative approaches in advancing financial relation extraction, ultimately contributing to the research value of making use of financial information and helping investors make better investment decisions.
\\~\\
\textbf{Textual Classification:}
Textual classification plays a crucial role in organizing and understanding large volumes of unstructured data within the financial domain. This classification task can be further categorized into several sub-tasks, such as industry/company classification and document/topic classification. By effectively classifying and organizing this information, businesses and researchers can extract valuable insights and make informed decisions. The utilization of these classification techniques, in conjunction with the establishment of financial relationships, is essential for leveraging the extracted information to enhance decision-making and analytical processes within the finance sector.

Company or industry classification involves grouping companies into distinct categories based on shared characteristics such as business activities and market performance, with the aim of creating coherent and differentiated groups. Identifying similar company profiles is a fundamental task in finance, with applications spanning investment portfolio construction, securities pricing, and financial risk attribution. Traditionally, financial analysts have relied on industry classification systems, such as the Global Industry Classification System (GICS), the Standard Industrial Classification (SIC), the North American Industry Classification System (NAICS), and the Fama French (FF) model, to identify companies with similar profiles \cite{rizinski2024comparative}. However, these systems do not provide a means to rank companies based on their degree of similarity and require time-consuming, effort-intensive manual analysis and data processing by domain experts \cite{rizinski2024comparative}.

Recently, a team at BlackRock \cite{vamvourellis2023company} explores a novel approach to company classification using LLMs. They investigated the use of pre-trained and fine-tuned LLMs to generate company embeddings based on business descriptions from SEC filings. Their study aimed to assess the embeddings’ ability to reproduce GICS classifications, benchmark LLM performance on various downstream financial tasks, and examine the impact of factors such as pre-training objective, fine-tuning, and model size on embedding quality. The results showed that LLM-generated embeddings, particularly those from fine-tuned Sentence-BERT models, could accurately reproduce GICS sector and industry classifications and outperform them on tasks like identifying similar companies based on return correlations and explaining cross-sectional equity returns.

Interestingly, knowledge graphs can also be used to enrich industry classification and improve the performance of domain-specific text classification tasks. \citet{wang2021enriching} propose a novel Knowledge Graph Enriched BERT (KGEB) model that integrates external knowledge from a local knowledge graph with word representations. They demonstrated the effectiveness of their approach by constructing a large dataset based on companies listed on the Chinese National Equities Exchange and Quotations (NEEQ) and showing that the KGEB model outperforms competitive baselines, including graph convolutional network, Logistic Regression, TextCNN, BERT, and K-BERT, achieving an accuracy of 91.98\% and an F1 score of 90.89\%.

Document or topic classification is another crucial sub-task within the broader scope of textual classification in the financial domain. This task involves categorizing financial documents or texts, such as news articles \cite{mishra2023esg}, \cite{nugroho2021large} or company filings \cite{arslan2021comparison}, \cite{loukas2023making}, into predefined topics or themes. \citet{alias2023financial} propose a novel approach that utilizes the FinBERT model to extract and categorize relevant topics of Key Audit Matters (KAM) from the annual reports of publicly listed companies in Bursa Malaysia. Similarly, \citet{burke2023using} fine-tune the FinBERT model to classify accounting topics within three unlabelled financial disclosures, including custom notes to the financial statements, the Management's Discussion and Analysis section, and the risk factor section.

Another important classification task in the financial domain involves categorizing Environmental, Social, and Governance (ESG) information. This task requires identifying and classifying ESG-related data, such as carbon emissions, diversity and inclusion, and corporate governance practices, from multiple sources including corporate sustainability reports, news articles, and social media posts. In a recent study, \citet{lee2023esg} propose an ESG classifier that can discriminate ESG information by fine-tuning a pre-trained language model. The classifier was trained on a manually labeled dataset constructed from sustainability reports of Korean companies across five sectors and achieved a classification accuracy of 86.66\% for the four-class classification problem (Environment, Social, Governance, and Neutral). Similarly, \citet{mehra2022esgbert} develop a domain-specific language model called ESGBERT to enhance the classification of ESG-related text by fine-tuning BERT’s pre-trained weights using ESG-specific text and further fine-tuning the model for classification tasks.

Textual classification techniques, including industry/company classification and document/topic classification, play a vital role in organizing and understanding large volumes of unstructured data within the financial domain. Recent advancements in LLMs and knowledge graph integration have significantly improved the accuracy and efficiency of these classification tasks. The successful application of these techniques can further provide valuable insights and support informed decision-making in various financial contexts, such as investment portfolio construction, risk assessment, and ESG analysis.

\subsection{Sentiment Analysis}
\label{sec:sentiment}

Sentiment analysis emerges as a crucial component within the domain of NLP and is one of the most important tasks in financial applications. It involves the quantitative exploration of opinions, sentiments, subjectivity, and emotions articulated in textual data \cite{tan2023survey}, \cite{bordoloi2023sentiment}. This task acquires particular significance within financial applications, where the interpretation of market sentiment can lead to impactful forecasting and actions \cite{mishev2020evaluation}. Its evolution mirrors the broader advancements in NLP, transitioning from rule-based systems to sophisticated machine learning models and, more recently, to deep learning approaches that leverage large pre-trained language models. 

\subsubsection{Pre-LLM Sentiment Analysis}
First, we outline the significant milestones in sentiment analysis in this section, leading up to the era before LLMs like ChatGPT and BERT revolutionized the field. Additionally, it highlights key applications within the financial domain, demonstrating the impact of sentiment analysis on various applications. 
\\~\\
\textbf{Lexicon-Based Methods:}
Early sentiment analysis relied on lexicon-based approaches, where the sentiment of a text was inferred based on the presence of predefined words associated with positive or negative sentiments. These methods, simple yet effective for certain applications, include the General inquirer \cite{stone1966general}, the Linguistic Inquiry and Word Count (LIWC) lexicons \cite{pennebaker2001lexicon}, SO-CAL \cite{taboada2011lexicon}, and Loughran and McDonald’s (LM) word lists \cite{loughran2011liability}. 

One of the strengths of lexicon-based methods is their simplicity and interpretability. However, their performance can be limited by the context-dependency of sentiment expressions and the inability to capture the sentiment expressed by complex linguistic constructs such as sarcasm or irony. Despite these limitations, lexicon-based methods have been effectively applied in finance, particularly in analyzing investor sentiment from financial news or social media content \cite{sohangir2018financial}, \cite{yekrangi2021financial}, \cite{consoli2022fine}. 
\\~\\
\textbf{Machine Learning Methods:} 
With the advent of machine learning, financial sentiment analysis (FSA) experienced significant advancements. ML-based methods can be broadly categorized into supervised and unsupervised learning. When doing FSA, supervised learning approaches require labeled data and include techniques such as Support Vector Machines (SVM) \cite{chiong2018sentiment}, Naive Bayes \cite{kalra2019efficacy}, KNN (K-Nearest Neighbor) \cite{kirange2016sentiment}, Random Forests \cite{dickinson2015sentiment} and Multi-layer perceptrons (MLPs) \cite{valencia2019price}. Unsupervised learning, in contrast, does not require labeled data and typically involves clustering techniques to discern sentiment \cite{yadav2020sentiment}. 
%Besides, deep learning models also play an important role.

% \begin{figure}[t]
%     \centering
%     %\includegraphics[width=\linewidth]{example-image} % Uncomment and replace with actual image
%     \fbox{\rule{0pt}{2in} \rule{0.9\linewidth}{0pt}} % Placeholder box
%     \caption{Figure for Sentiment Analysis}
%     \label{fig:placeholder}
% \end{figure}

\tikzstyle{my-box}=[
    rectangle,
    draw=hidden-draw,
    rounded corners,
    text opacity=1,
    minimum height=1.5em,
    minimum width=5em,
    inner sep=2pt,
    align=center,
    fill opacity=.5,
    line width=0.8pt,
]
\tikzstyle{leaf}=[my-box, minimum height=1.5em,
    fill=hidden-pink!80, text=black, align=left, font=\normalsize,
    inner xsep=2pt,
    inner ysep=4pt,
    line width=0.8pt,
]
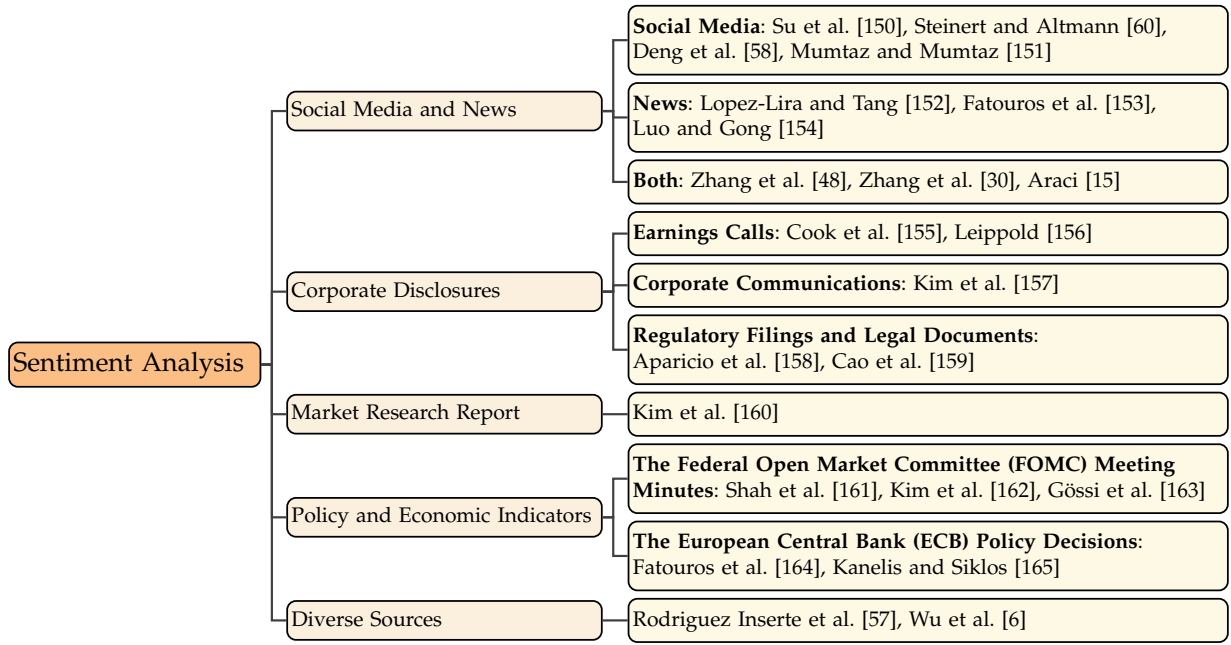
\begin{figure*}[!t]
    \centering
    \resizebox{0.9\textwidth}{!}
    {
        \begin{forest}
            forked edges,
            for tree={
                fill=level0!80,
                grow=east,
                reversed=true,
                anchor=base west,
                parent anchor=east,
                child anchor=west,
                base=left,
                font=\large,
                rectangle,
                draw=hidden-draw,
                rounded corners,
                align=left,
                minimum width=4em,
                edge+={darkgray, line width=1pt},
                s sep=3pt,
                inner xsep=2pt,
                inner ysep=3pt,
                line width=0.8pt,
                text width=11.5em,
                ver/.style={rotate=90, child anchor=north, parent anchor=south, anchor=center},
            },
            where level=1{text width=14.5em,font=\normalsize,fill=level1!80,}{},
            [Sentiment Analysis
                [Social Media and News
                    [\textbf{Social Media}: \citet{su2022improving}{,} \citet{steinert2023linking}{,} \\ \citet{deng2023llms}{,} \citet{mumtaz2023potential}, leaf, text width=28em]
                    [\textbf{News}: \citet{lopez2023can}{,} \citet{fatouros2023transforming}{,} \\ \citet{luo2024pre}, leaf, text width=28em]
                    [\textbf{Both}: \citet{zhang2023enhancing}{,} \citet{zhang2023instruct}{,} \citet{araci2019finbert}, leaf, text width=28em]
                ]
                [Corporate Disclosures
                    [\textbf{Earnings Calls}: \citet{cook2023evaluating}{,} \citet{leippold2023sentiment}, leaf, text width=28em]
                    [\textbf{Corporate Communications}: \citet{kim2023bloated}, leaf, text width=28em]
                    [\textbf{Regulatory Filings and Legal Documents}: \\ \citet{aparicio2024biofinbert}{,} \citet{cao2023talk}, leaf, text width=28em]
                ]
                [Market Research Report
                    [\citet{kim2023llms}, leaf, text width=28em]
                ]
                [Policy and Economic Indicators
                    [\textbf{The Federal Open Market Committee (FOMC) Meeting} \\ \textbf{Minutes}: \citet{shah2023trillion}{,} \citet{kim2023analyzing}{,} \citet{gossi2023finbert}, leaf, text width=28em]
                    [\textbf{The European Central Bank (ECB) Policy Decisions}: \\ \citet{fatouros2024can}{,} \citet{kanelis2024ecb}, leaf, text width=28em]
                ]
                [Diverse Sources
                    [\citet{rodriguez2024large}{,} \citet{wu2023bloomberggpt}, leaf, text width=28em]
                ]
            ]
        \end{forest}
    }    
\vspace{-0.5em}
\caption{Selected representative papers for sentiment analysis tasks in finance, categorized by various data sources.}
\vspace{-1em}
\end{figure*}

In finance, ML has been used to predict market movements based on sentiment in financial news and social media, showcasing its ability to capture financial sentiment nuances \cite{renault2020sentiment}. Machine learning methods offer the advantage of being able to capture complex patterns in data that are not apparent to lexicon-based methods. However, they require large datasets for training, and the versatility is limited on a specific domain. 
\\~\\
\textbf{Embedding-Based Methods:}
% The introduction of word embeddings marked a significant milestone in sentiment analysis. (Remark: adding detailed illustration of Word2Vec \cite{mikolov2013word2vec}) Embedding-based methods represent words in a high-dimensional space where semantically similar words are closer together. This representation captures not only the sentiment but also the context of words, leading to improved performance in sentiment analysis tasks. Techniques such as GloVe\cite{pennington2014glove}, FastText \cite{bojanowski2017fasttext} and ELMo \cite{sarzynska2021elmo} are prominent examples based on word embeddings. 
% Besides word embeddings, there are some other
The introduction of word embeddings marked a significant milestone in general sentiment analysis. Embedding-based methods represent textual information in a high-dimensional space where semantically similar words are closer together. This representation captures not only the sentiment but also the context of words, leading to improved performance in sentiment analysis tasks. 
\citeauthor{mikolov2013word2vec}'s introduction of Word2Vec \cite{mikolov2013word2vec} in 2013 was a pioneering development in this domain. Word2Vec employs neural networks to learn word associations from large datasets, generating embeddings that capture a wide array of linguistic relationships and nuances. The innovative aspect of Word2Vec lies in its ability to learn high-quality word vectors from vast datasets efficiently. It offers two architectures for this purpose: Continuous Bag of Words (CBOW) and Skip-gram. CBOW predicts target words from context words, while Skip-gram does the opposite, predicting context words from a target word, making it particularly effective for capturing semantic and syntactic word relationships.

Subsequent to Word2Vec, several other embedding models have emerged, further advancing the field. Notable among these are Global Vectors for Word Representation (GloVe) \cite{pennington2014glove}, which introduces an unsupervised learning algorithm for obtaining vector representations of words through aggregating global word-word co-occurrence statistics from a corpus; FastText \cite{bojanowski2017fasttext}, which extends Word2Vec to consider subword information, thereby enhancing the representation of rare words; and Embeddings from Language Models (ELMo) \cite{sarzynska2021elmo}, which leverages bidirectional language models to generate contextually enriched word embeddings. 

Beyond word-level embeddings, there has been a push towards capturing longer contextual dependencies. An exemplar in this area is Doc2Vec, also known as Paragraph Vector, introduced by \citet{le2014doc2vec}. Doc2Vec extends the Word2Vec paradigm to support document-level embeddings, enabling the capture of document-wide contextual information which is crucial for tasks requiring comprehension of extended textual content. By learning fixed-length feature representations from variable-length pieces of texts, Doc2Vec facilitates a deeper understanding of document semantics, thereby broadening the applicability of embedding techniques in sentiment analysis and beyond.

Embedding-based methods have the advantage of capturing contextual complexity and semantic relationships between words, significantly improving the accuracy of sentiment analysis. This has made them popular in FSA as well. \citet{sohangir2018bigdata} highlight the effectiveness of these methods in financial domains, demonstrating their ability to extract sentiment from large volumes of unstructured financial data with high accuracy.

However, they are not without drawbacks. A notable limitation is their dependency on large datasets for training, which might not always be feasible in specialized domains. Additionally, while adept at semantic understanding, they may overlook slight differences in syntax and require retraining to adapt to new language uses or vocabularies. Pre-trained embeddings can also perpetuate biases present in their training data, leading to potential issues in fairness and representation. Despite these challenges, embedding-based methods are crucial in advancing natural language understanding and have paved the way for large language models like BERT and GPT-3, which build on these embeddings to achieve state-of-the-art NLP performance.

%Mention relationship with LLM.

% \begin{itemize}
%     \item Summarize the milestones of general Sentiment Analysis (not limited to finance) during the development of NLP, e.g. word2vec or more.
%     \item Summarize some representative work of financial sentiment analysis during this period.
% \end{itemize}

\subsubsection{Sentiment Analysis with LLMs}
The advent of ChatGPT and other LLMs represents a pivotal milestone in the domain of FSA. Nowadays, these models have demonstrated their effectiveness in numerous tasks and offer several unique advantages for FSA applications.

Firstly, LLMs excel in deciphering the complexities of financial language, adeptly navigating informal expressions, emojis, memes, and specialized terminology across social media and financial blogs \cite{deng2023llms}, \cite{steinert2023linking}, \cite{chen2023understanding}, \cite{vamossy2023emtract}, \cite{jeong2024fine}, \cite{mumtaz2023potential}. Their proficiency in identifying subtleties like irony, sarcasm, and sector-specific jargon is vital for accurately analyzing sentiments across various formats, from tweets to comprehensive financial reports \cite{buaroiu2023capable}, \cite{wu2023bloomberggpt}. 

Secondly, LLMs' ability and great potential to process multimodal data, including images, audio, and video, are essential for comprehensive sentiment analysis in financial contexts like earnings calls \cite{cook2023evaluating} and FOMC Meetings \cite{curti2023let}. This capability allows for the integration of non-verbal cues and visual data into the sentiment analysis process \cite{bhatia2024fintral}.

Thirdly, LLMs' ability to process extensive documents enables thorough analysis of detailed financial reports and lengthy articles, ensuring no sentiment-bearing information is overlooked. This feature is particularly beneficial for evaluating the sentiments expressed in annual reports, earnings transcripts, and extensive financial narratives \cite{kim2023bloated}.

Moreover, LLMs exhibit enhanced resilience to adversarial attacks or deceptive information tactics that could be encountered in FSA tasks. Their advanced algorithms and broader contextual understanding help in identifying and mitigating misleading or manipulative sentiment indicators, enhancing the reliability of sentiment analysis outcomes.
\citet{leippold2023sentiment} highlights the contrast between traditional keyword-based sentiment analysis methods and LLMs in the face of adversarial attacks. The research involved using GPT-3 to substitute negative words with synonyms to assess model robustness, showcasing FinBERT's enhanced resilience against adversarial attacks over traditional keyword-based methods.
%\citet{leippold2023sentiment} highlights the contrast between traditional keyword-based sentiment analysis methods and LLMs in the face of adversarial attacks. Recent advanced method like GPT-3 could  By manipulating sentiment with advanced method like GPT-3, the author demonstrates that keyword-based approaches are highly vulnerable. In contrast, advanced models like FinBERT, employing context-aware techniques, show remarkable resilience against these attacks. 

%LLM calibrate forecasters

\subsubsection{Data-driven Applications}
We further delve into the recent advancements in the integration of LLMs within FSA, categorically analyzing the impact and contributions according to diverse data sources. We embark on this exploration by categorizing the data into four key segments: Social Media and News, Corporate Disclosures, Market Research Reports, and Policy and Economic Indicators.
%Social Media, Events and News, Policy and Economic Indicators, Analyst Reports and Investment Research, Earnings Calls and Corporate Communications, and Regulatory Filings and Legal Documents. 
This structured approach enables a comprehensive understanding of how LLMs have revolutionized the domain of FSA, offering unprecedented insights and analytics capabilities.
\\~\\
\textbf{Social Media and News:} 
Social media platforms like Twitter, general online forums like Reddit, and finance-specific forums such as StockTwits, along with financial blogs and microblogs, have become rich sources of data for FSA. These platforms are crucial due to their rich repositories of real-time, unstructured textual content that mirrors public sentiment regarding financial markets, specific stocks, and the overall economic environment. The immediacy and public nature of the discussions on these platforms make them an invaluable resource for capturing the mood of the market, which can be predictive of future market movements. 
\citet{su2022improving} leverage BERT for extracting sentiment and semantic insights from Twitter, facilitating improved covariance estimation and enhancing portfolio optimization. The integration of text-derived covariance data into mean-variance optimization resulted in superior performance in this work, especially during COVID-19 crash period. 
Furthermore, \citet{steinert2023linking} employ GPT-4 for sentiment analysis of microblogging messages on the Stocktwits platform, outperforming the naive buy-and-hold strategy for Apple and Tesla stocks by a significant margin, which underscores the potential of LLMs in predicting stock price movements through sentiment analysis.
Despite the efficacy of LLMs in sentiment analysis, social media sources present unique challenges, including the vast volume of information, the colloquial language often used, possible selective bias, and the presence of misinformation or inaccuracies in the messages shared, which complicate the task of accurately capturing and interpreting market sentiments \cite{ebrahimi2017challenges}. 

News represents another crucial data source, which shares many similarities with social media in terms of rapid dissemination and broad reach, but it generally focuses more on objective events. Contrary to the often subjective and personal nature of social media, news content typically originates from more prestigious and established media outlets, including the renowned newspapers like \emph{The New York Times}, television broadcasters like \emph{CNN} and \emph{BBC}, as well as finance-specific publications such as \emph{The Economist}. The credibility and professionalism of journalists and writers in these outlets lend a higher trustworthiness to the content, despite sometimes at the cost of timeliness. 
Evidence increasingly supports the advantages of post-ChatGPT LLMs over earlier approaches, particularly in analyzing the sentiment of news headlines. \citet{lopez2023can} investigate ChatGPT's effectiveness in predicting stock market returns, illustrating its capability to accurately assign sentiment scores to headlines and outperform earlier models such as GPT-2 and BERT. Besides, \citet{fatouros2023transforming} reveal that GPT-3.5 offers considerable improvements over FinBERT in analyzing forex-related news headlines. Similarly, \citet{luo2024pre} report noteworthy success with the open-source Llama2-7B model \cite{touvron2023llama}, achieving performance that exceeded previous BERT-based approaches and conventional methods like LSTM with ELMo. These researches underscore the significance of advanced LLMs in decision-making and quantitative trading.
%There are several evidences on the advantages of post-ChatGPT LLMs against the previous methods, especially when regarding doing sentiment analysis of news headlines. \citet{lopez2023can} explores ChatGPT's capability in forecasting stock market returns, showing that ChatGPT can effectively assign sentiment scores to headlines, outperforming more basic models such as GPT-2, and BERT, emphasizing the value of more advanced LLMs in decision-making and quantitative trading.  Besides, \citet{fatouros2023transforming} demonstrated that GPT-3.5 has considerable advancements over FinBERT using forex-related news headlines. \citet{luo2024pre} also achieving good performance with the open-source Llama2-7B \cite{touvron2023llama}, a popular pretrained LLM, that surpassed those of prior BERT-based models and traditional techniques like LSTM with ELMo. 

%\citet{luo2024pre} utilized the open-source Llama2-7B \cite{touvron2023llama}, a popular pretrained LLM, for classifying news headlines, achieving results that surpassed those of prior BERT-based models and traditional techniques like LSTM with ELMo. 

In this digital age, the phenomenon of real-time news is becoming increasingly prevalent. Distributed via live broadcasts or online platforms, these news sources manage to strike a balance between accuracy and immediacy, offering timely insights into market conditions and public events that could influence financial sentiments \cite{arvanitis2017real}. \citet{chen2022expected} investigate using advanced LLMs like BERT, RoBERTa, and OPT for sentiment analysis and stock prediction. These models significantly outperform traditional methods such as Word2vec by capturing complex text information and providing a more accurate contextual understanding. It also demonstrates that LLM-based models achieve higher Sharpe ratios and better performance. Crucially, the research reveals that news information is incorporated into stock prices with a delay due to limits-to-arbitrage, creating opportunities for real-time trading strategies to exploit these inefficiencies. This underscores the potential of LLMs in real-time financial text mining.
% \textbf{Corporate Disclosures:} 
% \begin{itemize}
%     \item Earnings Calls 
%     \item Corporate Communications 
%     \item Regulatory Filings and Legal Documents (e.g., SEC filings)
% \end{itemize}
\\~\\
\textbf{Corporate Disclosures:} 
Corporate disclosures is increasingly recognized for its significance in FSA. This section delves into three primary categories of corporate disclosures: Earnings Calls, Corporate Communications, and Regulatory Filings and Legal Documents (e.g., SEC filings), each highlighted for its importance and accompanied by pertinent studies.

Earnings calls are crucial for providing insights into a company's financial health, strategic direction, and management's perspective on performance and future prospects. The sentiment analysis of earnings calls transcripts can reveal underlying tones and sentiments that may influence investor decisions and market perceptions. \citet{cook2023evaluating} evaluate the performance of local LLMs in interpreting financial texts, particularly focusing on analyzing the tone and content of bank earnings calls during the post-pandemic era and the early 2023 banking stress. They show that local LLMs are effective for analyzing financial communications, demonstrating a shift in bank earnings call content towards more homogeneity and less positive sentiment during periods of increased banking stress. \citet{leippold2023sentiment} demonstrates the susceptibility of financial sentiment analysis to adversarial attacks using GPT-3, highlighting the need for LLMs to ensure the reliability of AI in financial text processing.

Corporate communications encompass a wide range of official statements, press releases, and announcements made by a company to its stakeholders. The sentiment embedded within these communications can significantly affect how stakeholders perceive the company's current state and future outlook. LLMs can process these communications to assess the sentiment and identify potential market-moving information. For instance, \citet{kim2023bloated} illustrate that ChatGPT can significantly streamline and clarify corporate disclosures for investors by reducing the length and amplifying the sentiment of the content, while also revealing the prevalent issue of 'bloat'—excessive, redundant, or irrelevant information in financial reports—that can obscure the true insights needed for informed investment decisions.

Regulatory Filings and Legal Documents are essential for compliance, governance, and transparency, providing a wealth of information on a company's operations, risks, and financial condition. LLMs can process these complex documents and identify sentiment-related information, such as litigation risks, accounting irregularities, and management changes. \citet{aparicio2024biofinbert} introduce BioFinBERT, a fine-tuned language model that utilizes sentiment analysis of regulatory filings and legal documents, such as 10-Q, 10-K, 6-K, and 20-F reports, along with biotech company press releases, to execute market orders and predict stock price movements in the biotechnology sector. Another paper \cite{cao2023talk} investigates how corporations adjust their regulatory disclosures to be more machine-readable in the age of AI, influencing both the sentiment expressed and the speed of information dissemination in financial markets.
\\~\\
\textbf{Market Research Reports:}
Market research reports, which encompass a wide range of data including economic indicators, industry analysis, and consumer behavior, are crucial for informed decision-making in finance. The significance of analyst reports and investment research lies in their detailed analysis and recommendations on securities, offering a profound understanding of market trends and potential investment opportunities. 
Analyst ratings, such as "buy," "hold," or "sell" recommendations, provide another concise evaluation of a security's future performance, serving as a valuable guide for investors. These ratings are based on rigorous financial analysis and are closely monitored by investors to assess market sentiment and make strategic investment choices \cite{kim2023llms}.
\\~\\
\textbf{Policy and Economic Indicators:} 
In the field of financial sentiment analysis, particularly with respect to policy and economic indicators, a significant focus has been placed on the analysis of Federal Open Market Committee (FOMC) meeting minutes, European Central Bank (ECB) policy decisions, and other key indicators like non-farm payroll data, unemployment rates, inflation rates, and GDP growth. These sources are critically important for understanding the market dynamics and guiding investment decisions based on the sentiment derived from policy decisions and economic reports.

The FOMC meeting minutes are a important source of information for understanding the U.S. Federal Reserve's monetary policy stance \cite{rosa2013financial}, \cite{smales2017does}. These minutes provide a detailed account of the discussions and deliberations that take place during FOMC meetings, shedding light on the economic outlook, inflation expectations, and potential interest rate changes \cite{shah2023trillion}. Researchers have employed LLMs to analyze the sentiment and tone of FOMC meeting minutes. \citet{kim2023analyzing} investigate that while FinBERT outperforms traditional techniques in predicting negative sentiment within FOMC statements, there is a need for further enhancements and exploration of alternative approaches to optimize the analysis of FOMC texts and gain more comprehensive economic insights. \citet{gossi2023finbert} present a fine-tuned FinBERT model with a Sentiment Focus method, which significantly improves the sentiment analysis accuracy of complex financial sentences in FOMC Minutes, particularly those containing conjunctions with contradicting sentiments. 

The ECB is responsible for setting monetary policy for the eurozone, and its policy decisions have a significant impact on financial markets \cite{klejdysz2023shifts}. ECB policy decisions, including interest rate changes and asset purchase programs, are closely monitored by investors and analysts \cite{anastasiou2023bank}, \cite{mody2024central}. Recent research has utilized LLMs to analyze the sentiment and impact of ECB policy decisions on financial markets \cite{fatouros2024can}. Utilizing the FinBERT model, \citet{kanelis2024ecb} reveal that sentiments from monetary policy speeches explain the tone of press conference statements, while financial stability speeches offer little explanatory power, highlighting the LLM's ability to provide detailed sentiment analysis in economic communication.

%restart
In addition to FOMC meeting minutes and ECB policy decisions, several other economic indicators and research papers are relevant to FSA. Non-farm payroll data and unemployment rates provide insights into the labor market and can have a significant impact on market sentiment \cite{nia2022cross}. Inflation rates and GDP growth are also closely watched indicators, as they reflect the overall health of the economy \cite{shapiro2022measuring}, \cite{biswas2020scope}. Applying LLMs to analyze the sentiment and impact of these economic indicators on financial markets deserves further exploration for future research.

\subsection{Financial Time Series Analysis}

\subsubsection{LLMs for Time Series}

Deep learning has revolutionized time series analysis, offering powerful tools for modeling and forecasting sequential data \cite{lim2021time}, \cite{wang2023st}, \cite{wang2023stgin}. Prominent deep learning models like LSTM networks and CNNs have demonstrated significant effectiveness in capturing temporal dependencies and anomalies in time series data  \cite{pan2024structural}, \cite{wang2022novel}, \cite{chen2022time}. 

With the recent surge in popularity of LLMs, these tools are increasingly being utilized to assist in time series tasks \cite{jiang2024empowering}, \cite{zhang2024large}. They offer a multitude of auxiliary functions such as generating additional features from textual data and producing descriptive statistics, as we have discussed in Section \ref{sec:linguistic} and \ref{sec:sentiment}, which can enhance the accuracy of time series models by tapping into a broader spectrum of information beyond the original data.

Beyond these supportive roles, LLMs are also being employed to directly analyze time series data \cite{jin2024position}, \cite{pan2024textbf}, a development supported by several factors. This is primarily attributed to LLMs' ability to understand and process sequential data, which is a common trait between text and time series. Also, the prevalent Transformer architecture underlying most LLMs has proven effective in various time series tasks \cite{zhou2022fedformer}, \cite{nie2022time}, \cite{wen2022transformers}. Furthermore, LLMs exhibit remarkable multimodal capabilities, suggesting that their pre-training on vast datasets, even if solely text-based, imparts general inference and reasoning abilities beyond the specific data modality \cite{zhu2023minigpt}. This characteristic not only serves as supportive evidence to the direct application of LLMs in time series analysis but also paves the way for future multimodal foundation models \cite{zhang2023meta}.

Several notable works have demonstrated the efficacy of LLMs in time series analysis. Pioneering efforts by \citet{zhou2024one} demonstrate the versatility of LLMs across tasks such as forecasting, anomaly detection, classification, and imputation. Using a GPT-2 backbone, they establish the potential for LLMs to process and model time series data effectively. \citet{gruver2024large} further explore the zero-shot capabilities of pretrained LLMs for time series forecasting. Through appropriate tokenization of time series data, they found that LLMs can implicitly understand temporal patterns and generate forecasts without explicit training. \citet{jin2023time} apply the concept of \textit{reprogramming} to enhance LLM performance in time series analysis. This technique translates time series data into representations more readily understood by LLMs, leading to state-of-the-art forecasting results. Beyond direct LLM applications, researchers are focusing on developing foundation models specifically for time series analysis \cite{jin2023large}, \cite{liang2024foundation}. These efforts aim to establish a new paradigm for time series modeling, leveraging the power of techniques in LLMs to capture complex temporal dependencies.

\begin{figure}[t]
    \centering
    \includegraphics[width=0.8\linewidth]{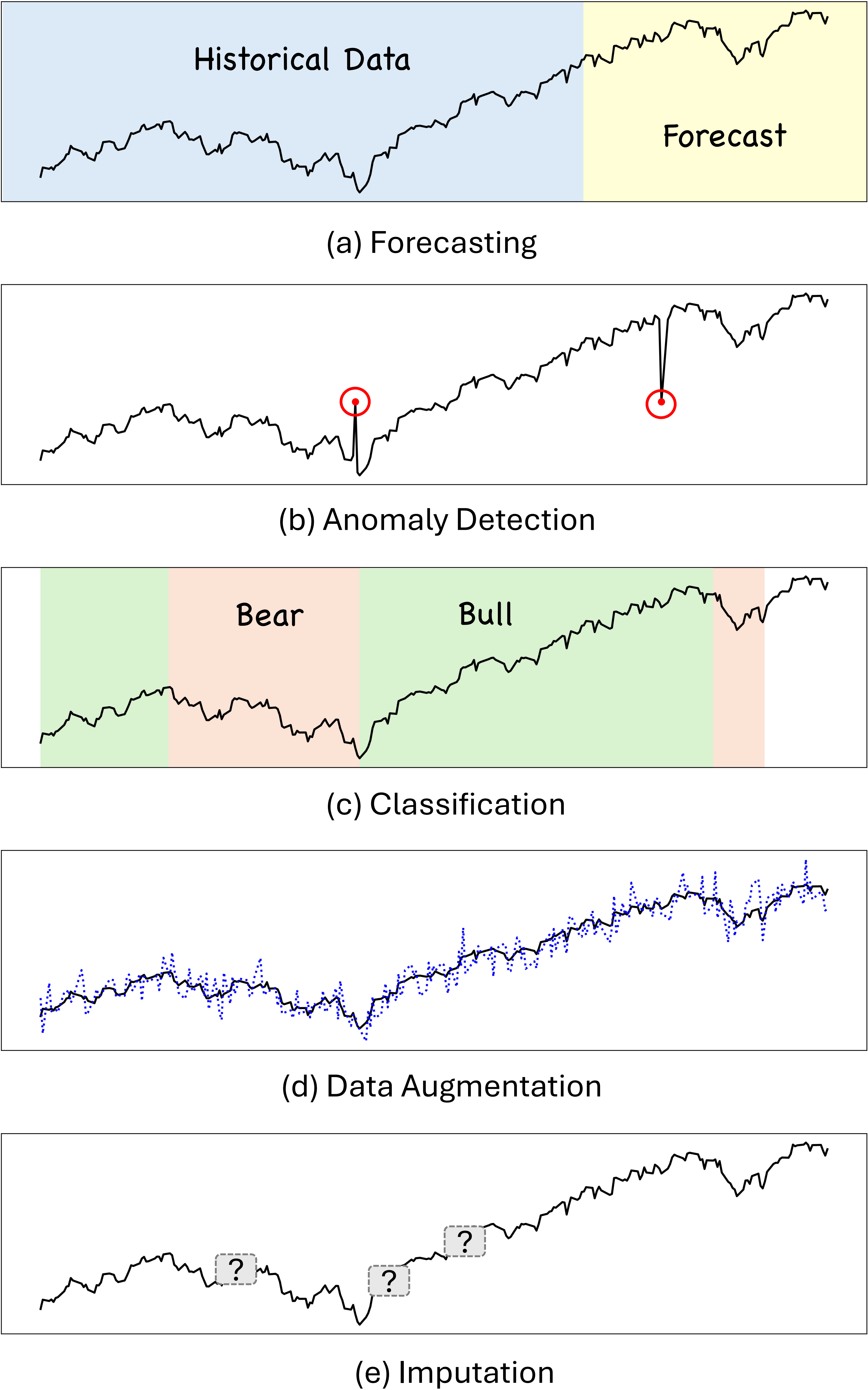} % Uncomment and replace with actual image
    %\fbox{\rule{0pt}{2in} \rule{0.9\linewidth}{0pt}} % Placeholder box
    \caption{Illustration of financial time series analysis.}
    \label{fig:time series}
    % \vspace{-2em}
\end{figure}

\subsubsection{Forecasting}
%Investigating the predictive power of LLMs in forecasting financial markets and economic indicators, leveraging historical data to anticipate future trends.
Recent research has explored the utility of LLMs in the domain of financial time series forecasting, demonstrating both the potential and the limitations of these advanced computational tools. This section reviews key studies that have contributed to our understanding of how LLMs can be applied to predict stock market movements and other financial indicators. 

LLMs can be directly used for stock forecasting, as stated in \cite{yu2023temporal}. Their research explores NASDAQ-100 stock prediction with LLMs, and demonstrates that by integrating diverse data sources, LLMs not only provide robust predictions but also enhance the explainability. The study emphasizes the importance of instruction-based fine-tuning and chain of thought reasoning, which have been shown to significantly improve the performance of LLMs over traditional statistical models in this field. Another way is to integrate LLMs to enhance the other neural networks. \citet{chen2023chatgpt} introduced a framework that leverages ChatGPT to enhance graph neural networks (GNN) for stock movement prediction. Their approach adeptly extracts evolving network structures from textual data and incorporates these networks into GNNs for predictive tasks. The experimental results indicate that this model consistently outperforms state-of-the-art deep learning-based benchmarks with higher annualized cumulative returns and reduced volatility.          

Besides, LLMs are notable for their capability to be integrated in multimodal data analysis as discussed in the previous section, which can be crucial when analyzing alternative data. For instance, \citet{wimmer2023leveraging} introduce innovative models that leverage both textual and visual data to forecast market movements. Utilizing CLIP-based models, their research shows significant outperformance against established benchmarks in predicting the trends of the German share index. Metrics such as Precision, F1 Score, Balanced Accuracy, and others show the effectiveness of these multimodal approaches. Another noteworthy study is the RiskLabs framework, which combines various types of financial data, including textual and vocal information from earnings conference calls, market-related time series data, and contextual news data \cite{cao2024risklabs}. The framework's multi-stage process starts with extracting and analyzing these data using LLMs, followed by processing time-series data to model risk over different timeframes. RiskLabs employs multimodal fusion techniques to combine these varied data features for comprehensive multi-task financial risk prediction. The empirical results demonstrate the framework's effectiveness in forecasting both volatility and variance in financial markets, indicating the potential of LLMs in financial risk assessment.

However, the application of LLMs in financial forecasting is not without challenges. \citet{xie2023wall} specifically assess ChatGPT’s performance in zero-shot multimodal stock movement prediction tasks and find that it underperforms when compared to both traditional machine learning models and other state-of-the-art techniques. Their findings highlight the necessity for ongoing research to enhance the predictive capabilities of LLMs in complex financial environments. On the other hand, \citet{lopez2023can} examine how well these models, particularly GPT-4, can predict stock market returns using news headlines as input. Their results indicate that advanced LLMs significantly outperform both traditional models and earlier versions of LLMs. Notably, the models show higher efficacy especially after negative news and for smaller stocks, a phenomenon explained through theories of information diffusion, arbitrage limitations, and investor sophistication. The debate on the effectiveness of LLMs in financial forecasting remains open, with evidence supporting both their limitations and potential.

Though early challenges exist, research reveals considerable promise for LLMs in financial time series forecasting. Explainability, comprehensive understanding of news, and multimodal integration stand out as compelling areas for future investigation and refinement. However, they also mark the challenges and the necessity for further research to fully realize the potential of LLMs in this domain.
\subsubsection{Anomaly Detection}
\label{sec: anomaly}

Anomaly detection is a fundamental task in various domains, particularly in finance where identifying unusual patterns or outliers is crucial \cite{chandola2009anomaly}. For instance, identifying fraudulent transactions or unusual account activity is a top priority for financial institutions. Anomaly detection algorithms can flag potentially fraudulent behavior, preventing financial losses \cite{zojaji2016survey}. Besides, market manipulation schemes, such as pump-and-dump tactics, can be detected through anomaly detection in trading volumes and price patterns \cite{chen2019detecting}. Anomaly detection is also valuable in risk assessment and mitigation strategies, since anomalies in market trends or macroeconomic indicators can signal potential risks.

Financial time series data, like stock prices, can be highly complex, characterized by volatility, seasonality, and non-linear relationships. 
Traditional statistical approaches, though robust, often struggle to encapsulate the full spectrum of these complexities, thereby constraining their anomaly detection capabilities. 
The development of deep learning has catalyzed a fundamental transformation, offering novel methodologies that hold great promise for this domain \cite{darban2022deep}, \cite{crepey2022anomaly}.
Particularly, LLMs have emerged as a pivotal method, demonstrating remarkable efficacy in anomaly detection across a myriad of tasks, as evidenced by recent scholarly works \cite{darban2022deep}, \cite{zhu2024llms}.
For instance, \citet{park2024enhancing} introduces an LLM-based multi-agent framework that synergizes traditional statistical methods with AI-driven analytics. This innovative fusion is exemplified through an application to the S\&P 500 index, showcasing a marked enhancement in the efficiency, precision, and automation of anomaly detection in financial markets, thereby diminishing the dependency on manual interventions. 
The integration of LLMs into financial time series anomaly detection will likely become increasingly valuable, which has the potential to not only address the limitations of conventional techniques, but also reduce manual processes and enhance algorithmic trading systems that capitalize on market anomalies, paving the way for more sophisticated and automated trading systems.

% Anomaly detection can be integrated into algorithmic trading systems to identify and capitalize on profitable market opportunities. Traditional statistical methods often fall short in capturing these complexities, limiting their effectiveness in anomaly detection. 

% The recent advancements in deep learning have opened up promising avenues for enhancing anomaly detection techniques in financial time series analysis \cite{darban2022deep}, \cite{crepey2022anomaly}. 

% Besides, LLMs also shown to be effective in anomaly detection on various tasks \cite{darban2022deep} \cite{zhu2024llms}

% \cite{park2024enhancing}: This paper presents an LLM-based multi-agent framework that enhances anomaly detection in financial markets by integrating traditional statistical techniques with AI-driven analysis, using the S\&P 500 index as a case study to demonstrate improved efficiency, accuracy, and reduced reliance on manual processes.

\subsubsection{Other Time Series Tasks}

Besides forecasting and anomaly detection, the capabilities of LLMs offer promising potential within several other domains of financial time series analysis. 
\\~\\
\textbf{Classification: } 
Financial time series can be classified into various categories based on trends, volatility, or other characteristics. LLMs can learn these complex patterns and assign labels accordingly. For instance, they could classify stocks as "growth" or "value," or identify different market regimes (bullish, bearish, etc.) \cite{bosancic2024regime}. LLMs can efficiently classify financial time series data by understanding and predicting patterns that are indicative of specific financial behaviors. This includes the applications of sentiment analysis (section \ref{sec:sentiment}) and anomaly detection (section \ref{sec: anomaly}) that we have already discussed.
\\~\\
\textbf{Data Augmentation:} 
%The limited size and variability of financial datasets can sometimes hinder machine learning models. LLMs offer a path toward data augmentation, which involves generating synthetic data that can be used for training machine learning models, ensuring robustness despite the original limitations of the dataset. By simulating various market scenarios, LLMs can help in creating a richer, more diverse dataset that aids in building more accurate predictive models.
The limited size and variability of financial datasets can sometimes hinder machine learning models. Generative AI offers a path toward data augmentation, which involves generating synthetic data that can be used for training machine learning models, ensuring robustness despite the original limitations of the dataset. A recent paper by \citet{nagy2023generative} introduces a generative AI model for end-to-end limit order book modeling, demonstrating the use of a token-level autoregressive generative model to produce realistic order flow in financial markets. This model utilizes structured state-space layers to efficiently handle long sequences of order book states and tokenized messages. The model shows promising performance in approximating data distribution and forecasting mid-price returns, suggesting potential applications in high-frequency financial reinforcement learning. While this work focuses on generative AI rather than directly employing LLMs, its approach and insights are relevant for augmenting financial time series data, highlighting the versatility of generative models in this domain. By simulating various market scenarios, LLMs can help in creating a richer, more diverse dataset that aids in building more accurate predictive models \cite{ding2024data}.
\\~\\
\textbf{Imputation:} 
Financial time series data often suffers from missing values due to errors or unavailability. Imputation refers to the method of filling in missing or incomplete data points in financial time series. LLMs have a good potential to fill these missing values based on their superior generative capability \cite{zhao2023survey}. This is particularly useful in maintaining the quality and continuity of financial data analysis. Accurate imputation helps in avoiding biases or inaccuracies that might occur due to gaps in the data, thus ensuring more reliable financial assessments and forecasts.
\\~\\
In summary, LLMs demonstrate significant potential in financial time series analysis, offering capabilities in forecasting, anomaly detection, pattern classification, data augmentation, imputation, and more. Their ability to process and understand complex financial data opens avenues for novel approaches to market analysis. As LLM research progresses, we can anticipate continued advancements in the application of these models within the financial time series domain.

\subsection{Financial Reasoning}
Another key application of LLMs in finance is to support financial reasoning. As previously discussed, LLMs are capable of processing and synthesizing vast amounts of financial data from various sources, including market reports, financial news, and historical pricing data. This comprehensive understanding of the financial landscape and market dynamics may enable LLMs to support strategic financial planning, generate investment recommendations, provide advisory services, and assist in financial decision-making. 

The use of LLMs in financial reasoning offers several key advantages. Firstly, they can \textbf{enhance data analysis} by processing vast amounts of financial information, identifying patterns and trends that help inform better decision-making. Secondly, LLMs can be used for \textbf{predictive modeling}, allowing them to forecast market conditions and asset performance, which may lead to robust investment recommendations. Additionally, LLMs could offer \textbf{personalized advisory services}. They can analyze a person’s or organization’s financial situation, goals, and risk tolerance to provide customized advice. Another benefit could be \textbf{real-time monitoring and alerts}, where LLMs can monitor financial market trends and news, providing timely updates and alerts to help users adjust their strategies as needed. Moreover, LLMs may \textbf{improve accessibility and engagement}. By integrating these models into user-friendly interfaces like chatbots, financial planning and advisory become more accessible and engaging, where individuals can take control of their own financial well-being.

In this section, we will explore these applications through the literature, potentially inspiring further innovations.

\begin{figure*}[t]
    \centering
    \includegraphics[width=.9\linewidth]{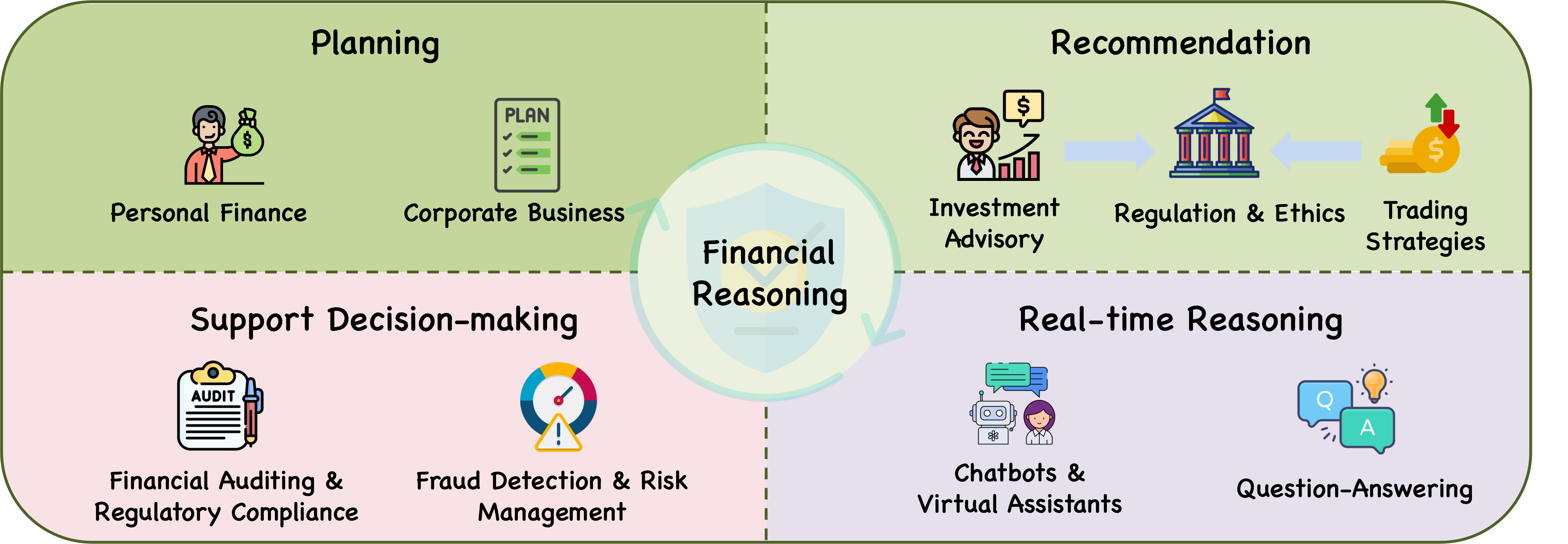} % Uncomment and replace with actual image
    % \fbox{\rule{0pt}{2in} \rule{0.9\linewidth}{0pt}} % Placeholder box
    \caption{Illustration of various financial reasoning tasks.}
    \label{fig:financial reasoning}
    \vspace{-1em}
\end{figure*}

\subsubsection{Planning}
Financial planning involves setting financial goals, assessing current financial situations, and devising strategies to achieve those goals. This process includes analyzing income, expenses, investments, and risk management to create a comprehensive plan for long-term financial stability and growth.

In a corporate context, LLMs can be utilized to support various aspects of financial planning. For instance, LLMs can analyze market trends and competitor data to help organizations develop business strategies. \citet{nguyen2023generative} examine the use of generative AI models, such as GPT-4 and other transformer-based models, for business strategy development. By employing Named Entity Recognition (NER) and Zero-shot Classifiers (ZSC) to automatically extract and classify relationships between companies, they created dynamic signed business networks that reflect the competitive and collaborative market landscape. This method provides business stakeholders with insights into market conditions and supports strategic decision-making. 

Moreover, LLMs can streamline financial planning processes, as demonstrated by \citet{ludwig2023streamlining}. By integrating ChatGPT into financial planning practices, they illustrate how financial planners could leverage this AI model to enhance client communication and provide immediate, semi-personalized responses to common financial concerns, such as preparing for economic recessions. They also highlight ChatGPT’s role in client education and its ability to simplify complex financial concepts for better understanding. Despite these benefits, the authors emphasize the need for human oversight to ensure the accuracy and quality of the advice provided, addressing potential limitations of the models. 

In personal financial planning, LLMs can help individuals create customized strategies for long-term financial well-being. A recent study by \citet{lakkaraju2023can} evaluates the performance of LLM-based chatbots, ChatGPT and Bard, in providing personal financial advice. The study covers various aspects of personal finance, including decisions related to bank accounts, credit cards, and certificates of deposits (CDs). It assesses how these models handle complex financial interactions and make recommendations across different languages and dialects, such as English, African American Vernacular English, and Telugu. Their findings reveal that while ChatGPT often provides more personalized and accurate responses, both models face challenges, including mathematical errors, lack of visual aids to support explanations, and difficulties in processing non-English queries effectively. The paper emphasizes the need for improvements in these LLMs to enhance their reliability and inclusivity when applied to financial planning, a topic that will be further discussed in section \ref{sec: challenges}.

Additionally, LLMs can optimize budgeting strategies by incorporating AI-driven recommendations into individual and household financial models. \citet{de2023optimized} present an optimization framework for individual budget allocation to maximize savings and extend this approach to household finances, addressing the complexities of multiple incomes and shared expenses. In high-net-worth contexts, LLMs can also be used to simulate various tax scenarios, identify optimal tax strategies, and provide proactive advice based on changing tax law to minimize tax liabilities and maximize financial growth \cite{fava2023future}. 

The integration of LLMs in financial planning has the potential to transform how individuals and businesses approach their financial objectives. By leveraging the data processing and analysis capabilities of LLMs, financial planning can become more efficient, accurate, and personalized. As research and development in this field continue to progress, LLMs are poised to become vital tools in the financial planning environment, allowing users to make educated and strategic decisions. The examples discussed in this section highlight the diverse range of applications and the potential for LLMs to revolutionize financial planning practices for both corporate and personal contexts.

\subsubsection{Recommendation}
LLMs are revolutionizing investment recommendations and wealth management by analyzing financial data, forecasting market trends, and optimizing portfolios. They provide personalized advice based on individual risk profiles and preferences, which improves robo-advisors and investment strategies. However, the integration of LLMs in wealth management needs regulatory frameworks to assure fairness, effectiveness and informed decision-making in conjunction with human expertise. 
\\~\\
\textbf{LLM in Investment Advisory:} 
LLMs play a crucial role in enhancing the capabilities of robo-advisors by providing personalized and automated investment recommendations. For example, \citet{huang2024research} highlight the effectiveness of platforms like Wealthfront and Betterment, which employ AI algorithms to deliver customized asset management plans aimed at optimizing investment returns based on individual user profiles. The study emphasizes the importance of consistent use, transparency, and user-centric design in maximizing the benefits of intelligent advisors. To build user trust and improve the overall effectiveness of robo-advisors, the authors recommend focusing on key areas such as enhancing transparency, designing intuitive user interfaces, and offering tailored financial services for individual needs. 

Similarly, \citet{lu2023chatgpt} explore the potential of ChatGPT in generating investment portfolio recommendations. Using textual data from the Wall Street Journal and Chinese policy announcements, the researchers evaluate ChatGPT’s ability to generate portfolios that outperform the market. Through fine-tuning and performance measurements, the study demonstrates that ChatGPT can achieve a monthly three-factor alpha of up to 3\%, particularly when analyzing policy-related news. They highlight the importance of model parameters, such as the “temperature” setting, in influencing the recommendations’ creativity and accuracy, indicating that generative AI, with appropriate tuning, can be a valuable tool for financial advisors. 

Another development in the field is the Cogniwealth system, introduced by \citet{ramyadevi2024cogniwealth}. This platform utilizes the Llama 2 model as a financial advisor. The system leverages NLP and machine learning techniques to assist both professional fund researchers and laymen investors by providing personalized investment recommendations and financial insights. Cogniwealth’s ability to handle user-provided data and deliver human-like responses through an intuitive interface ensures high levels of adaptability, user-friendliness, and engagement. 
\\~\\
\textbf{Impact on Investment Strategies:} 
LLMs are transforming the landscape of investment strategies, offering the potential to deliver more accurate, diverse, and accessible investment advice. A prime example of this is demonstrated in \citeauthor{ko2024can}'s study \cite{ko2024can}, where they showcase ChatGPT’s ability to construct portfolios with superior diversity and performance compared to randomly selected ones. This finding highlights the potential of LLMs to serve as valuable advisory tools for both professional portfolio managers and individual investors, democratizing access to advanced investment strategies.

LLMs can also impact the development of algorithmic trading strategies by automating the creation of accurate and executable code for technical indicators. The study conducted by \citet{noguer2024evaluating} compares the capabilities of various LLMs, such as GPT-4-Turbo, Gemini-Pro, Mistral, Llama 2, and Codellama, in generating code that runs correctly and matches baseline implementations. The study emphasizes the importance of well-designed prompts and the models’ ability to handle complex financial calculations for successful code generation. 

Recently, \citet{kim2024financial} investigates the capability of an LLM, specifically GPT-4 Turbo, to perform financial statement analysis comparable to that of professional human analysts. By providing standardized and anonymous financial statements, the study examines the model's ability to predict future earnings without any narrative or industry-specific context. The findings reveal that the LLM not only outperforms human analysts in predicting earnings changes, particularly in challenging scenarios, but also matches the performance of specialized state-of-the-art machine learning models. The authors claim that the model's predictions derive not from its training memory but from generating useful narrative insights about a company’s future performance, thus eliminating look-ahead bias. To address this bias, the research design uses a consistent anonymized format for financial statements across firms, making it virtually impossible for the model to infer a firm’s identity. Additionally, the statements do not contain any dates and use relative years, mitigating concerns about the model leveraging macroeconomic trends from specific years. Furthermore, trading strategies based on the LLM's predictions demonstrate higher Sharpe ratios and alphas compared to those based on other models.

Another promising application of LLMs in investment strategies is the analysis of annual reports to extract valuable insights, thereby enhancing stock investment strategies. \citet{gupta2023gpt} introduces a framework that utilizes GPT-3.5 to streamline the process of analyzing comprehensive 10-K filings of a company. By combining the generated insights with historical stock data, the study demonstrates that machine learning models trained on these LLM-generated features can outperform traditional market benchmarks, such as the S\&P 500 index. This approach highlights the potential of integrating LLMs with historical data to improve the accuracy of stock predictions and enhance investment strategies. 

Moreover, \citet{zhang2024breakgpt} introduce BreakGPT for detecting financial breakouts. BreakGPT’s multi-stage structure improves the accuracy and stability of detecting true and false breakouts in financial markets by systematically analyzing price movements and order flows. The model’s superior performance compared to ChatGPT-3.5 and ChatGPT-4 makes it a valuable tool for traders and investors in detecting financial breakouts. 

However, despite these promising developments, \citet{chuang2022buy} raise an important concern regarding the implicit biases present in pre-trained language models, such as BERT and FinBERT. The study reveals that these models exhibit significant biases towards certain stocks and industry sectors, which can influence the quality and fairness of investment recommendations. They emphasize the need for awareness and mitigation of such biases in financial decision-making systems to ensure more reliable and fair investment advice. This research highlights the importance of careful model training and evaluation in financial contexts to develop robust and accountable financial advisory systems. 
\\~\\
\textbf{Regulatory and Ethical Considerations:} 
The application of LLMs in financial advisory services has raised significant regulatory and ethical concerns. \citet{caspi2023generative} examine the regulatory landscape, highlighting key concerns such as maintaining fiduciary duties, ensuring transparency, and preventing conflicts of interest. They discuss potential regulatory strategies to address the challenges posed by generative AI, emphasizing the need for effective regulation that balances innovation with consumer protection.                                          
Moreover, \citet{niszczota2023gpt} investigate the financial literacy of GPT models, revealing GPT-4’s near-perfect score on financial literacy tests. However, they also find that individuals with lower financial knowledge tend to rely more heavily on GPT’s advice.

\citet{lakkaraju2023llms} also compare the effectiveness and fairness of LLM-based chatbots (ChatGPT and Bard) to a rule-based chatbot (SafeFinance) in providing personal financial advice. They find that while ChatGPT and Bard generate fluent responses, they exhibit inconsistencies and biases across different user groups and languages. In contrast, SafeFinance provides reliable answers, albeit with limited generalization. The study demonstrates the need for improvements in LLM-based systems to ensure fairness and accuracy in financial advisement. 

While LLMs have demonstrated potential in transforming financial advisory services, their application raises important regulatory and ethical considerations. Effective regulation should balance innovation with consumer protection, while educating users about the limitations and potential biases of AI-driven financial recommendations is essential to promote informed decision-making.

\subsubsection{Support Decision-making}
Operational risk management and compliance are critical components in the financial sector, as they help safeguard the integrity of financial institutions, protect consumers, and maintain the stability of the entire financial system. However, the increasing complexity of financial products, ever-changing regulations, and the constant threat of fraudulent activities pose significant challenges for financial institutions. LLMs are emerging as powerful tools that enhance these processes by providing sophisticated analytical capabilities. By leveraging LLMs, financial institutions can improve the accuracy of audits, streamline compliance verification, and detect inconsistencies more efficiently. This enables financial institutions to make informed decisions in critical areas such as Financial Auditing and Regulatory Compliance, and Fraud Detection and Risk Management, ultimately enhancing their operational resilience and ensuring compliance with regulatory requirements.
\\~\\
\textbf{Financial Auditing and Regulatory Compliance:} 
Financial auditing involves the systematic examination of financial records and statements to ensure accuracy and compliance with regulations. LLMs are increasingly being used to enhance these processes by improving the accuracy and efficiency of text matching and regulatory interpretation \cite{berger2023towards}. A study conducted by \citet{hillebrand2023improving} introduces ZeroShotALI, which stands for Zero-Shot Automated List Inspection. It combines GPT-4 and a domain-specific SentenceBERT model to enhance the matching of text segments from financial reports with specific legal requirements. This system significantly improves the efficiency and accuracy of financial audits compared to traditional methods. 

Moreover, another study conducted by \citet{cao2024large} examines the use of LLMs (such as GPT-4, GPT-3.5, Claude-3-Opus, Gemini-1.5-Pro) for interpreting complex financial regulations, specifically focusing on Basel III capital requirements. Effective prompt design and document loading methods guide LLMs in translating regulatory texts into a concise mathematical framework, aiming to significantly enhance regulatory interpretation accuracy. 

In addition, by analyzing firms’ public narrative disclosures with GPT-4, \citet{Choi2024FirmLevelTax} develop a novel measure of tax audit periods at the firm level. Their measure shows high conformity with data released by the IRS (the Internal Revenue Service) and reveals that tax audits lead to reduced tax avoidance, decreased capital investments, and increased stock volatility. 

LLMs have shown potential in uncovering inconsistencies and contradictions in financial reports. A study conducted by \citet{deusser2023uncovering} develops an innovative approach to identifying discrepancies in financial reports by leveraging the power of LLMs such as GPT-4 and Llama. The study employs embedding-based paragraph clustering to efficiently detect contradictions across various datasets, including both annotated and unannotated financial reports. By utilizing sentence-pair data, document-level data, and intelligent bucketing systems, the researchers optimize the query process for the LLMs, enabling them to effectively pinpoint inconsistencies and contradictions. The results of this study demonstrate significant enhancements in the accuracy and efficiency of financial audits, ultimately reducing the time and effort required to conduct thorough and reliable financial report audits. 
\\~\\
\textbf{Fraud Detection and Risk Management:} 
Fraud detection and risk management are critical components of maintaining financial integrity and stability. LLMs offer advanced capabilities to detect fraudulent activities and manage risks through sophisticated data analysis and pattern recognition. A study conducted by \citet{feng2023empowering} highlights the potential of LLMs to revolutionize credit scoring and risk assessment. By instruction tuning, LLMs can match or surpass traditional credit scoring models, leading to more inclusive and comprehensive evaluations. However, the study also emphasizes the need to address biases within LLMs to ensure fair financial decision-making. 

Furthermore, \citet{cao2024risklabs} present a novel framework named RiskLabs that leverages LLMs to predict financial risk by integrating data from various sources. By processing and fusing features from diverse data types, including textual and vocal information from Earnings Conference Calls (ECCs), market-related time series data, and contextual news data surrounding ECC release dates, RiskLabs outperforms traditional methods and existing models in forecasting financial risks, providing a more comprehensive understanding of market dynamics. 

Several papers explore the application of LLMs in fraud detection. \citet{zhao2023generative} introduce an innovative GPT-based model for identifying fraudulent activities in payment systems, which excels in capturing detailed behavioral sequences through temporal and contextual analysis. \citet{yang2023finchain} introduce the FinChain-BERT model, which enhances fraud detection accuracy by focusing on key financial terms and optimizing model performance. Similarly, \citet{bhattacharya2024accounting} demonstrate the effectiveness of the BERT model in detecting accounting fraud in financial reports by fine-tuning the model on Management Discussion and Analysis sections of annual 10-K reports from the Securities and Exchange Commission (SEC) database, outperforming existing benchmark models.

While LLMs have shown great potential in fraud detection and risk management, it is crucial to acknowledge and address the inherent biases that may exist within these models. Biases in LLMs can lead to unfair and discriminatory practices in financial decision-making. Ongoing research and development efforts are necessary to mitigate these biases and ensure the responsible and ethical deployment of LLMs in the financial sector.

\subsubsection{Real-time Reasoning}
Real-time reasoning enables instant and dynamic interactions between users and AI-powered systems. By leveraging the vast knowledge and understanding of LLMs, financial institutions can deploy chatbots, virtual assistants, and question-answering systems that provide accurate, relevant, and timely information to customers and stakeholders. These real-time applications streamline customer support, simplify complex financial transactions, and offer immediate access to financial insights and advice. 
\\~\\
\textbf{Chatbots and Virtual Assistants:}
Chatbots and virtual assistants are changing the way financial institutions interact with customers and streamline internal processes. By leveraging the capabilities of LLMs, these AI-driven tools can further provide more personalized and effective assistant, thereby enhancing customer satisfaction and boosting organizational efficiency. For instance, \citet{aggarwal2023multi} present a multi-purpose NLP chatbot that incorporates LLM models, including ChatGPT, BERT, and DistilBERT. The proposed system incorporates emotion recognition, multilingual support, and voice conversion. The chatbot demonstrates exceptional performance in providing personalized financial advice, understanding and responding to human emotions, and maintaining functionality in offline modes.

In another study, \citet{yue2023gptquant} introduce GPTQuant, a conversational AI chatbot designed to facilitate investment research. GPTQuant leverages few-shot learning and LangChain’s integration to generate Python code for backtesting and strategy analysis. The chatbot uses prompt templates to activate GPT-3’s capabilities, demonstrating efficacy in portfolio construction, rebalancing, and factor score querying. 

Lastly, \citet{yadav2024generative} introduce a virtual assistant that utilizes LLMs to enhance the financial reconciliation process. The assistant automates the generation of SQL queries from natural language inputs, streamlining and expediting reconciliation, research, and validation processes for accountants. Utilizing a retrieve-and-refine strategy with retrieval-augmented generation (RAG) and few-shot prompting, the virtual assistant achieved 95\% accuracy in generating correct SQL queries for real-world questions related to account reconciliation. This integration of LLMs significantly improves the accuracy and efficiency of generating SQL queries, demonstrating the potential of LLMs to automate repetitive and time-consuming tasks in financial reconciliation.
\\~\\
\textbf{Question-Answering:} 
Question-answering systems powered by LLMs have shown remarkable progress in understanding and responding to complex queries related to financial documents. Recent studies have focused on enhancing the numerical reasoning capabilities of these systems, enabling them to handle multi-step calculations and extract relevant information from various data sources. For example, \citet{arun2023numerical} develop a pipeline utilizing fine-tuned LLMs, such as Llama-2-7B and T5, to analyze financial reports and answer numerical reasoning questions. By extracting and serializing tables from PDFs, generating embeddings, and training on the FinQA dataset, the authors demonstrated the potential for real-time analysis of financial reports. The study concludes that with appropriate fine-tuning and methodologies, LLMs could significantly enhance the efficiency and accuracy of financial data analysis, enabling swift and informed decision-making in dynamic market environments through rapid extraction and interpretation of crucial data points. 

Furthermore, \citet{phogat2023zero} introduce zero-shot prompting techniques (ZS-FinPYT and ZS-FinDSL) for LLMs including GPT-3, GPT-3.5-turbo, and GPT-4 to perform complex numerical reasoning over financial documents. By encoding reasoning into Python/DSL(domain-specific languages) programs, these techniques mitigate arithmetic limitations. Evaluations on datasets such as FinQA, ConvFinQA, and TATQA demonstrate superior performance compared to baselines, particularly in table/text data, multi-step reasoning, and numerical questions.

In a related study, \citet{srivastava2024assessing} investigate the mathematical reasoning capabilities of LLMs on financial documents. They introduce a novel prompting strategy, EEDP (Elicit-Extract-Decompose-Predict), designed to enhance LLM performance in scenarios requiring multi-step numerical reasoning. Extensive experimentation with multiple LLMs across financial datasets reveals that EEDP outperforms baseline strategies like Direct Prompting, Chain of Thought (CoT), and Program of Thoughts (PoT). The study highlights the potential of structured prompting strategies in improving LLM performance for complex reasoning tasks and identified common error types, emphasizing the need for precise information extraction.

Moreover, \citet{xue2023weaverbird} propose a cutting-edge dialogue system designed specifically for the finance sector, named WeaverBird. It leverages a LLM with GPT architecture fine-tuned on extensive financial corpora. This enables WeaverBird to understand and provide informed responses to complex financial queries, such as investment strategies during inflation. The system's performance is further enhanced by integrating a local knowledge base and search engine, allowing it to retrieve relevant information and generate responses conditioned on web search results, complete with proper source references for enhanced credibility. Comparative evaluations across a broad spectrum of financial question-answering tasks demonstrate WeaverBird's superior performance compared to other models, positioning it as a powerful tool for financial dialogue and decision support.

\subsection{Agent-based Modeling}

Agent-based modeling (ABM) represents a significant advancement in simulating complex systems, particularly in finance. The core principle of ABM involves creating autonomous agents that interact within a defined environment, allowing the emergence of complex phenomena from the bottom up. Unlike traditional models that assume uniform behavior among agents and equilibrium states, ABM captures the diversity of behaviors and adaptive strategies that characterize real-world financial markets. This flexibility makes ABM a powerful tool for understanding market dynamics, investor behavior, and the impact of various external factors on financial systems.

In recent years, the integration of LLMs with agent-based modeling has opened new avenues for research and application \cite{guo2024large}, \cite{xi2023rise}, \cite{ma2024computational}. With their advanced NLP capabilities, LLMs enhance the cognitive functions of agents, allowing them to interpret and react to vast amounts of unstructured data such as financial news, reports, and social media posts. This synergy between LLMs and ABM leads to more realistic and adaptive simulations, which are crucial for developing robust trading and investment strategies \cite{zhang2024llm}.

Traditional applications of ABM in finance have focused on modeling the interactions between different types of market participants, such as institutional investors, individual traders, and regulatory bodies \cite{epstein1999agent}. These models have been used to study the impact of regulatory changes, market shocks, and behavioral biases on market dynamics. For instance, agent-based models have been employed to simulate the effects of high-frequency trading, the propagation of financial crises, and the formation of asset bubbles. The addition of LLMs to these models further enhances their predictive power and accuracy by enabling agents to process and respond to real-time information in a manner similar to human analysts.

In this section, we explore the integration of LLMs with agent-based modeling in various contexts. We discuss how LLM-based trading and investment agents enhance decision-making and strategy formulation. We also examine the use of LLMs in simulating markets and economic activities, highlighting their impact on policy analysis and market predictions. Additionally, we review the role of multi-agent systems in improving financial process automation and monitoring, emphasizing the potential of these advanced models in revolutionizing financial analysis and strategy development.

\begin{figure*}[t]
    \centering
    \includegraphics[width=.9\linewidth]{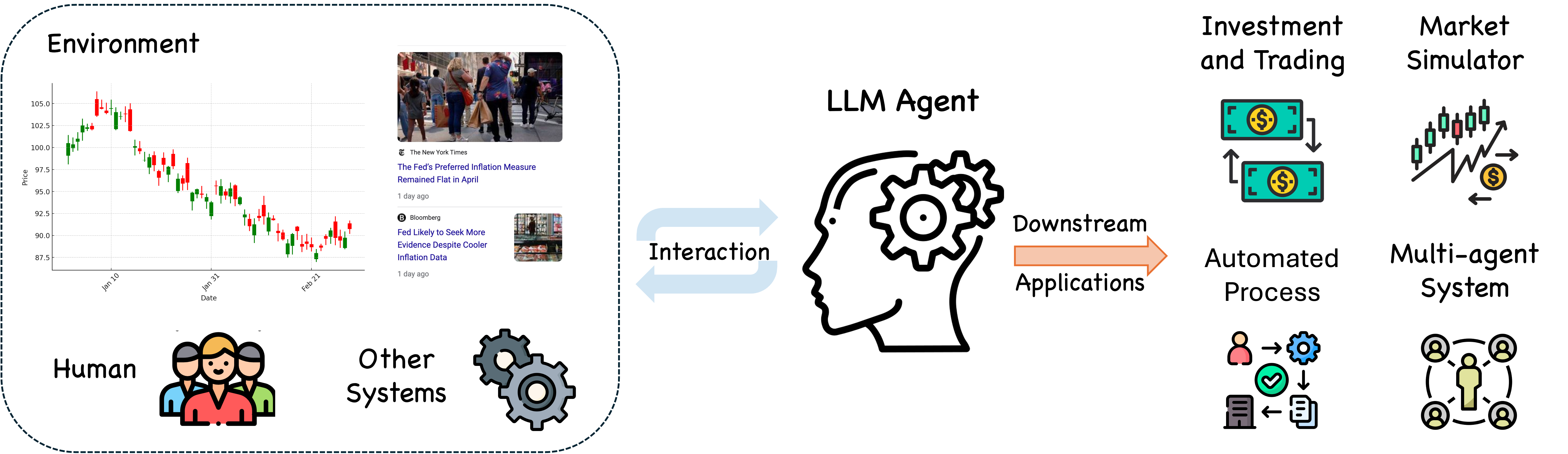} % Uncomment and replace with actual image
    %\fbox{\rule{0pt}{2in} \rule{0.9\linewidth}{0pt}} % Placeholder box
    \caption{Illustration of financial tasks related to agent-based modeling,}
    \label{fig:agent}
    \vspace{-1em}
\end{figure*}

\subsubsection{Trading and Investments}
The financial markets are dynamic and complex, requiring advanced tools to navigate effectively. LLMs have proven to be powerful allies in this domain by enabling the creation of intelligent trading agents that can process vast amounts of data and execute trades with high precision. These agents leverage LLMs' NLP capabilities to interpret and synthesize financial news, market reports, and historical data, significantly improving market predictions and trading strategies. \textbf{StockAgent} \cite{liuai}, for instance, explores the potential of AI-driven trading systems to simulate and analyze stock market behaviors under various external influences. It is a multi-agent system powered by LLMs designed to mimic real investor behaviors and assess the impact of external factors like macroeconomic events, policy changes, and financial reports on trading activities. The study finds that different LLMs, such as GPT-3.5 Turbo and Gemini, exhibit distinct trading behaviors and preferences, with GPT agents showing more diverse and independent trading styles compared to the more homogeneous and trend-following behavior of Gemini agents. This variation suggests that LLM-based systems can offer personalized investment strategies and insights. The research also highlights that removing financial information or communication channels like BBS (Bulletin Board System) can significantly alter trading behaviors and market dynamics, underscoring the complexity and interdependence of factors influencing stock trading.

A notable advancement in LLM applications is integrating multimodal data—textual, numerical, and visual—into trading agents. \textbf{FinAgent} \cite{zhang2024finagent} exemplifies this by combining these data types to support quantitative and high-frequency trading including stocks and Crypto. Its diversified memory retrieval system and tool augmentation features enable FinAgent to interact with various data sources and tools, enhancing adaptability and performance in dynamic trading environments.

LLM-based trading agents excel in continuous learning and adaptation as well. \textbf{FINMEM} \cite{yu2024finmem} introduces a layered memory and character design, enhancing the agent's ability to process hierarchical financial data and convert insights into trading decisions. The memory module of FINMEM, inspired by human cognitive processes, includes working memory and layered long-term memory components. This design allows FINMEM to categorize and prioritize information based on its relevance and timeliness, retaining critical insights longer and enabling agile responses to new investment cues. Through real-world testing and continuous learning, FINMEM evolves its trading strategies, demonstrating improved decision-making and adaptability in volatile financial environments. Similarly, \textbf{QuantAgent} \cite{wang2024quantagent} focuses on self-improvement through a two-layer loop system. The inner loop refines responses using a knowledge base, while the outer loop involves real-world testing and knowledge enhancement. This iterative approach enables QuantAgent to autonomously extract financial signals and uncover viable trading opportunities, showcasing LLMs' dynamic potential.

Integrating human expertise with AI capabilities is another significant advancement. The \textbf{Alpha-GPT} series, including Alpha-GPT \cite{wang2023alpha} and Alpha-GPT 2.0 \cite{yuan2024alpha}, emphasizes human-AI interaction in the alpha mining process. Alpha-GPT 2.0 further introduces a human-in-the-loop framework for iterative refinement of investment strategies. These agents interpret trading ideas and translate them into effective strategies, providing insightful and actionable alphas. By leveraging both human expertise and AI capabilities, this approach enhances the efficiency and creativity of the alpha mining process, leading to more effective investment decisions.

% \textit{StockAgent} uses LLMs for stock trading, employing the Gemini-pro model to analyze market trends and execute trades autonomously. This approach demonstrates the potential of LLMs to enhance trading strategies and improve market predictions.

% \textit{FinAgent}, a multimodal foundation agent, integrates textual, numerical, and visual data to support quantitative and high-frequency trading. This agent's diversified memory retrieval system and incorporation of established trading strategies highlight the versatility of LLMs in managing complex financial tasks.

% \textit{FINMEM} features a layered memory and character design, enhancing its ability to process hierarchical financial data and convert insights into trading decisions. This system allows for continuous evolution of trading knowledge, improving decision-making over time.

% \textit{QuantAgent} focuses on mining financial signals through a self-improving LLM framework. It employs a two-layer loop system for refining responses and real-world testing, showcasing the agent's ability to uncover viable trading signals and enhance forecast accuracy.

% The \textit{Alpha-GPT} series, including Alpha-GPT and Alpha-GPT 2.0, demonstrates the application of LLMs in interactive alpha mining for quantitative investment. These systems enhance human-AI interaction, translating trading ideas into effective strategies and improving the efficiency and creativity of the alpha mining process.

\subsubsection{Simulating Markets and Economic Activities}

Simulating markets and economic activities has long been a critical aspect of financial research and policy analysis. Traditional simulators, typically grounded in econometric models and system dynamics, have been the cornerstone of this effort. These simulators rely on historical data and established economic theories to predict future market behaviors. For instance, models like the Vector Autoregression (VAR) model and the Dynamic Stochastic General Equilibrium (DSGE) model are widely used for economic forecasting and policy analysis \cite{sims1980macroeconomics}, \cite{smets2003estimated}. While they offer a structured and mathematically rigorous approach, traditional simulators often struggle with the complexity and dynamism inherent in real-world economic systems. They are generally static, assuming rational behavior and equilibrium, which can limit their accuracy and adaptability to unforeseen economic shocks or behavioral intricacies.

In contrast, agent-based simulators represent a significant advancement in the simulation of economic activities. These models consist of autonomous agents, each with distinct behaviors and decision-making processes. These agents interact within a defined environment, allowing for the emergence of complex macroeconomic phenomena from the bottom up. The primary advantage of ABM lies in their flexibility and ability to model heterogeneous agents with varying strategies and interactions. This approach can capture the non-linear dynamics of markets, such as feedback loops, market sentiments, and adaptive behaviors \cite{tesfatsion2006agent}.

However, agent-based simulators are not without their challenges. One significant drawback is the computational complexity, as simulating numerous agents with intricate interactions demands substantial processing power. Additionally, the development of realistic agent behaviors and interaction rules requires deep domain expertise and can be time-consuming. Moreover, while agent-based simulators can model emergent phenomena, the validation of these models against real-world data remains a challenging task, often requiring extensive calibration and sensitivity analysis \cite{farmer2009economy}.

The integration of LLMs with agent-based simulators represents a cutting-edge development in economic simulations. With their advanced NLP capabilities, LLMs can enhance the perception, reflection, and decision-making processes of agents within simulators. This hybrid approach leverages the strengths of both technologies: the detailed and adaptive behaviors modeled by agent-based simulators and the comprehensive data processing and learning capabilities of LLMs.

Research by \citet{li2023large2} exemplifies the potential of this integration by demonstrating the ability to simulate complex macroeconomic activities. Their study, \textbf{EconAgent}, shows how LLM-empowered agents can realistically model economic activities by processing economic data through advanced mechanisms. These agents can simulate human-like decision-making processes, providing a comprehensive understanding of how different economic factors interact. This enables more accurate predictions of economic trends and the effects of policy changes. Equipped with layered memory systems, these agents can adapt their strategies based on real-time data inputs and historical analysis, making them highly effective for forecasting and policy simulation.

Similarly, \textbf{\citet{horton2023large}} explores the use of LLMs as computational models for economic simulations. By endowing LLMs with preferences and decision-making frameworks, their approach allows the simulation of human-like economic behavior. These simulations are particularly valuable for social science experiments and exploring economic scenarios, providing insights that can inform policy and strategy. The study introduces Homo Silicus agents, designed to emulate human economic agents by incorporating principles of behavioral economics. This enables the agents to make decisions based on a mix of rational analysis and emotional factors, providing a more realistic simulation of economic activities and market behaviors.

Furthermore, \citet{zhao2023competeai} investigate the competitive behaviors of LLM-based agents in a simulated environment, demonstrating how competition among agents can lead to the emergence of innovative strategies and enhanced performance. They propose \textbf{CompeteAI}, a framework that simulates a virtual town where restaurant agents compete for customers, revealing how competition drives agents to continually adapt and improve their strategies, aligning with established sociological and economic theories.

The evolution from traditional simulators to agent-based models and now to LLM-empowered agents marks a significant stride in the field of economic simulation. The integration of LLMs with ABM offers a promising avenue for more realistic and adaptive modeling of economic activities, capturing the complex interplay of factors that drive markets and economies. This hybrid approach not only enhances our understanding of economic dynamics but also provides a powerful tool for forecasting and policy analysis.

\subsubsection{Automated Financial Processes}
The integration of LLMs into financial processes has reformed the way financial tasks are automated, offering enhanced capabilities for workflow generation and strategic planning. These applications streamline operations and provide robust solutions for complex financial tasks.

One notable application is \textbf{FlowMind} \cite{zeng2023flowmind}, which presents an innovative approach to automating financial workflows using LLMs. FlowMind leverages the capabilities of models like GPT to generate workflows on-the-fly, addressing the limitations of traditional robotic process automation that relies on predefined tasks. The system uses a structured lecture recipe to ground LLM reasoning with reliable APIs, mitigating issues such as hallucinations and ensuring data privacy by avoiding direct interaction with proprietary code. FlowMind includes a feedback loop that allows users to inspect high-level descriptions of the generated workflows and provide adjustments, enhancing the system's adaptability. This approach is demonstrated using the NCEN-QA dataset, a benchmark for evaluating workflow generation in financial question-answering tasks, where FlowMind significantly outperforms traditional methods. This framework showcases the potential of LLMs to automate complex, spontaneous tasks in financial services while maintaining data integrity and security.

Another application is \textbf{AUCARENA} \cite{chen2023put}, which evaluates strategic planning and execution in auction environments to assess the strategic reasoning capabilities of LLM agents. In the ascending-price auctions, LLM agents like GPT-4 compete, managing budgets and adapting strategies in real-time. Utilizing a Belief-Desire-Intention model, agents update beliefs, adjust desires, and replan based on auction developments. This setup allows for a detailed analysis of how LLM agents manage resources, adhere to goals, and adapt to new information in competitive contexts. The study shows that LLM agents, especially GPT-4, are effective in strategic planning and resource management, though sometimes outperformed by simpler methods, highlighting areas for further improvement in LLM design. AUCARENA demonstrates the potential of LLMs to enhance decision-making processes in complex, competitive scenarios.

\subsubsection{Multi-agent Systems}

The use of multi-agent systems in financial analysis leverages the strengths of LLMs to enhance the robustness and accuracy of financial strategies. 
Multi-agent systems improve trading performance by simulating various agent interactions and providing a more comprehensive analysis of the tasks. \textbf{TradingGPT} \cite{li2023tradinggpt} exemplifies this approach with its innovative multi-agent framework designed for financial trading. It organizes memory into three distinct layers: short-term, medium-term, and long-term, each governed by a custom decay mechanism that matches human cognitive processes. In TradingGPT, agents can engage in inter-agent communication and debate, enhancing their decision-making capabilities. Each agent is equipped with individualized trading characters, such as risk-seeking, risk-neutral, and risk-averse profiles, which enrich the diversity of perspectives and improve decision-making robustness. By leveraging layered memory processing and consistent information exchange, this framework demonstrates augmented adaptability to historical trades and real-time market cues, significantly enhancing automated trading outcomes. 
Aside from trading tasks, \textbf{SocraPlan} \cite{tsao2023multi} leverages multi-agent reasoning with LLMs for effective corporate planning. This framework conducts comprehensive market research, customer profiling, product usage analysis, and sales strategy formulation. By combining human insights with AI capabilities, SocraPlan enhances corporate planning, enabling businesses to devise strategies that are both innovative and grounded in detailed market analysis. SocraPlan employs a multi-agent architecture where each agent specializes in a different aspect of corporate planning, such as competitive analysis, customer segmentation, or trend forecasting. These specialized agents collaborate to provide a holistic view of the market, which helps businesses make informed strategic decisions.

Multi-agent systems also benefit the in analyzing financial sentiments or textual information, which is a critical component of market analysis and strategy formulation as we have discussed in Section \ref{sec:linguistic} and \ref{sec:sentiment}. An example is \textbf{HAD} \cite{xing2024designing}, which indicates heterogeneous agent discussion, employing specialized agents focused on different types of errors common in FSA. This framework ensures that each of the agents focuses on particular errors, such as sarcasm, aspect mismatches, and temporal expressions, making the system robust against common pitfalls in sentiment analysis. The HAD framework has shown significant improvements in accuracy and F-1 scores across multiple datasets, proving its efficacy in refining sentiment analysis for financial texts. Another example is \cite{wan2024enhancing}, which introduces a multi-agent framework that automates the verification of information between loan applications and bank statements, powered by both open-sourced models like Llama 3 and close-sourced models such as GPT-4. Despite higher operational costs, this approach is more economical and faster than manual reviews, offering a reliable solution for structured finance auditing and compliance.

Moreover, multi-agent systems can be used for monitoring and anomaly detection in financial markets. 
%Enhancing Anomaly Detection in Financial Markets with an LLM-Based Multi-Agent Framework by 
\textbf{\citet{park2024enhancing}} introduces a sophisticated multi-agent framework designed to improve the validation and interpretation of financial data anomalies. The framework employs a network of specialized LLM agents, each focusing on distinct tasks such as data conversion, web-based expert analysis, utilization of institutional knowledge, cross-checking, and report consolidation. This collaborative approach enhances the efficiency and accuracy of anomaly detection, reducing the need for manual verification. By applying this framework to the S\&P 500 index, the study demonstrates significant improvements in anomaly detection, showing that LLM-based agents can autonomously and accurately identify and interpret anomalies in financial market data, thereby supporting more effective financial market monitoring and decision-making

Besides the multi-agent systems, agents can interact with itself in an autonomous way as well \cite{su2022competitive}. The Self-Reflective LLM framework \textbf{SEP} \cite{koa2024learning}, which means Summarize-Explain-Predict, addresses this need by enabling the generation of explainable stock predictions. SEP combines verbal self-reflective agents with Proximal Policy Optimization (PPO) to provide autonomous and explainable predictions. This framework allows agents to self-reflect on their decision-making processes, ensuring that the predictions are not only accurate but also interpretable. By enhancing the explainability of stock predictions, SEP improves accuracy, transparency, and trustworthiness among investors and analysts.
% \textit{TradingGPT} utilizes a multi-agent system with layered memory and distinct characters, facilitating enhanced financial trading performance. This system's ability to categorize and prioritize memory events closely aligns with human cognitive processes, improving the robustness of trading strategies.
% \textit{Heterogeneous LLM Agents} are designed for financial sentiment analysis, employing specialized agents to focus on different types of errors in sentiment analysis. This approach enhances the accuracy of sentiment analysis by leveraging insights from multiple agents.
% The \textit{multi-agent framework} for structured finance aims to improve the efficiency and accuracy of underlying asset reviews. By automating the verification of information between loan applications and bank statements, this system reduces manual errors and streamlines due diligence processes.
% \textit{Self-Reflective Large Language Models} in the SEP framework enable the generation of explainable stock predictions. This approach combines verbal self-reflective agents and Proximal Policy Optimization (PPO) to provide autonomous, explainable predictions, improving transparency and trust in financial forecasting.
\\~\\
In summary, the integration of LLMs into agent-based modeling in finance offers significant advancements in trading, investment, financial analysis, and economic simulation. These applications demonstrate the versatility and effectiveness of LLMs in enhancing decision-making, strategy formulation, and market analysis. Future research in this area promises to further refine these systems, improving their accuracy, efficiency, trustworthiness, and adaptability in the ever-evolving financial landscape \cite{sharma2024investigating}, \cite{zhang2024survey}, \cite{hua2024trustagent}.

% \subsection{Agent-based Modeling}
% \subsubsection{Simulation}
% \begin{itemize}
%     \item A detailed look at how LLMs simulate the actions and interactions of agents within a market, providing insights into market dynamics and potential future states.
% \end{itemize}

% \subsubsection{Tool user, organizer...}

% \subsubsection{Application}
% \begin{itemize}
%     \item Exploring specific applications of agent-based simulations powered by LLMs, including stress testing, scenario analysis, and the exploration of hypothetical market conditions.
% \end{itemize}

% \subsection{Other Applications}
% All other applications that not listed above.

\subsection{Other Applications}

Cloud computing could be integrated with LLMs to enhance scalability, efficiency, and cost-effectiveness across financial sectors. As discussed in previous sections, LLMs’ advanced NLP capabilities are being utilized to automate complex processes, improve customer interactions, and support decision-making in banking. The use of serverless architecture within cloud computing frameworks could provide a scalable and efficient platform for deploying these AI models, eliminating the need for traditional server management \cite{kathiriyaserverless}. By leveraging the synergy between LLMs and serverless computing, financial institutions can enhance operational resilience, ensure regulatory compliance, and maintain vendor independence. Practical implementations, such as Kore.AI and the Devin framework, have demonstrated the transformative impact of this integration. As the financial sector continues to evolve, the strategic use of LLMs in cloud computing has the potential to drive significant innovation, operational efficiency, and customer-centricity \cite{georgetransforming}.

\begin{table*}[]
    \caption{Benchmarks of LLMs on Financial Applications.}
    \label{tab: benchmark}
    \centering
    \begin{threeparttable}
    \begin{tabular}{ccccccc}
        \hline
        & \\[-1.5ex]
        Name & Year & Task & Modality & Model & Language & Open Source \\
        \hline
        \\[-2ex]
        PIXIU \cite{xie2023pixiu,xie2024finben} & 2023 & \makecell{Multiple financial NLP tasks, \\ Stock prediction} & \makecell{Text, Table, \\ Time-series} & FinMA & Chinese, English & Yes\textsuperscript{[1]} \\
        \hline
        \\[-2ex]
        FLUE \cite{shah2022flue} & 2022 & Multiple financial NLP tasks & Text & FLANG & English & Yes\textsuperscript{[2]} \\
        \hline
        \\[-2ex]
        AlphaFin \cite{li2024alphafin} & 2024 & \makecell{Financial question answering, \\ Stock prediction} & Text & Stock-Chain & Chinese, English & Yes\textsuperscript{[3]} \\
        \hline
        \\[-2ex]
        \citet{li2023chatgpt} & 2023 & Multiple financial NLP tasks & Text & - & English & - \\
        \hline
        \\[-2ex]
        BizBench \cite{koncel2023bizbench} & 2023 & \makecell{Multiple financial NLP tasks, \\ Program Synthesis} & \makecell{Text, Table, \\ Code} & - & English & Yes\textsuperscript{[4]} \\
        \hline
        \\[-2ex]
        DOCMATH-EVAL \cite{zhao2023docmath} & 2023 & Numerical reasoning & Text, Table & - & English & Yes\textsuperscript{[5]} \\
        \hline
        \\[-2ex]
        EconLogicQA \cite{quan2024econlogicqa} & 2024 & Financial question answering & Text & - & English & Yes\textsuperscript{[6]} \\
        \hline
        \\[-2ex]
        FINANCEBENCH \cite{islam2023financebench} & 2023 & Financial question answering & Text & - & English & Yes\textsuperscript{[7]} \\
        \hline
        \\[-2ex]
        \citet{lakkaraju2023llms} & 2023 & Financial advisement & Text & - & English & - \\
        \hline
        \\[-2ex]
        MultiLing 2019 \cite{el2019multiling} & 2019 & Financial narrative summarisation & Text & - & English & Yes\textsuperscript{[8]} \\
        \hline
        \\[-2ex]
        R-Judge \cite{yuan2024r} & 2024 & Safety judgment, Risk identification & Text & - & English & Yes\textsuperscript{[9]} \\
        \hline
        \\[-2ex]
        BBT-Fin \cite{lu2023bbt} & 2023 & Multiple financial NLP tasks & Text & BBT-FinT5 & Chinese & Yes\textsuperscript{[10]} \\
        \hline
        \\[-2ex]
        CFBenchmark \cite{lei2023cfbenchmark} & 2024 & Multiple financial NLP tasks & Text & - & Chinese & Yes\textsuperscript{[11]} \\
        \hline
        \\[-2ex]
         \citet{hirano2024construction} & 2024 & Multiple financial NLP tasks & Text & - & Japanese & Yes\textsuperscript{[12]} \\
        \hline
        \\[-2ex]
        FLARE-ES \cite{zhang2024d} & 2024 & Multiple financial NLP tasks & \makecell{Text, Table, \\ Time-series} & FinMA-ES & Spanish, English & Yes\textsuperscript{[1]} \\
        \hline
        \\[-2ex]
        FinEval \cite{zhang2023fineval} & 2023 & Financial domain knowledge & Text & - & Chinese & Yes\textsuperscript{[13]} \\
        \hline
        \\[-2ex]
        ICE-PIXIU \cite{hu2024no} & 2024 & Multiple financial NLP tasks & \makecell{Text, Table, \\ Time-series} & ICE-INTENT & Chinese, English & Yes\textsuperscript{[1]} \\
        \hline
        \\[-2ex]
        SuperCLUE-Fin \cite{xu2024superclue} & 2024 & Various financial tasks & Text & - & Chinese & Yes\textsuperscript{[14]} \\
        \hline
    \end{tabular}
    {
      \scriptsize
      \begin{tablenotes}
        \item[]
        [1] \href{https://github.com/The-FinAI/PIXIU}{https://github.com/The-FinAI/PIXIU}
        [2] \href{https://salt-nlp.github.io/FLANG/}{https://salt-nlp.github.io/FLANG/}
        [3] \href{https://github.com/AlphaFin-proj/AlphaFin}{https://github.com/AlphaFin-proj/AlphaFin}
        \item[]
        [4] \href{https://huggingface.co/datasets/kensho/BizBench}{https://huggingface.co/datasets/kensho/BizBench}
        [5] \href{https://github.com/yale-nlp/DocMath-Eval}{https://github.com/yale-nlp/DocMath-Eval}
        \item[]
        [6] \href{https://huggingface.co/datasets/yinzhu-quan/econlogicqa}{https://huggingface.co/datasets/yinzhu-quan/econlogicqa}
        [7] \href{https://github.com/patronus-ai/financebench}{https://github.com/patronus-ai/financebench}
        \item[]
        [8] \href{http://multiling.iit.demokritos.gr/pages/view/1754/multiling-2019}{http://multiling.iit.demokritos.gr/pages/view/1754/multiling-2019}
        [9] \href{https://github.com/Lordog/R-Judge}{https://github.com/Lordog/R-Judge}
        \item[]
        [10] \href{https://github.com/ssymmetry/BBT-FinCUGE-Application}{https://github.com/ssymmetry/BBT-FinCUGE-Application}
        [11] \href{https://github.com/TongjiFinLab/CFBenchmark}{https://github.com/TongjiFinLab/CFBenchmark}
        \item[]
        [12] \href{https://github.com/pfnet-research/japanese-lm-financial-benchmark}{https://github.com/pfnet-research/japanese-lm-financial-benchmark}
        \item[]
        [13] \href{https://github.com/SUFE-AIFLM-Lab/FinEval}{https://github.com/SUFE-AIFLM-Lab/FinEval}
        [14] \href{https://www.CLUEbenchmarks.com}{https://www.CLUEbenchmarks.com}
      \end{tablenotes}
    }
   \end{threeparttable}
\end{table*}

\section{Datasets, Code and Benchmark}
\label{sec: data}

\addtocontents{toc}{\protect\setcounter{tocdepth}{1}}
\subsection{Datasets}
\addtocontents{toc}{\protect\setcounter{tocdepth}{2}}

The datasets used in this survey paper cover a wide range of financial domains and tasks. These datasets are crucial for training and evaluating models on specific financial tasks such as sentiment analysis, question answering, relation extraction, and numerical reasoning. Several widely-used datasets include:

\begin{itemize}
    \item \textbf{Financial PhraseBank (FPB)} \cite{malo2014good}: This is a dataset consisting of financial phrases annotated with sentiment labels. It is widely used for sentiment analysis in financial contexts due to its detailed and domain-specific annotations.
    \item \textbf{Financial Question Answering and Opinion Mining (FiQA)} \cite{maia201818}: This dataset focuses on aspect-based sentiment analysis and opinion-based question answering. It includes financial news headlines and microblogs, annotated for sentiment and aspect categories. The dataset is designed to challenge models with tasks that require fine-grained sentiment and opinion extraction from financial texts.
    \item \textbf{FinQA} \cite{chen2021finqa}: A dataset designed for numerical reasoning over financial data. FinQA includes questions that require understanding and manipulating numerical information from financial reports. It emphasizes the need for models to perform complex reasoning tasks involving financial metrics and calculations.
\end{itemize}

Other datasets such as \textbf{ECTSum} \cite{mukherjee2022ectsum}, \textbf{FiNER} \cite{shah2023finer}, \textbf{FinRED} \cite{sharma2022finred}, \textbf{REFinD} \cite{kaur2023refind}, \textbf{FinSBD} \cite{au2021finsbd} and \textbf{CFLUE} \cite{zhu2024benchmarking} contribute to various specific financial NLP tasks. These include earnings call summarization, named entity recognition, relation extraction, and financial language understanding evaluations. Collectively, these datasets provide a robust foundation for developing and benchmarking LLMs in financial applications.

\addtocontents{toc}{\protect\setcounter{tocdepth}{1}}
\subsection{Benchmarks and Code}
\addtocontents{toc}{\protect\setcounter{tocdepth}{2}}

Furthermore, we outline the comprehensive benchmark used to assess LLM performance in the financial domain. Robust benchmarks are vital as they provide standardized measures to objectively compare models, ensuring reliability and accuracy in financial text understanding and prediction. This systematic evaluation fosters transparency, reproducibility, and continuous improvement in LLM applications. Sharing code and methodologies promotes collaboration, driving innovation and practical implementation in real-world financial scenarios.

%Two significant contributions to this field are the "WHEN FLUE MEETS FLANG: Benchmarks and Large Pre-trained Language Model for Financial Domain" and "PIXIU: A Large Language Model, Instruction Data and Evaluation Benchmark for Finance." These benchmarks are among the most popular and have been instrumental in advancing the capabilities of LLMs for financial applications.

A notable work in this field is \textbf{FLUE} \cite{shah2022flue}, which denotes Financial Language Understanding Evaluation, addressing the unique challenges posed by financial texts. It is a comprehensive suite of benchmarks designed to assess the performance of language models on various financial NLP tasks.
FLUE consists of five tasks: financial sentiment analysis using the FPB dataset, news headline classification based on the gold news headline dataset, named entity recognition with data from financial agreements, structure boundary detection using the FinSBD dataset, and question answering with data from the FiQA challenge. 
Besides, this paper introduces FLANG-BERT and FLANG-ELECTRA, two models specifically trained on financial data using a novel pre-training methodology that incorporates financial keywords and phrases for better masking, as well as span boundary and in-filing objectives. which we have introduced in Section \ref{sec: models}. 
These benchmarks cover a range of tasks critical for financial NLP, providing a robust platform to evaluate the effectiveness of financial language models.

\textbf{PIXIU} \cite{xie2023pixiu} represents a more recent development in the field, introducing a comprehensive framework that includes a financial LLM called FinMA, a large-scale multi-task instruction dataset, and a holistic evaluation benchmark named FLARE (Financial Language Understanding And Prediction Evaluation Benchmark). 
PIXIU is characterized by its open resources, making all components, including the model, instruction tuning data, and benchmarks, publicly available to promote transparency and further research. The instruction tuning data in PIXIU covers various financial tasks and modalities, including text, tables, and time-series data, ensuring comprehensive model training. The FLARE benchmark evaluates models on four financial NLP tasks (sentiment analysis, news headline classification, named entity recognition, and question answering) and one financial prediction task (stock movement prediction), covering nine datasets in total. This broad assessment allows for a thorough evaluation of a model's capabilities in handling diverse financial data, providing a more holistic benchmark compared to previous ones focused solely on NLP.

\vspace{5pt}

In addition, various other benchmarks have been developed to evaluate LLMs on a wide range of financial tasks. These benchmarks are closely related to the real-world applications that we discussed in the previous sections, including linguistic tasks, sentiment analysis, numerical reasoning, and comprehensive financial analysis. For example, \textbf{\citet{li2023chatgpt}} explore the effectiveness of LLMs in financial text analytics. 
\textbf{MultiLing 2019} \cite{el2019multiling} and \textbf{BizBench}  \cite{koncel2023bizbench} evaluate models on their ability to summarize financial narratives and perform quantitative reasoning in business and finance contexts.
For interpretable financial prediction, benchmarks like \textbf{AlphaFin} \cite{li2024alphafin} and \textbf{FinanceBench} \cite{islam2023financebench} assess models on stock trend prediction and financial question answering. Numerical reasoning capabilities are evaluated using benchmarks like \textbf{DocMath-Eval} \cite{zhao2023docmath}, which tests models on interpreting and calculating complex financial data from long documents. 
Comprehensive benchmarks like \textbf{R-Judge} \cite{yuan2024r} and \textbf{EconLogicQA} \cite{quan2024econlogicqa} focus on assessing risk awareness, safety in financial decision-making, and sequential reasoning within economic contexts. 
Together, these benchmarks provide a promising development for evaluating the diverse capabilities of LLMs in financial applications, ensuring models are tested across a broad spectrum of tasks.
%%%%%%%%%%%%%%%%%%%%%%%%%%%%%%%%%%%%%%%%%%%%%%%%%%
\\~\\
\textbf{Impact of Language:}
Besides the aforementioned benchmarks, the language impact on the performance of financial LLMs has become another topic of interest and has been extensively explored. This research often focuses on creating benchmarks for specific languages or comparing model performance across different languages to understand their effectiveness in diverse linguistic contexts.

Several benchmarks have been developed to evaluate models on tasks such as sentiment analysis, named entity recognition, relation extraction, and financial news summarization in the Chinese financial domain. Benchmarks like \textbf{BBT-Fin} \cite{lu2023bbt} and \textbf{CFBenchmark} \cite{lei2023cfbenchmark} are designed to provide comprehensive datasets and evaluation frameworks tailored to the linguistic and financial nuances of Chinese texts. Similarly, \textbf{FinEval} \cite{zhang2023fineval} and \textbf{SuperCLUE-Fin} \cite{xu2024superclue} focus on a broader range of financial tasks, advancing Chinese financial NLP by addressing both theoretical knowledge and practical applications such as compliance, risk management, and investment analysis. 
%On the other hand, the \textbf{DISC-FinLLM} \cite{chen2023disc} benchmark emphasizes multi-expert fine-tuning to enhance model performance across various financial tasks, providing a detailed assessment of models' capabilities in handling complex financial scenarios.

In the Japanese context, benchmarks such as the one developed by \textbf{\citet{hirano2024construction}} evaluate models on tasks like sentiment analysis, auditing tasks from the Japanese CPA (Certified Public Accountant) exam, and financial planner exam questions. This benchmark provides a robust framework to assess models' proficiency in Japanese financial texts.

Furthermore, there are several researches that explore bilingual capabilities to examine the performance of financial LLMs between different languages. \textbf{\citet{zhang2024d}} focuses on the comparison between Spanish and English, highlighting the challenges and effectiveness of models in processing and understanding financial texts across these languages.  \textbf{\citet{hu2024no}} extends this comparison to Chinese and English, providing insights into the models' generalization and adaptation capabilities across diverse linguistic contexts.

\vspace{10pt}

These language-specific benchmarks and comparative studies are crucial for understanding the linguistic impact on financial LLMs. They ensure that models are capable of accurately processing and interpreting financial information in various major languages, thereby broadening their applicability and effectiveness in global financial markets.

\section{Challenges and Opportunities}
\label{sec: challenges}
Despite the numerous advantages of integrating LLMs in financial applications, as discussed in section \ref{sec: why llm}, it is crucial to acknowledge and address the challenges that come with this innovative approach. Alongside these challenges, there exist significant opportunities for further development and refinement of LLMs in financial applications. In this section, we will explore the key challenges and opportunities associated with the use of LLMs in the financial sector, focusing on how researchers and practitioners can collaborate to overcome obstacles and unlock new possibilities for data-driven decision-making.

%\addtocontents{toc}{\protect\setcounter{tocdepth}{1}}
\subsection{Data Issues}
%\addtocontents{toc}{\protect\setcounter{tocdepth}{2}}

\textbf{Handle High-Dimensional Financial Data:}
While LLMs have demonstrated remarkable proficiency in processing and understanding contextual information within long text sequences, their performance in handling high-dimensional financial time series data remains uncertain. The unique challenges posed by the complex and highly dimensional nature of financial data present an opportunity for further research and exploration. By investigating the potential of hybrid models that combine the contextual understanding of LLMs with specialized techniques for handling high-dimensional data, domain-specific pre-training strategies, and the integration of LLMs with other machine learning techniques, researchers can develop powerful AI models tailored to analyze and understand financial time series. These advancements could ultimately enhance the performance and applicability of AI in the financial sector, leading to more accurate predictions, better risk management, and improved decision-making processes.
\\~\\
\textbf{Data Pollution:}
Data pollution could be a multifaceted challenge that can significantly impact the performance and reliability of these LLM models. The first aspect of data pollution involves the presence of inaccurate, misleading, or irrelevant data, such as spam advertisements or deliberately spreading misinformation that has been fed into this LLM model. This type of data pollution can severely degrade the performance of LLMs, leading to poor decision-making and compromised integrity of financial models, particularly when using cloud-based LLMs like ChatGPT, as the contamination can spread throughout the entire training environment.

Besides, a second and increasingly important aspect of data pollution lies in the growing trend of data being generated by LLMs themselves, rather than by humans. This phenomenon raises concerns about the quality and relevance of the data used to train these models. For instance, if financial reports are generated by LLMs, the models are essentially learning from their own output, which can result in increasingly rigid and inflexible learning. The models may fail to capture the true intentions and nuances of human expression, leading to a deterioration in the quality of the generated content.

To address this issue, major companies are strongly emphasizing the collection of high-quality, diverse datasets that include real human interactions. One potential solution to mitigate the impact of LLM-generated data pollution is to develop evaluation methods to assess the meaningfulness of the data created by LLMs. In this case, we can enhance the performance and reliability of these models, ultimately leading to more accurate and trustworthy financial analyses and predictions.
\\~\\
\textbf{Signal Decay:}
The widespread adoption of LLMs for generating trading strategies in the rapidly evolving financial world poses a unique challenge: signal decay. As more market participants employ LLMs, the effectiveness of these strategies may diminish over time, leading to a depletion of profitable market signals. However, this challenge also presents an opportunity to develop adaptive LLMs that continuously learn from new data and evolve alongside market conditions. By leveraging their ability to process vast amounts of financial information and identify emerging patterns, these models can maintain their effectiveness over time through continuous retraining and validation. 

%\addtocontents{toc}{\protect\setcounter{tocdepth}{1}}
\subsection{Modeling Issues}
%\addtocontents{toc}{\protect\setcounter{tocdepth}{2}}

\textbf{Inference Speed and Cost:}
Balancing the need for fast and cost-effective model inference with performance requirements is a significant challenge due to the high computational demands of LLMs. This can sometimes lead to high inference costs and slower speeds, particularly when processing large datasets. However, advances in model optimization and hardware offer exciting opportunities to reduce these costs and improve speeds. This makes LLMs more accessible and practical for various financial applications, promoting more efficient resource use and wider adoption of LLM technologies in the finance industry.

For instance, a hybrid inference approach, as discussed in the work by \citet{ding2024hybrid}, proposes using a router to dynamically allocate queries to either a small or large model based on the predicted query difficulty and the required quality level. This method aims to balance the trade-off between cost and performance effectively. The router can be fine-tuned to ensure that simpler queries are handled by the smaller, less expensive models, while the more complex queries are directed to the larger, more powerful models. This approach can lead to significant cost savings—up to 40\% fewer calls to the large model—without compromising the response quality. Such optimization can make LLMs more economically viable for financial applications, where precision and speed are critical, thereby enhancing their adoption and utility across various financial services and operations. 
\\~\\
\textbf{Future Lookahead Bias in Financial Backtesting:}
Backtesting financial models using LLMs presents a significant challenge due to the risk of future lookahead bias \cite{sarkar2024lookahead}. This bias occurs when the model inadvertently incorporates information from the future during the training process, leading to overly optimistic and misleading backtesting results. Consequently, the model’s reliability and predictive accuracy come into question, as it may not perform as well on unseen, real-time data. Addressing this issue requires careful handling of data and the implementation of robust validation techniques to ensure the integrity of the backtesting process.

Despite the challenges posed by future lookahead bias, researchers can explore innovative solutions to address this issue and design LLMs that effectively mitigate its impact. One of the straightforward methods, as addressed by \citet{kim2024financial}, is to use anonymized data that cannot be identified by the LLM. This ensures that the LLM cannot leverage its pre-trained memory when dealing with specific firm questions. However, robust validation techniques should still be implemented. The authors perform formal analyses to further rule out concerns about look-ahead bias.

Similarly, recent work in \cite{drinkall2024time} specifically designed a series of point-in-time LLMs called \textbf{T}ime\textbf{M}achine\textbf{GPT} (TiMaGPT). These models are trained on datasets that maintain temporal integrity, ensuring that they remain uninformed about future factual information and linguistic changes. By avoiding the incorporation of future information during training, TiMaGPT models can provide more accurate and reliable insights for time-series forecasting and other dynamic contexts in financial modeling. The availability of both the models and training datasets further enhances the transparency and reproducibility of the results.
\\~\\
\textbf{Hallucinations in LLM Financial Outputs:}
The use of LLM-generated content in various financial tasks raises significant concerns about legality and reliability. Financial reports are subject to strict legal and regulatory standards, and inaccuracies can lead to severe consequences for organizations. A primary issue with LLMs is their potential to generate fake, hallucinated, or factually incorrect statements due to their training on vast amounts of data. Ensuring LLM-generated content adheres to legal standards and is error-free is complex and requires careful consideration and monitoring, especially when the output may not undergo the same scrutiny as a full financial report. The lack of standardized frameworks and guidelines for robot-generated financial content could further complicate this process. 

To address the challenge of ensuring accuracy and trustworthiness in LLM-generated financial content, leveraging advanced tools like GenAudit \cite{krishna2024genaudit} presents significant opportunities. GenAudit is designed to assist in fact-checking LLM responses for document-grounded tasks. It suggests edits by revising or removing unsupported claims and presents evidence for supported facts. Comprehensive evaluations by human raters demonstrate GenAudit’s effectiveness in detecting errors across various LLM outputs from diverse domains. The system is designed to increase error recall while minimizing the impact on precision, ensuring that most errors are flagged and corrected.
\\~\\
\textbf{Uncertainty Estimation for LLM responses:}
Estimating the uncertainty and providing confidence intervals for model predictions is critical in finance because LLM outputs are essentially sampled from a distribution, rather than being deterministic. This means that asking the LLM the same question multiple times may yield different responses, with some samples potentially having significant errors. For financial decision-making or forecasting, relying on a single sample can be misleading. Moreover, when applying these predictions practically, the error range remains unknown, making risk control challenging. Therefore, to manage risk, it is necessary to perform uncertainty estimation on LLM outputs and establish confidence intervals for their predictions. This approach helps control errors and mitigate risks. Developing sophisticated methods for uncertainty quantification can provide more reliable confidence intervals, thereby enhancing risk management and decision-making in finance. It allows stakeholders to make more informed and confident decisions based on LLM predictions.

%\addtocontents{toc}{\protect\setcounter{tocdepth}{1}}
\subsection{Benchmarking}
%\addtocontents{toc}{\protect\setcounter{tocdepth}{2}}

\textbf{Evaluation of Trading Strategies:}
In addition to the aforementioned signal decay caused by the widespread adoption of LLM models, another significant challenge in constructing trading strategies using LLMs lies in the evaluation process. The difficulty arises from the fact that the benchmarks currently used to test LLM-generated signals were constructed before the emergence of LLMs. As a result, the environment has changed, making it challenging to effectively evaluate the effectiveness of these LLM-generated signals. The benchmarks that were once suitable for evaluating trading signals in a pre-LLM environment may no longer be applicable, as the widespread availability of LLMs has altered the landscape. This change in the environment is not a gradual decay but rather a fundamental shift that requires a new approach to evaluation. To address this issue, it is crucial to develop new benchmarks that are adaptable to LLMs and aligned with the current state of the market. Without such benchmarks, it becomes difficult to accurately assess the performance of LLM-generated signals, leading to uncertainty regarding their effectiveness. Therefore, in addition to the traditional signal decay problem, the evaluation difficulty posed by the changed environment should also be recognized and addressed to effectively leverage LLMs in constructing trading strategies.
\\~\\
\textbf{Interpretability:}
The lack of interpretability in LLMs used within the finance industry presents a significant challenge. Stakeholders require a clear understanding of how these models arrive at their decisions to establish trust and effectively utilize their outputs. Developing methods to enhance the transparency and interpretability of LLMs is an ongoing effort \cite{yu2023temporal, lopez2023can}. By investing in research aimed at improving the interpretability of LLMs, financial institutions can build trust and transparency in their AI-driven processes, leading to better decision-making and increased adoption of LLMs in the financial sector. As described in PloutosGPT \cite{tong2024ploutos}, two quantifiable metrics—faithfulness and informativeness—are employed to verify the quality of the interpretability of the generated rationales. Faithfulness measures whether the facts in the model’s response are based on or can be inferred from the given knowledge, while informativeness measures the amount of information contained in the model’s response. The development of tools that explain model decisions can help stakeholders comprehend and effectively use the insights generated by AI.

%\addtocontents{toc}{\protect\setcounter{tocdepth}{1}}
\subsection{Ethical Issues}
%\addtocontents{toc}{\protect\setcounter{tocdepth}{2}}
\textbf{Benign Alignment:}
Ensuring that LLMs output content that aligns with social values and avoids harmful recommendations is a key concern \cite{yao2024survey}. This involves making sure that the outputs conform not only to ethical standards but also to legal regulations, avoiding suggestions that could lead to negative actions. This issue intersects with both attack prevention and safety measures. The challenge lies in aligning the objectives of LLMs with benign and ethical goals, as misaligned models can produce unintended and potentially harmful consequences. Therefore, it is crucial to ensure that LLMs operate within ethical boundaries and adhere to regulatory standards. The opportunity here is to proactively align LLM objectives with ethical standards to mitigate risks and ensure that these models contribute positively, particularly in the financial sector. This includes developing frameworks for ethical AI \cite{cao2023defending, he2024s} use in finance, which can foster trust and compliance.
\\~\\
\textbf{Legal Responsibility:}
As LLMs continue to play an increasingly significant role in financial decision-making, the issues of legal responsibility and accountability become more prominent. The complexity of these models and their potential for misuse present unique challenges in determining accountability when things go wrong. It is crucial to establish clear frameworks and regulations to address these concerns. The development of a well-defined legal framework for the use of LLMs in the financial sector is essential to provide certainty and foster confidence among stakeholders. By clarifying the lines of responsibility and accountability, such a framework can promote the widespread adoption of these technologies while ensuring their responsible use. This framework should allocate liability in cases where LLMs are misused or produce unintended consequences, establish standards for the development, testing, and deployment of LLMs in financial applications, and provide mechanisms for red flags and compensation in cases where LLMs cause financial harm. 
\\~\\
\textbf{Safety and Privacy:}
The security and privacy of financial data are extremely important, given the significant threats posed by data breaches and compliance violations. Deploying LLMs in the financial sector presents unique challenges in maintaining robust data protection measures and safeguarding sensitive information. However, advancements in cybersecurity can bolster the security and privacy of financial data used by LLMs. By implementing strong security protocols, we can mitigate the risks of data leaks and ensure adherence to privacy regulations, thereby building trust and protecting sensitive information. To further prevent data leakage, especially with cloud-based GPT models, it is essential to process confidential data in a local environment. This approach minimizes the risk of breaches while still leveraging the capabilities of LLMs. With the growing availability of open-source models, organizations can now utilize LLMs within their local infrastructure, ensuring the security and privacy of their financial data while benefiting from the advanced features these models offer.
\\~\\
\textbf{Understanding Incentives:}
The highly competitive nature of the financial industry, coupled with the massive amounts of capital outcomes, necessitates a careful examination of the incentives driving the development and application of LLMs. As LLMs become increasingly prevalent in various domains, including finance, it is crucial to consider their potential impact on individuals and organizations, including government agencies. 

Ethical concerns surrounding AI are growing. Professional organizations like the Association for Computing Machinery (ACM) \cite{acm_code_of_ethics} have developed codes of ethics and conduct to guide the development and use of AI technologies. However, unlike regulated professions such as medicine, law, or engineering, where practitioners are bound by professional designations and can face consequences for violating ethical standards, LLM developers are not subject to similar oversight. This lack of formal accountability mechanisms poses challenges in ensuring that LLM developers adhere to established ethical guidelines. Furthermore, LLMs themselves, when performing reasoning and decision-making, are likely to operate in an opaque manner, creating barriers to uncover and understand all their potential incentives, particularly those that may lead to negative ethical implications.

To address this, there is a pressing need for greater transparency regarding the incentives behind LLM recommendations. For instance, the fund industry has been moving towards clear reporting of management fees for fund managers. A similar approach should be adopted for LLMs to systematically evaluate their impact on stakeholders. Europe has taken proactive steps with the A.I. Act \cite{eu_ai_regulation}, adopting a "risk-based approach" to regulate high-risk applications and mitigate potential harms such as racial biases. This framework highlights the challenge of balancing effective regulation with fostering innovation. 

As LLMs continue to evolve and integrate into the financial industry, understanding and aligning incentives will be critical to ensuring their responsible and beneficial application. This may involve a combination of approaches, including developing and enforcing industry-specific ethical guidelines and best practices, understanding breakdowns of what data was used to train the systems, promoting transparency in LLM development and recommendation processes, implementing accountability mechanisms to verify compliance with ethical standards, encouraging collaboration between LLM developers, domain experts, ethicists, and regulators, and educating stakeholders about the capabilities, limitations, and potential risks of LLMs.

In the end, natural language takes place in various situations: to inform, to persuade, to entertain, to educate, and so on.   Thus, we would expect LLMs to be employed within these constructs. While humans have exquisite talents with situational awareness, it will be interesting to see if LLMs will be able to develop their own skills in this regard. As the financial industry increasingly adopts LLMs, a proactive and collaborative approach to addressing ethical concerns, aligning incentives, and ensuring responsible application will be essential to harnessing the benefits of this transformative technology while mitigating potential harms.

% \begin{itemize}
%     \item Future Lookahead Bias on Backtest (opputunities: generate/ track some indicators which originally have laggs)
%     \item Financial Report Generation: Legal Concerns and Reliability of Robot-Generated Content
%     \item Data pollution
%     \item Signal Decay and Effectiveness of Strategies in the Age of LLM
%     \item Fairness in Accessing and Using LLM
%     \item Inference speed and cost.
%     \item Uncertainty estimate / confidence interval.
%     \item High dimensional data vs low dimensional data.
%     \item Benign Alignment
%     \item Interpretability
%     \item Legal responsibility
%     \item Safety and Privacy (risk of leaking data)
%     \item Mathematical reasoning
%     \item Real-time monitoring
%     \item database
% \end{itemize}

\section{Conclusion}
\label{sec: conclusion}

This survey provides a comprehensive overview of the application of LLMs in the financial domain, highlighting their capabilities in enhancing various financial tasks such as linguistic tasks, sentiment analysis, financial time series analysis, financial reasoning, and agent-based modeling. LLMs demonstrate remarkable potential in improving the efficiency and accuracy of financial processes through advanced contextual understanding and real-time analysis.

Despite their promising capabilities, challenges such as data privacy, interpretability, and computational costs need to be addressed to ensure the responsible and effective deployment of LLMs in finance. By summarizing the current state, advantages, and limitations of LLMs in financial applications, this survey aims to inspire further research and innovation. As research continues to evolve, it is our hope that this review will encourage more exploration and discussion on the potential and limitations of LLMs, advancing their integration into the financial sector for more strategic investment and efficient decision-making.

%\appendices
%\section{Proof of the First Zonklar Equation}
%Appendix one text goes here.

% you can choose not to have a title for an appendix
% if you want by leaving the argument blank
%\section{}
%Appendix two text goes here.

% use section* for acknowledgment
\ifCLASSOPTIONcompsoc
  % The Computer Society usually uses the plural form
  \section*{Acknowledgments}
\else
  % % regular IEEE prefers the singular form
  % \section*{Acknowledgment}
\fi
This work was supported in part by a grant from Princeton Language and Intelligence.
X.D. acknowledges support from the Oxford-Man Institute of Quantitative Finance.
% Can use something like this to put references on a page
% by themselves when using endfloat and the captionsoff option.
\ifCLASSOPTIONcaptionsoff
  \newpage
\fi

\bibliographystyle{IEEEtranN}
\bibliography{references}

% Generated by IEEEtranN.bst, version: 1.14 (2015/08/26)
\begin{thebibliography}{318}
\providecommand{\natexlab}[1]{#1}
\providecommand{\url}[1]{#1}
\csname url@samestyle\endcsname
\providecommand{\newblock}{\relax}
\providecommand{\bibinfo}[2]{#2}
\providecommand{\BIBentrySTDinterwordspacing}{\spaceskip=0pt\relax}
\providecommand{\BIBentryALTinterwordstretchfactor}{4}
\providecommand{\BIBentryALTinterwordspacing}{\spaceskip=\fontdimen2\font plus
\BIBentryALTinterwordstretchfactor\fontdimen3\font minus \fontdimen4\font\relax}
\providecommand{\BIBforeignlanguage}[2]{{%
\expandafter\ifx\csname l@#1\endcsname\relax
\typeout{** WARNING: IEEEtranN.bst: No hyphenation pattern has been}%
\typeout{** loaded for the language `#1'. Using the pattern for}%
\typeout{** the default language instead.}%
\else
\language=\csname l@#1\endcsname
\fi
#2}}
\providecommand{\BIBdecl}{\relax}
\BIBdecl

\bibitem[Mulvey et~al.(2022)Mulvey, Gu, Holen, and Nie]{mulvey2022applications}
J.~M. Mulvey, J.~Gu, M.~Holen, and Y.~Nie, ``Applications of machine learning in wealth management,'' \emph{Journal of Investment Consulting}, vol.~21, no.~1, pp. 66--82, 2022.

\bibitem[Lee et~al.(2024)Lee, Stevens, Han, and Song]{lee2024survey}
J.~Lee, N.~Stevens, S.~C. Han, and M.~Song, ``A survey of large language models in finance {(FinLLMs)},'' \emph{arXiv preprint arXiv:2402.02315}, 2024.

\bibitem[Li et~al.(2023{\natexlab{a}})Li, Wang, Ding, and Chen]{li2023large}
Y.~Li, S.~Wang, H.~Ding, and H.~Chen, ``Large language models in finance: A survey,'' in \emph{Proceedings of the Fourth ACM International Conference on AI in Finance}, 2023, pp. 374--382.

\bibitem[Dong et~al.(2023)Dong, Stratopoulos, and Wang]{dong2023scoping}
M.~M. Dong, T.~C. Stratopoulos, and V.~X. Wang, ``A scoping review of {ChatGPT} research in accounting and finance,'' \emph{Theophanis C. and Wang, Victor Xiaoqi, A Scoping Review of ChatGPT Research in Accounting and Finance (December 30, 2023)}, 2023.

\bibitem[Zhao et~al.(2024)Zhao, Liu, Wu, Li, Yang, Shu, Xu, Dai, Zhao, Mai, et~al.]{zhao2024revolutionizing}
H.~Zhao, Z.~Liu, Z.~Wu, Y.~Li, T.~Yang, P.~Shu, S.~Xu, H.~Dai, L.~Zhao, G.~Mai \emph{et~al.}, ``Revolutionizing finance with {LLMs}: An overview of applications and insights,'' \emph{arXiv preprint arXiv:2401.11641}, 2024.

\bibitem[Wu et~al.(2023)Wu, Irsoy, Lu, Dabravolski, Dredze, Gehrmann, Kambadur, Rosenberg, and Mann]{wu2023bloomberggpt}
S.~Wu, O.~Irsoy, S.~Lu, V.~Dabravolski, M.~Dredze, S.~Gehrmann, P.~Kambadur, D.~Rosenberg, and G.~Mann, ``{BloombergGPT}: A large language model for finance,'' \emph{arXiv preprint arXiv:2303.17564}, 2023.

\bibitem[Liu et~al.(2023{\natexlab{a}})Liu, Zhu, Wu, Yang, You, Wang, Lu, Liu, Zheng, Sun, et~al.]{liu2023medical}
F.~Liu, T.~Zhu, X.~Wu, B.~Yang, C.~You, C.~Wang, L.~Lu, Z.~Liu, Y.~Zheng, X.~Sun \emph{et~al.}, ``A medical multimodal large language model for future pandemics,'' \emph{NPJ Digital Medicine}, vol.~6, no.~1, p. 226, 2023.

\bibitem[Wang et~al.(2024{\natexlab{a}})Wang, Xu, Li, Zhang, Liang, Tang, Yu, and Wen]{wang2024large}
S.~Wang, T.~Xu, H.~Li, C.~Zhang, J.~Liang, J.~Tang, P.~S. Yu, and Q.~Wen, ``Large language models for education: A survey and outlook,'' \emph{arXiv preprint arXiv:2403.18105}, 2024.

\bibitem[Radford et~al.(2018)Radford, Narasimhan, Salimans, Sutskever, et~al.]{radford2018improving}
A.~Radford, K.~Narasimhan, T.~Salimans, I.~Sutskever \emph{et~al.}, ``Improving language understanding by generative pre-training,'' 2018.

\bibitem[Radford et~al.(2019)Radford, Wu, Child, Luan, Amodei, Sutskever, et~al.]{radford2019language}
A.~Radford, J.~Wu, R.~Child, D.~Luan, D.~Amodei, I.~Sutskever \emph{et~al.}, ``Language models are unsupervised multitask learners,'' \emph{OpenAI blog}, vol.~1, no.~8, p.~9, 2019.

\bibitem[Brown et~al.(2020)Brown, Mann, Ryder, Subbiah, Kaplan, Dhariwal, Neelakantan, Shyam, Sastry, Askell, et~al.]{brown2020language}
T.~Brown, B.~Mann, N.~Ryder, M.~Subbiah, J.~D. Kaplan, P.~Dhariwal, A.~Neelakantan, P.~Shyam, G.~Sastry, A.~Askell \emph{et~al.}, ``Language models are few-shot learners,'' \emph{Advances in Neural Information Processing Systems}, vol.~33, pp. 1877--1901, 2020.

\bibitem[Achiam et~al.(2023)Achiam, Adler, Agarwal, Ahmad, Akkaya, Aleman, Almeida, Altenschmidt, Altman, Anadkat, et~al.]{achiam2023gpt}
J.~Achiam, S.~Adler, S.~Agarwal, L.~Ahmad, I.~Akkaya, F.~L. Aleman, D.~Almeida, J.~Altenschmidt, S.~Altman, S.~Anadkat \emph{et~al.}, ``{GPT-4} technical report,'' \emph{arXiv preprint arXiv:2303.08774}, 2023.

\bibitem[Tong et~al.(2024)Tong, Li, Wu, Gong, Zhang, and Zhang]{tong2024ploutos}
H.~Tong, J.~Li, N.~Wu, M.~Gong, D.~Zhang, and Q.~Zhang, ``Ploutos: Towards interpretable stock movement prediction with financial large language model,'' \emph{arXiv preprint arXiv:2403.00782}, 2024.

\bibitem[Devlin et~al.(2018)Devlin, Chang, Lee, and Toutanova]{devlin2018bert}
J.~Devlin, M.-W. Chang, K.~Lee, and K.~Toutanova, ``{BERT}: Pre-training of deep bidirectional transformers for language understanding,'' \emph{arXiv preprint arXiv:1810.04805}, 2018.

\bibitem[Araci(2019)]{araci2019finbert}
D.~Araci, ``{FinBERT}: Financial sentiment analysis with pre-trained language models,'' \emph{arXiv preprint arXiv:1908.10063}, 2019.

\bibitem[Yang et~al.(2020)Yang, Uy, and Huang]{yang2020finbert}
Y.~Yang, M.~C.~S. Uy, and A.~Huang, ``{FinBERT}: A pretrained language model for financial communications,'' \emph{arXiv preprint arXiv:2006.08097}, 2020.

\bibitem[Liu et~al.(2021)Liu, Huang, Huang, Li, and Zhao]{liu2021finbert}
Z.~Liu, D.~Huang, K.~Huang, Z.~Li, and J.~Zhao, ``{FinBERT}: A pre-trained financial language representation model for financial text mining,'' in \emph{Proceedings of the Twenty-Ninth International Conference on International Joint Conferences on Artificial Intelligence}, 2021, pp. 4513--4519.

\bibitem[Liu et~al.(2019)Liu, Ott, Goyal, Du, Joshi, Chen, Levy, Lewis, Zettlemoyer, and Stoyanov]{liu2019roberta}
Y.~Liu, M.~Ott, N.~Goyal, J.~Du, M.~Joshi, D.~Chen, O.~Levy, M.~Lewis, L.~Zettlemoyer, and V.~Stoyanov, ``{RoBERTa}: A robustly optimized bert pretraining approach,'' \emph{arXiv preprint arXiv:1907.11692}, 2019.

\bibitem[Zhang et~al.(2021)Zhang, Zhang, Chen, Guo, Hua, Wang, and Zhou]{zhang2021mengzi}
Z.~Zhang, H.~Zhang, K.~Chen, Y.~Guo, J.~Hua, Y.~Wang, and M.~Zhou, ``Mengzi: Towards lightweight yet ingenious pre-trained models for {Chinese},'' 2021.

\bibitem[Raffel et~al.(2020)Raffel, Shazeer, Roberts, Lee, Narang, Matena, Zhou, Li, and Liu]{raffel2020exploring}
C.~Raffel, N.~Shazeer, A.~Roberts, K.~Lee, S.~Narang, M.~Matena, Y.~Zhou, W.~Li, and P.~J. Liu, ``Exploring the limits of transfer learning with a unified text-to-text transformer,'' \emph{Journal of Machine Learning Research}, vol.~21, no. 140, pp. 1--67, 2020.

\bibitem[Lu et~al.(2023{\natexlab{a}})Lu, Wu, Liang, Xu, He, Geng, Han, Xin, and Xiao]{lu2023bbt}
D.~Lu, H.~Wu, J.~Liang, Y.~Xu, Q.~He, Y.~Geng, M.~Han, Y.~Xin, and Y.~Xiao, ``{BBT-Fin}: Comprehensive construction of {Chinese} financial domain pre-trained language model, corpus and benchmark,'' \emph{arXiv preprint arXiv:2302.09432}, 2023.

\bibitem[Clark et~al.(2020)Clark, Luong, Le, and Manning]{clark2020electra}
K.~Clark, M.-T. Luong, Q.~V. Le, and C.~D. Manning, ``{ELECTRA}: Pre-training text encoders as discriminators rather than generators,'' \emph{arXiv preprint arXiv:2003.10555}, 2020.

\bibitem[Shah et~al.(2022)Shah, Chawla, Eidnani, Shah, Du, Chava, Raman, Smiley, Chen, and Yang]{shah2022flue}
R.~S. Shah, K.~Chawla, D.~Eidnani, A.~Shah, W.~Du, S.~Chava, N.~Raman, C.~Smiley, J.~Chen, and D.~Yang, ``{WHEN FLUE MEETS FLANG}: Benchmarks and large pre-trained language model for financial domain,'' \emph{arXiv preprint arXiv:2211.00083}, 2022.

\bibitem[Le~Scao et~al.(2023)Le~Scao, Fan, Akiki, Pavlick, Ili{\'c}, Hesslow, Castagn{\'e}, Luccioni, Yvon, Gall{\'e}, et~al.]{le2023bloom}
T.~Le~Scao, A.~Fan, C.~Akiki, E.~Pavlick, S.~Ili{\'c}, D.~Hesslow, R.~Castagn{\'e}, A.~S. Luccioni, F.~Yvon, M.~Gall{\'e} \emph{et~al.}, ``{BLOOM}: A 176{B}-parameter open-access multilingual language model,'' 2023.

\bibitem[Zhang and Yang(2023)]{zhang2023xuanyuan}
X.~Zhang and Q.~Yang, ``{XuanYuan 2.0}: A large {Chinese} financial chat model with hundreds of billions parameters,'' in \emph{Proceedings of the 32nd ACM International Conference on Information and Knowledge Management}, 2023, pp. 4435--4439.

\bibitem[Touvron et~al.(2023{\natexlab{a}})Touvron, Lavril, Izacard, Martinet, Lachaux, Lacroix, Rozi{\`e}re, Goyal, Hambro, Azhar, et~al.]{touvron2023llama}
H.~Touvron, T.~Lavril, G.~Izacard, X.~Martinet, M.-A. Lachaux, T.~Lacroix, B.~Rozi{\`e}re, N.~Goyal, E.~Hambro, F.~Azhar \emph{et~al.}, ``{LLaMA}: Open and efficient foundation language models,'' \emph{arXiv preprint arXiv:2302.13971}, 2023.

\bibitem[Xie et~al.(2023{\natexlab{a}})Xie, Han, Zhang, Lai, Peng, Lopez-Lira, and Huang]{xie2023pixiu}
Q.~Xie, W.~Han, X.~Zhang, Y.~Lai, M.~Peng, A.~Lopez-Lira, and J.~Huang, ``{PIXIU}: A large language model, instruction data and evaluation benchmark for finance,'' \emph{arXiv preprint arXiv:2306.05443}, 2023.

\bibitem[William~Todt(2023)]{Fin-LLAMA}
P.~B. William~Todt, Ramtin~Babaei, ``{Fin-LLAMA}: Efficient finetuning of quantized {LLMs} for finance,'' \url{https://github.com/Bavest/fin-llama}, 2023.

\bibitem[Yu(2023)]{Cornucopia-LLaMA-Fin-Chinese}
Y.~Yu, ``{Cornucopia-LLaMA-Fin-Chinese},'' \url{https://github.com/jerry1993-tech/Cornucopia-LLaMA-Fin-Chinese}, 2023.

\bibitem[Zhang et~al.(2023{\natexlab{a}})Zhang, Yang, and Liu]{zhang2023instruct}
B.~Zhang, H.~Yang, and X.-Y. Liu, ``{Instruct-FinGPT}: Financial sentiment analysis by instruction tuning of general-purpose large language models,'' \emph{arXiv preprint arXiv:2306.12659}, 2023.

\bibitem[Yang et~al.(2023{\natexlab{a}})Yang, Tang, and Tam]{yang2023investlm}
Y.~Yang, Y.~Tang, and K.~Y. Tam, ``{InvestLM}: A large language model for investment using financial domain instruction tuning,'' \emph{arXiv preprint arXiv:2309.13064}, 2023.

\bibitem[Touvron et~al.(2023{\natexlab{b}})Touvron, Martin, Stone, Albert, Almahairi, Babaei, Bashlykov, Batra, Bhargava, Bhosale, et~al.]{touvron2023llama2}
H.~Touvron, L.~Martin, K.~Stone, P.~Albert, A.~Almahairi, Y.~Babaei, N.~Bashlykov, S.~Batra, P.~Bhargava, S.~Bhosale \emph{et~al.}, ``{LLAMA 2}: Open foundation and fine-tuned chat models,'' \emph{arXiv preprint arXiv:2307.09288}, 2023.

\bibitem[Yang et~al.(2023{\natexlab{b}})Yang, Liu, and Wang]{yang2023fingpt}
H.~Yang, X.-Y. Liu, and C.~D. Wang, ``{FinGPT}: Open-source financial large language models,'' \emph{arXiv preprint arXiv:2306.06031}, 2023.

\bibitem[Konstantinidis et~al.(2024)Konstantinidis, Iacovides, Xu, Constantinides, and Mandic]{konstantinidis2024finllama}
T.~Konstantinidis, G.~Iacovides, M.~Xu, T.~G. Constantinides, and D.~Mandic, ``{FinLlama}: Financial sentiment classification for algorithmic trading applications,'' \emph{arXiv preprint arXiv:2403.12285}, 2024.

\bibitem[Yu et~al.(2024{\natexlab{a}})Yu, Huber, and Tang]{yu2024greedllama}
J.~Yu, M.~Huber, and K.~Tang, ``{GreedLlama}: Performance of financial value-aligned large language models in moral reasoning,'' \emph{arXiv preprint arXiv:2404.02934}, 2024.

\bibitem[{Meta AI}(2024)]{meta2024llama3}
\BIBentryALTinterwordspacing
{Meta AI}, ``Introducing {Meta} {Llama 3}: The most capable openly available {LLM} to date,'' 2024. [Online]. Available: \url{https://ai.meta.com/blog/meta-llama-3/}
\BIBentrySTDinterwordspacing

\bibitem[Bhatia et~al.(2024)Bhatia, Nagoudi, Cavusoglu, and Abdul-Mageed]{bhatia2024fintral}
G.~Bhatia, E.~M.~B. Nagoudi, H.~Cavusoglu, and M.~Abdul-Mageed, ``{FinTral}: A family of {GPT-4} level multimodal financial large language models,'' \emph{arXiv preprint arXiv:2402.10986}, 2024.

\bibitem[Jiang et~al.(2023{\natexlab{a}})Jiang, Sablayrolles, Mensch, Bamford, Chaplot, Casas, Bressand, Lengyel, Lample, Saulnier, et~al.]{jiang2023mistral}
A.~Q. Jiang, A.~Sablayrolles, A.~Mensch, C.~Bamford, D.~S. Chaplot, D.~d.~l. Casas, F.~Bressand, G.~Lengyel, G.~Lample, L.~Saulnier \emph{et~al.}, ``{Mistral 7B},'' \emph{arXiv preprint arXiv:2310.06825}, 2023.

\bibitem[Zhou et~al.(2024{\natexlab{a}})Zhou, Li, Tian, Ni, Liu, Ye, and Chai]{zhou2024silversight}
Y.~Zhou, Z.~Li, S.~Tian, Y.~Ni, S.~Liu, G.~Ye, and H.~Chai, ``{SilverSight}: A multi-task {Chinese} financial large language model based on adaptive semantic space learning,'' \emph{arXiv preprint arXiv:2404.04949}, 2024.

\bibitem[Bai et~al.(2023)Bai, Bai, Chu, Cui, Dang, Deng, Fan, Ge, Han, Huang, et~al.]{bai2023qwen}
J.~Bai, S.~Bai, Y.~Chu, Z.~Cui, K.~Dang, X.~Deng, Y.~Fan, W.~Ge, Y.~Han, F.~Huang \emph{et~al.}, ``Qwen technical report,'' \emph{arXiv preprint arXiv:2309.16609}, 2023.

\bibitem[Chen et~al.(2023{\natexlab{a}})Chen, Wang, Long, Zhang, Lu, Li, Wang, Xu, Bai, Huang, et~al.]{chen2023disc}
W.~Chen, Q.~Wang, Z.~Long, X.~Zhang, Z.~Lu, B.~Li, S.~Wang, J.~Xu, X.~Bai, X.~Huang \emph{et~al.}, ``{DISC-FinLLM}: A {Chinese} financial large language model based on multiple experts fine-tuning,'' \emph{arXiv preprint arXiv:2310.15205}, 2023.

\bibitem[{Baichuan-inc}(2023)]{baichuan2023}
\BIBentryALTinterwordspacing
{Baichuan-inc}, ``{Baichuan-13B},'' 2023. [Online]. Available: \url{https://github.com/baichuan-inc/Baichuan-13B}
\BIBentrySTDinterwordspacing

\bibitem[Li et~al.(2023{\natexlab{b}})Li, Bian, Wang, Lei, Cheng, Ding, and Jiang]{li2023cfgpt}
J.~Li, Y.~Bian, G.~Wang, Y.~Lei, D.~Cheng, Z.~Ding, and C.~Jiang, ``{CFGPT}: {Chinese} financial assistant with large language model,'' \emph{arXiv preprint arXiv:2309.10654}, 2023.

\bibitem[{InternLM}(2024)]{internlm}
\BIBentryALTinterwordspacing
{InternLM}, ``{InternLM},'' 2024. [Online]. Available: \url{https://github.com/InternLM}
\BIBentrySTDinterwordspacing

\bibitem[Wang et~al.(2023{\natexlab{a}})Wang, Li, Wu, Soon, and Zhang]{wang2023finvis}
Z.~Wang, Y.~Li, J.~Wu, J.~Soon, and X.~Zhang, ``{FinVis-GPT}: A multimodal large language model for financial chart analysis,'' \emph{arXiv preprint arXiv:2308.01430}, 2023.

\bibitem[Liu et~al.(2023{\natexlab{b}})Liu, Li, Wu, and Lee]{liu2023llava}
H.~Liu, C.~Li, Q.~Wu, and Y.~J. Lee, ``Visual instruction tuning,'' 2023.

\bibitem[Wei et~al.(2021)Wei, Bosma, Zhao, Guu, Yu, Lester, Du, Dai, and Le]{wei2021finetuned}
J.~Wei, M.~Bosma, V.~Y. Zhao, K.~Guu, A.~W. Yu, B.~Lester, N.~Du, A.~M. Dai, and Q.~V. Le, ``Finetuned language models are zero-shot learners,'' \emph{arXiv preprint arXiv:2109.01652}, 2021.

\bibitem[Zhang et~al.(2023{\natexlab{b}})Zhang, Yang, Zhou, Ali~Babar, and Liu]{zhang2023enhancing}
B.~Zhang, H.~Yang, T.~Zhou, M.~Ali~Babar, and X.-Y. Liu, ``Enhancing financial sentiment analysis via retrieval augmented large language models,'' in \emph{Proceedings of the Fourth ACM International Conference on AI in Finance}, 2023, pp. 349--356.

\bibitem[Hu et~al.(2021)Hu, Shen, Wallis, Allen-Zhu, Li, Wang, Wang, and Chen]{hu2021lora}
E.~J. Hu, Y.~Shen, P.~Wallis, Z.~Allen-Zhu, Y.~Li, S.~Wang, L.~Wang, and W.~Chen, ``{LoRA}: Low-rank adaptation of large language models,'' \emph{arXiv preprint arXiv:2106.09685}, 2021.

\bibitem[Ma et~al.(2024{\natexlab{a}})Ma, Wang, Ma, Wang, Wang, Huang, Dong, Wang, Xue, and Wei]{ma2024era}
S.~Ma, H.~Wang, L.~Ma, L.~Wang, W.~Wang, S.~Huang, L.~Dong, R.~Wang, J.~Xue, and F.~Wei, ``The era of 1-bit {LLMs}: All large language models are in 1.58 bits,'' \emph{arXiv preprint arXiv:2402.17764}, 2024.

\bibitem[Dettmers et~al.(2024)Dettmers, Pagnoni, Holtzman, and Zettlemoyer]{dettmers2024qlora}
T.~Dettmers, A.~Pagnoni, A.~Holtzman, and L.~Zettlemoyer, ``{QLoRA}: Efficient finetuning of quantized {LLMs},'' \emph{Advances in Neural Information Processing Systems}, vol.~36, 2024.

\bibitem[Liu et~al.(2024)Liu, Zhang, Wang, Tong, and Walid]{liu2024fingpt}
X.-Y. Liu, J.~Zhang, G.~Wang, W.~Tong, and A.~Walid, ``{FinGPT-HPC}: Efficient pretraining and finetuning large language models for financial applications with high-performance computing,'' \emph{arXiv preprint arXiv:2402.13533}, 2024.

\bibitem[Pavlyshenko(2023)]{pavlyshenko2023financial}
B.~M. Pavlyshenko, ``Financial news analytics using fine-tuned {Llama 2 GPT} model,'' \emph{arXiv preprint arXiv:2308.13032}, 2023.

\bibitem[Yakar et~al.(2019)Yakar, Nie, Wada, Agarwal, and Ercan]{yakar2019energy}
O.~Yakar, Y.~Nie, K.~Wada, A.~Agarwal, and {\.I}.~Ercan, ``Energy efficiency of microring resonator {(MRR)}-based binary decision diagram {(BDD)} circuits,'' in \emph{2019 IEEE International Conference on Rebooting Computing (ICRC)}.\hskip 1em plus 0.5em minus 0.4em\relax IEEE, 2019, pp. 1--8.

\bibitem[Nie and Yuan(2021)]{nie2021neural}
Y.~Nie and H.~Yuan, ``Neural network is heterogeneous: Phase matters more,'' \emph{arXiv preprint arXiv:2111.02014}, 2021.

\bibitem[Wang et~al.(2022{\natexlab{a}})Wang, Sun, and Boukerche]{wang2022novel2}
Z.~Wang, P.~Sun, and A.~Boukerche, ``A novel time efficient machine learning-based traffic flow prediction method for large scale road network,'' in \emph{ICC 2022-IEEE International Conference on Communications}.\hskip 1em plus 0.5em minus 0.4em\relax IEEE, 2022, pp. 3532--3537.

\bibitem[Rodriguez~Inserte et~al.(2024)Rodriguez~Inserte, Nakhl{\'e}, Qader, Caillaut, and Liu]{rodriguez2024large}
P.~Rodriguez~Inserte, M.~Nakhl{\'e}, R.~Qader, G.~Caillaut, and J.~Liu, ``Large language model adaptation for financial sentiment analysis,'' \emph{arXiv e-prints}, pp. arXiv--2401, 2024.

\bibitem[Deng et~al.(2023)Deng, Bashlovkina, Han, Baumgartner, and Bendersky]{deng2023llms}
X.~Deng, V.~Bashlovkina, F.~Han, S.~Baumgartner, and M.~Bendersky, ``What do {LLMs} know about financial markets? a case study on {Reddit} market sentiment analysis,'' in \emph{Companion Proceedings of the ACM Web Conference 2023}, 2023, pp. 107--110.

\bibitem[Wei et~al.(2022)Wei, Tay, Bommasani, Raffel, Zoph, Borgeaud, Yogatama, Bosma, Zhou, Metzler, et~al.]{wei2022emergent}
J.~Wei, Y.~Tay, R.~Bommasani, C.~Raffel, B.~Zoph, S.~Borgeaud, D.~Yogatama, M.~Bosma, D.~Zhou, D.~Metzler \emph{et~al.}, ``Emergent abilities of large language models,'' \emph{arXiv preprint arXiv:2206.07682}, 2022.

\bibitem[Steinert and Altmann(2023)]{steinert2023linking}
R.~Steinert and S.~Altmann, ``Linking microblogging sentiments to stock price movement: An application of {GPT-4},'' \emph{arXiv preprint arXiv:2308.16771}, 2023.

\bibitem[Jiang et~al.(2023{\natexlab{b}})Jiang, Kelly, and Xiu]{jiang2023re}
J.~Jiang, B.~Kelly, and D.~Xiu, ``{(Re-) Imag (in) ing} price trends,'' \emph{The Journal of Finance}, vol.~78, no.~6, pp. 3193--3249, 2023.

\bibitem[Wang et~al.(2023{\natexlab{b}})Wang, Sun, Lei, Zhu, and Sun]{wang2023sst}
Z.~Wang, Y.~Sun, Z.~Lei, X.~Zhu, and P.~Sun, ``{SST}: A simplified swin transformer-based model for taxi destination prediction based on existing trajectory,'' in \emph{2023 IEEE 26th International Conference on Intelligent Transportation Systems (ITSC)}.\hskip 1em plus 0.5em minus 0.4em\relax IEEE, 2023, pp. 1404--1409.

\bibitem[Lipton et~al.(2015)Lipton, Berkowitz, and Elkan]{lipton2015critical}
Z.~C. Lipton, J.~Berkowitz, and C.~Elkan, ``A critical review of recurrent neural networks for sequence learning,'' \emph{arXiv preprint arXiv:1506.00019}, 2015.

\bibitem[Staudemeyer and Morris(2019)]{staudemeyer2019understanding}
R.~C. Staudemeyer and E.~R. Morris, ``Understanding {LSTM}--a tutorial into long short-term memory recurrent neural networks,'' \emph{arXiv preprint arXiv:1909.09586}, 2019.

\bibitem[Zmandar et~al.(2021)Zmandar, Singh, El-Haj, and Rayson]{zmandar2021joint}
N.~Zmandar, A.~Singh, M.~El-Haj, and P.~Rayson, ``Joint abstractive and extractive method for long financial document summarization,'' in \emph{Proceedings of the 3rd Financial Narrative Processing Workshop}, 2021, pp. 99--105.

\bibitem[Hadi et~al.(2023)Hadi, Qureshi, Shah, Irfan, Zafar, Shaikh, Akhtar, Wu, Mirjalili, et~al.]{hadi2023large}
M.~U. Hadi, R.~Qureshi, A.~Shah, M.~Irfan, A.~Zafar, M.~B. Shaikh, N.~Akhtar, J.~Wu, S.~Mirjalili \emph{et~al.}, ``Large language models: a comprehensive survey of its applications, challenges, limitations, and future prospects,'' \emph{Authorea Preprints}, 2023.

\bibitem[Raiaan et~al.(2024)Raiaan, Mukta, Fatema, Fahad, Sakib, Mim, Ahmad, Ali, and Azam]{raiaan2024review}
M.~A.~K. Raiaan, M.~S.~H. Mukta, K.~Fatema, N.~M. Fahad, S.~Sakib, M.~M.~J. Mim, J.~Ahmad, M.~E. Ali, and S.~Azam, ``A review on large language models: Architectures, applications, taxonomies, open issues and challenges,'' \emph{IEEE Access}, 2024.

\bibitem[Abdaljalil and Bouamor(2021)]{abdaljalil2021exploration}
S.~Abdaljalil and H.~Bouamor, ``An exploration of automatic text summarization of financial reports,'' in \emph{Proceedings of the Third Workshop on Financial Technology and Natural Language Processing}, 2021, pp. 1--7.

\bibitem[La~Quatra and Cagliero(2020)]{la2020end}
M.~La~Quatra and L.~Cagliero, ``End-to-end training for financial report summarization,'' in \emph{Proceedings of the 1st Joint Workshop on Financial Narrative Processing and MultiLing Financial Summarisation}, 2020, pp. 118--123.

\bibitem[Ni et~al.(2023)Ni, Li, and Li]{ni2023unified}
X.~Ni, P.~Li, and H.~Li, ``Unified text structuralization with instruction-tuned language models,'' \emph{arXiv preprint arXiv:2303.14956}, 2023.

\bibitem[Xia et~al.(2022)Xia, Rawte, Zaki, Gupta, et~al.]{xia2022fetilda}
B.~Xia, V.~D. Rawte, M.~J. Zaki, A.~Gupta \emph{et~al.}, ``{FETILDA}: An effective framework for fin-tuned embeddings for long financial text documents,'' \emph{arXiv preprint arXiv:2206.06952}, 2022.

\bibitem[Vanetik et~al.(2023)Vanetik, Podkaminer, and Litvak]{vanetik2023summarizing}
N.~Vanetik, E.~Podkaminer, and M.~Litvak, ``Summarizing financial reports with positional language model,'' in \emph{2023 IEEE International Conference on Big Data (BigData)}.\hskip 1em plus 0.5em minus 0.4em\relax IEEE, 2023, pp. 2877--2883.

\bibitem[Yepes et~al.(2024)Yepes, You, Milczek, Laverde, and Li]{yepes2024financial}
A.~J. Yepes, Y.~You, J.~Milczek, S.~Laverde, and L.~Li, ``Financial report chunking for effective retrieval augmented generation,'' \emph{arXiv preprint arXiv:2402.05131}, 2024.

\bibitem[Shukla et~al.(2022)Shukla, Vaid, Katikeri, Keeriyadath, and Raja]{shukla2022dimsum}
N.~Shukla, A.~Vaid, R.~Katikeri, S.~Keeriyadath, and M.~Raja, ``{DiMSum}: Distributed and multilingual summarization of financial narratives,'' in \emph{Proceedings of the 4th Financial Narrative Processing Workshop@ LREC2022}, 2022, pp. 65--72.

\bibitem[Shukla et~al.(2023)Shukla, Katikeri, Raja, Sivam, Yadav, Vaid, and Prabhakararao]{shukla2023generative}
N.~K. Shukla, R.~Katikeri, M.~Raja, G.~Sivam, S.~Yadav, A.~Vaid, and S.~Prabhakararao, ``Generative {AI} approach to distributed summarization of financial narratives,'' in \emph{2023 IEEE International Conference on Big Data (BigData)}.\hskip 1em plus 0.5em minus 0.4em\relax IEEE, 2023, pp. 2872--2876.

\bibitem[Khanna et~al.(2022)Khanna, Ghodratnama, Beheshti, et~al.]{khanna2022transformer}
U.~Khanna, S.~Ghodratnama, A.~Beheshti \emph{et~al.}, ``Transformer-based models for long document summarisation in financial domain,'' in \emph{Proceedings of the 4th Financial Narrative Processing Workshop@ LREC2022}, 2022, pp. 73--78.

\bibitem[Beltagy et~al.(2020)Beltagy, Peters, and Cohan]{beltagy2020longformer}
I.~Beltagy, M.~E. Peters, and A.~Cohan, ``Longformer: The long-document transformer,'' \emph{arXiv preprint arXiv:2004.05150}, 2020.

\bibitem[Foroutan et~al.(2022)Foroutan, Romanou, Massonnet, Lebret, and Aberer]{foroutan2022multilingual}
N.~Foroutan, A.~Romanou, S.~Massonnet, R.~Lebret, and K.~Aberer, ``Multilingual text summarization on financial documents,'' in \emph{Proceedings of the 4th Financial Narrative Processing Workshop@ LREC2022}, 2022, pp. 53--58.

\bibitem[Suzuki et~al.(2023)Suzuki, Sakaji, Hirano, and Izumi]{suzuki2023constructing}
M.~Suzuki, H.~Sakaji, M.~Hirano, and K.~Izumi, ``Constructing and analyzing domain-specific language model for financial text mining,'' \emph{Information Processing \& Management}, vol.~60, no.~2, p. 103194, 2023.

\bibitem[Avramelou et~al.(2023)Avramelou, Passalis, Tsoumakas, and Tefas]{avramelou2023domain}
L.~Avramelou, N.~Passalis, G.~Tsoumakas, and A.~Tefas, ``Domain-specific large language model finetuning using a model assistant for financial text summarization,'' in \emph{2023 IEEE Symposium Series on Computational Intelligence (SSCI)}.\hskip 1em plus 0.5em minus 0.4em\relax IEEE, 2023, pp. 381--386.

\bibitem[Li and Zhang(2021)]{li2021unified}
Q.~Li and Q.~Zhang, ``A unified model for financial event classification, detection and summarization,'' in \emph{Proceedings of the Twenty-Ninth International Conference on International Joint Conferences on Artificial Intelligence}, 2021, pp. 4668--4674.

\bibitem[Sarmah et~al.(2023)Sarmah, Mehta, Pasquali, and Zhu]{sarmah2023towards}
B.~Sarmah, D.~Mehta, S.~Pasquali, and T.~Zhu, ``Towards reducing hallucination in extracting information from financial reports using large language models,'' in \emph{Proceedings of the Third International Conference on AI-ML Systems}, 2023, pp. 1--5.

\bibitem[Gupta(2023)]{gupta2023gpt}
U.~Gupta, ``{GPT-InvestAR}: Enhancing stock investment strategies through annual report analysis with large language models,'' \emph{arXiv preprint arXiv:2309.03079}, 2023.

\bibitem[Li et~al.(2023{\natexlab{c}})Li, Gao, Wu, and Vasarhelyi]{li2023extracting}
H.~Li, H.~H. Gao, C.~Wu, and M.~A. Vasarhelyi, ``Extracting financial data from unstructured sources: Leveraging large language models,'' \emph{Available at SSRN}, 2023.

\bibitem[Yue et~al.(2023)Yue, Xu, Ma, Du, Liu, Ding, Jiang, Han, and Zhang]{yue2023leveraging}
C.~Yue, X.~Xu, X.~Ma, L.~Du, H.~Liu, Z.~Ding, Y.~Jiang, S.~Han, and D.~Zhang, ``Leveraging {LLMs} for {KPIs} retrieval from hybrid long-document: A comprehensive framework and dataset,'' \emph{arXiv preprint arXiv:2305.16344}, 2023.

\bibitem[Wang et~al.(2023{\natexlab{c}})Wang, Raman, Sibue, Ma, Babkin, Kaur, Pei, Nourbakhsh, and Liu]{wang2023docllm}
D.~Wang, N.~Raman, M.~Sibue, Z.~Ma, P.~Babkin, S.~Kaur, Y.~Pei, A.~Nourbakhsh, and X.~Liu, ``{DocLLM}: A layout-aware generative language model for multimodal document understanding,'' \emph{arXiv preprint arXiv:2401.00908}, 2023.

\bibitem[Li et~al.(2020)Li, Sun, Han, and Li]{li2020survey}
J.~Li, A.~Sun, J.~Han, and C.~Li, ``A survey on deep learning for named entity recognition,'' \emph{IEEE transactions on knowledge and data engineering}, vol.~34, no.~1, pp. 50--70, 2020.

\bibitem[Ehrmann et~al.(2023)Ehrmann, Hamdi, Pontes, Romanello, and Doucet]{ehrmann2023named}
M.~Ehrmann, A.~Hamdi, E.~L. Pontes, M.~Romanello, and A.~Doucet, ``Named entity recognition and classification in historical documents: A survey,'' \emph{ACM Computing Surveys}, vol.~56, no.~2, pp. 1--47, 2023.

\bibitem[Swaileh et~al.(2020)Swaileh, Paquet, Adam, and Rojas~Camacho]{swaileh2020named}
W.~Swaileh, T.~Paquet, S.~Adam, and A.~Rojas~Camacho, ``A named entity extraction system for historical financial data,'' in \emph{Document Analysis Systems: 14th IAPR International Workshop, DAS 2020, Wuhan, China, July 26--29, 2020, Proceedings 14}.\hskip 1em plus 0.5em minus 0.4em\relax Springer, 2020, pp. 324--340.

\bibitem[Alvarado et~al.(2015)Alvarado, Verspoor, and Baldwin]{alvarado2015domain}
J.~C.~S. Alvarado, K.~Verspoor, and T.~Baldwin, ``Domain adaption of named entity recognition to support credit risk assessment,'' in \emph{Proceedings of the Australasian Language Technology Association Workshop 2015}, 2015, pp. 84--90.

\bibitem[Nadeau and Sekine(2007)]{nadeau2007survey}
D.~Nadeau and S.~Sekine, ``A survey of named entity recognition and classification,'' \emph{Lingvisticae Investigationes}, vol.~30, no.~1, pp. 3--26, 2007.

\bibitem[Eddy(1996)]{EDDY1996361}
\BIBentryALTinterwordspacing
S.~R. Eddy, ``Hidden markov models,'' \emph{Current Opinion in Structural Biology}, vol.~6, no.~3, pp. 361--365, 1996. [Online]. Available: \url{https://www.sciencedirect.com/science/article/pii/S0959440X9680056X}
\BIBentrySTDinterwordspacing

\bibitem[Quinlan(1986)]{Induction_of_decision_trees}
\BIBentryALTinterwordspacing
J.~R. Quinlan, ``Induction of decision trees,'' \emph{Machine Learning}, vol.~1, no.~1, pp. 81--106, 1986. [Online]. Available: \url{https://doi.org/10.1007/BF00116251}
\BIBentrySTDinterwordspacing

\bibitem[Hearst et~al.(1998)Hearst, Dumais, Osuna, Platt, and Scholkopf]{support_vector_machines}
M.~Hearst, S.~Dumais, E.~Osuna, J.~Platt, and B.~Scholkopf, ``Support vector machines,'' \emph{IEEE Intelligent Systems and their Applications}, vol.~13, no.~4, pp. 18--28, 1998.

\bibitem[Pakhale(2023)]{pakhale2023comprehensive}
K.~Pakhale, ``Comprehensive overview of named entity recognition: Models, domain-specific applications and challenges,'' \emph{arXiv preprint arXiv:2309.14084}, 2023.

\bibitem[Wang et~al.(2023{\natexlab{d}})Wang, Pan, Chen, and Cui]{wang2023named}
X.~Wang, X.~Pan, C.~Chen, and J.~Cui, ``A named entity recognition model based on {BERT} model and lexical fusion in the financial regulation field,'' in \emph{2023 5th International Conference on Frontiers Technology of Information and Computer (ICFTIC)}.\hskip 1em plus 0.5em minus 0.4em\relax IEEE, 2023, pp. 477--482.

\bibitem[Hillebrand et~al.(2022)Hillebrand, Deu{\ss}er, Dilmaghani, Kliem, Loitz, Bauckhage, and Sifa]{hillebrand2022kpi}
L.~Hillebrand, T.~Deu{\ss}er, T.~Dilmaghani, B.~Kliem, R.~Loitz, C.~Bauckhage, and R.~Sifa, ``{KPI-BERT}: A joint named entity recognition and relation extraction model for financial reports,'' in \emph{2022 26th International Conference on Pattern Recognition (ICPR)}.\hskip 1em plus 0.5em minus 0.4em\relax IEEE, 2022, pp. 606--612.

\bibitem[Loukas et~al.(2022)Loukas, Fergadiotis, Chalkidis, Spyropoulou, Malakasiotis, Androutsopoulos, and Paliouras]{loukas2022finer}
L.~Loukas, M.~Fergadiotis, I.~Chalkidis, E.~Spyropoulou, P.~Malakasiotis, I.~Androutsopoulos, and G.~Paliouras, ``{FiNER}: Financial numeric entity recognition for {XBRL} tagging,'' \emph{arXiv preprint arXiv:2203.06482}, 2022.

\bibitem[Covas(2023)]{covas2023named}
E.~Covas, ``Named entity recognition using {GPT} for identifying comparable companies,'' \emph{arXiv preprint arXiv:2307.07420}, 2023.

\bibitem[Zhao et~al.(2021)Zhao, Li, Zheng, and Zhang]{zhao2021bert}
L.~Zhao, L.~Li, X.~Zheng, and J.~Zhang, ``A {BERT} based sentiment analysis and key entity detection approach for online financial texts,'' in \emph{2021 IEEE 24th International Conference on Computer Supported Cooperative Work in Design (CSCWD)}.\hskip 1em plus 0.5em minus 0.4em\relax IEEE, 2021, pp. 1233--1238.

\bibitem[Gupta et~al.(2021)Gupta, Verma, Mashetty, and Mishra]{gupta2021context}
H.~Gupta, S.~Verma, S.~Mashetty, and S.~Mishra, ``{Context-NER}: Contextual phrase generation at scale,'' \emph{arXiv preprint arXiv:2109.08079}, 2021.

\bibitem[Zhou et~al.(2023)Zhou, Zhang, Gu, Chen, and Poon]{zhou2023universalner}
W.~Zhou, S.~Zhang, Y.~Gu, M.~Chen, and H.~Poon, ``{UniversalNER}: Targeted distillation from large language models for open named entity recognition,'' \emph{arXiv preprint arXiv:2308.03279}, 2023.

\bibitem[Xue et~al.(2023)Xue, Zhou, Xu, Zhao, Xie, Jiang, Zhang, Zhou, Xu, Xiu, et~al.]{xue2023weaverbird}
S.~Xue, F.~Zhou, Y.~Xu, H.~Zhao, S.~Xie, C.~Jiang, J.~Zhang, J.~Zhou, P.~Xu, D.~Xiu \emph{et~al.}, ``{WeaverBird}: Empowering financial decision-making with large language model, knowledge base, and search engine,'' \emph{arXiv preprint arXiv:2308.05361}, 2023.

\bibitem[van Zwam et~al.(2020)van Zwam, Khalili, Jessurun, Oberoi, Beerepoot, Fernandez, Bijman, Easton, and Karatas]{Deloitte2020KnowledgeGraphs}
\BIBentryALTinterwordspacing
M.~van Zwam, A.~Khalili, J.~Jessurun, S.~Oberoi, M.~Beerepoot, S.~Fernandez, J.~Bijman, A.~Easton, and I.~Karatas, ``Knowledge graphs for financial services: The path to unlock new insights from your data,'' 2020. [Online]. Available: \url{https://www2.deloitte.com/content/dam/Deloitte/de/Documents/operations/knowledge-graphs-pov.pdf}
\BIBentrySTDinterwordspacing

\bibitem[Jiang et~al.(2023{\natexlab{c}})Jiang, Xu, Shen, Sun, Tang, Wang, Chen, Wang, and Guo]{jiang2023evolution}
X.~Jiang, C.~Xu, Y.~Shen, X.~Sun, L.~Tang, S.~Wang, Z.~Chen, Y.~Wang, and J.~Guo, ``On the evolution of knowledge graphs: A survey and perspective,'' \emph{arXiv preprint arXiv:2310.04835}, 2023.

\bibitem[Pan et~al.(2023)Pan, Razniewski, Kalo, Singhania, Chen, Dietze, Jabeen, Omeliyanenko, Zhang, Lissandrini, et~al.]{pan2023large}
J.~Z. Pan, S.~Razniewski, J.-C. Kalo, S.~Singhania, J.~Chen, S.~Dietze, H.~Jabeen, J.~Omeliyanenko, W.~Zhang, M.~Lissandrini \emph{et~al.}, ``Large language models and knowledge graphs: Opportunities and challenges,'' \emph{arXiv preprint arXiv:2308.06374}, 2023.

\bibitem[Pan et~al.(2024{\natexlab{a}})Pan, Luo, Wang, Chen, Wang, and Wu]{pan2024unifying}
S.~Pan, L.~Luo, Y.~Wang, C.~Chen, J.~Wang, and X.~Wu, ``Unifying large language models and knowledge graphs: A roadmap,'' \emph{IEEE Transactions on Knowledge and Data Engineering}, 2024.

\bibitem[Trajanoska et~al.(2023)Trajanoska, Stojanov, and Trajanov]{trajanoska2023enhancing}
M.~Trajanoska, R.~Stojanov, and D.~Trajanov, ``Enhancing knowledge graph construction using large language models,'' \emph{arXiv preprint arXiv:2305.04676}, 2023.

\bibitem[Ouyang et~al.(2024)Ouyang, Liu, Li, Bao, Harimoto, and Sun]{ouyang2024modal}
K.~Ouyang, Y.~Liu, S.~Li, R.~Bao, K.~Harimoto, and X.~Sun, ``Modal-adaptive knowledge-enhanced graph-based financial prediction from monetary policy conference calls with {LLM},'' \emph{arXiv preprint arXiv:2403.16055}, 2024.

\bibitem[Wang et~al.(2022{\natexlab{b}})Wang, Sun, Chen, and Cui]{wang2022relation}
X.~Wang, Y.~Sun, C.~Chen, and J.~Cui, ``A relation extraction model based on {BERT} model in the financial regulation field,'' in \emph{2022 2nd International Conference on Computer Science, Electronic Information Engineering and Intelligent Control Technology (CEI)}.\hskip 1em plus 0.5em minus 0.4em\relax IEEE, 2022, pp. 496--501.

\bibitem[Cheng et~al.(2022)Cheng, Wu, Lukasiewicz, Sallinger, and Gottlob]{cheng2022democratizing}
Z.~Cheng, L.~Wu, T.~Lukasiewicz, E.~Sallinger, and G.~Gottlob, ``Democratizing financial knowledge graph construction by mining massive brokerage research reports.'' in \emph{EDBT/ICDT Workshops}, 2022.

\bibitem[Mackie and Dalton(2022)]{mackie2022query}
I.~Mackie and J.~Dalton, ``Query-specific knowledge graphs for complex finance topics,'' \emph{arXiv preprint arXiv:2211.04142}, 2022.

\bibitem[Liang et~al.(2024{\natexlab{a}})Liang, Tan, Xie, Tao, Wang, Lan, and Qian]{liang2024aligning}
Y.~Liang, K.~Tan, T.~Xie, W.~Tao, S.~Wang, Y.~Lan, and W.~Qian, ``Aligning large language models to a domain-specific graph database,'' \emph{arXiv preprint arXiv:2402.16567}, 2024.

\bibitem[Wang et~al.(2024{\natexlab{b}})Wang, Lipka, Rossi, Siu, Zhang, and Derr]{wang2024knowledge}
Y.~Wang, N.~Lipka, R.~A. Rossi, A.~Siu, R.~Zhang, and T.~Derr, ``Knowledge graph prompting for multi-document question answering,'' in \emph{Proceedings of the AAAI Conference on Artificial Intelligence}, vol.~38, no.~17, 2024, pp. 19\,206--19\,214.

\bibitem[Li(2023)]{li2023findkg}
X.~V. Li, ``{FinDKG}: Dynamic knowledge graph with large language models for global finance,'' \emph{Available at SSRN 4608445}, 2023.

\bibitem[Ok(2023)]{ok2023fintree}
H.~Ok, ``{FinTree}: Financial dataset pretrain transformer encoder for relation extraction,'' \emph{arXiv preprint arXiv:2307.13900}, 2023.

\bibitem[Kaur et~al.(2023)Kaur, Smiley, Gupta, Sain, Wang, Siddagangappa, Aguda, and Shah]{kaur2023refind}
S.~Kaur, C.~Smiley, A.~Gupta, J.~Sain, D.~Wang, S.~Siddagangappa, T.~Aguda, and S.~Shah, ``{REFinD}: Relation extraction financial dataset,'' in \emph{Proceedings of the 46th International ACM SIGIR Conference on Research and Development in Information Retrieval}, 2023, pp. 3054--3063.

\bibitem[Chai et~al.(2023)Chai, Chen, Wu, and Wang]{chai2023fin}
Y.~Chai, M.~Chen, H.~Wu, and S.~Wang, ``{Fin-EMRC}: An efficient machine reading comprehension framework for financial entity-relation extraction,'' \emph{IEEE Access}, 2023.

\bibitem[Tian et~al.(2019)Tian, Zhao, and Ren]{tian2019chinese}
C.~Tian, Y.~Zhao, and L.~Ren, ``A {Chinese} event relation extraction model based on {BERT},'' in \emph{2019 2nd International Conference on Artificial Intelligence and Big Data (ICAIBD)}.\hskip 1em plus 0.5em minus 0.4em\relax IEEE, 2019, pp. 271--276.

\bibitem[Ghosh et~al.(2023)Ghosh, Umrao, Chen, and Naskar]{ghosh2023mask}
S.~Ghosh, S.~Umrao, C.-C. Chen, and S.~K. Naskar, ``The mask one at a time framework for detecting the relationship between financial entities,'' in \emph{Proceedings of the 15th Annual Meeting of the Forum for Information Retrieval Evaluation}, 2023, pp. 40--43.

\bibitem[Rajpoot and Parikh(2023)]{rajpoot2023gpt}
P.~K. Rajpoot and A.~Parikh, ``{GPT-FinRE}: In-context learning for financial relation extraction using large language models,'' \emph{arXiv preprint arXiv:2306.17519}, 2023.

\bibitem[Wan et~al.(2023)Wan, Wan, Xiao, Hu, Liu, and Liu]{wan2023cfere}
Q.~Wan, C.~Wan, K.~Xiao, R.~Hu, D.~Liu, and X.~Liu, ``{CFERE}: Multi-type {Chinese} financial event relation extraction,'' \emph{Information Sciences}, vol. 630, pp. 119--134, 2023.

\bibitem[Rizinski et~al.(2024)Rizinski, Jankov, Sankaradas, Pinsky, Mishkovski, and Trajanov]{rizinski2024comparative}
M.~Rizinski, A.~Jankov, V.~Sankaradas, E.~Pinsky, I.~Mishkovski, and D.~Trajanov, ``Comparative analysis of {NLP}-based models for company classification,'' \emph{Information}, vol.~15, no.~2, p.~77, 2024.

\bibitem[Vamvourellis et~al.(2023)Vamvourellis, Toth, Bhagat, Desai, Mehta, and Pasquali]{vamvourellis2023company}
D.~Vamvourellis, M.~Toth, S.~Bhagat, D.~Desai, D.~Mehta, and S.~Pasquali, ``Company similarity using large language models,'' \emph{arXiv preprint arXiv:2308.08031}, 2023.

\bibitem[Wang et~al.(2021)Wang, Pan, Xu, Hu, and Wang]{wang2021enriching}
S.~Wang, Y.~Pan, Z.~Xu, B.~Hu, and X.~Wang, ``Enriching {BERT} with knowledge graph embedding for industry classification,'' in \emph{Neural Information Processing: 28th International Conference, ICONIP 2021, Sanur, Bali, Indonesia, December 8--12, 2021, Proceedings, Part VI 28}.\hskip 1em plus 0.5em minus 0.4em\relax Springer, 2021, pp. 709--717.

\bibitem[Mishra(2023)]{mishra2023esg}
S.~Mishra, ``{ESG} impact type classification: Leveraging strategic prompt engineering and {LLM} fine-tuning,'' in \emph{Proceedings of the Sixth Workshop on Financial Technology and Natural Language Processing}, 2023, pp. 72--78.

\bibitem[Nugroho et~al.(2021)Nugroho, Sukmadewa, and Yudistira]{nugroho2021large}
K.~S. Nugroho, A.~Y. Sukmadewa, and N.~Yudistira, ``Large-scale news classification using {BERT} language model: Spark {NLP} approach,'' in \emph{Proceedings of the 6th International Conference on Sustainable Information Engineering and Technology}, 2021, pp. 240--246.

\bibitem[Arslan et~al.(2021)Arslan, Allix, Veiber, Lothritz, Bissyand{\'e}, Klein, and Goujon]{arslan2021comparison}
Y.~Arslan, K.~Allix, L.~Veiber, C.~Lothritz, T.~F. Bissyand{\'e}, J.~Klein, and A.~Goujon, ``A comparison of pre-trained language models for multi-class text classification in the financial domain,'' in \emph{Companion Proceedings of the Web Conference 2021}, 2021, pp. 260--268.

\bibitem[Loukas et~al.(2023)Loukas, Stogiannidis, Diamantopoulos, Malakasiotis, and Vassos]{loukas2023making}
L.~Loukas, I.~Stogiannidis, O.~Diamantopoulos, P.~Malakasiotis, and S.~Vassos, ``Making {LLMs} worth every penny: Resource-limited text classification in banking,'' in \emph{Proceedings of the Fourth ACM International Conference on AI in Finance}, 2023, pp. 392--400.

\bibitem[Alias et~al.(2023)Alias, Fuad, Hoong, and Hin]{alias2023financial}
M.~S. Alias, M.~H. Fuad, X.~L.~F. Hoong, and E.~G.~Y. Hin, ``Financial text categorisation with {FinBERT} on key audit matters,'' in \emph{2023 IEEE Symposium on Computers \& Informatics (ISCI)}.\hskip 1em plus 0.5em minus 0.4em\relax IEEE, 2023, pp. 63--69.

\bibitem[Burke et~al.(2023)Burke, Hoitash, Hoitash, and Xiao]{burke2023using}
J.~Burke, R.~Hoitash, U.~Hoitash, and S.~X. Xiao, ``Using a large language model for accounting topic classification,'' \emph{Available at SSRN 4484489}, 2023.

\bibitem[Lee and Kim(2023)]{lee2023esg}
J.~Lee and M.~Kim, ``{ESG} information extraction with cross-sectoral and multi-source adaptation based on domain-tuned language models,'' \emph{Expert Systems with Applications}, vol. 221, p. 119726, 2023.

\bibitem[Mehra et~al.(2022)Mehra, Louka, and Zhang]{mehra2022esgbert}
S.~Mehra, R.~Louka, and Y.~Zhang, ``{ESGBERT}: Language model to help with classification tasks related to companies environmental, social, and governance practices,'' \emph{arXiv preprint arXiv:2203.16788}, 2022.

\bibitem[Tan et~al.(2023)Tan, Lee, and Lim]{tan2023survey}
K.~L. Tan, C.~P. Lee, and K.~M. Lim, ``A survey of sentiment analysis: Approaches, datasets, and future research,'' \emph{Applied Sciences}, vol.~13, no.~7, p. 4550, 2023.

\bibitem[Bordoloi and Biswas(2023)]{bordoloi2023sentiment}
M.~Bordoloi and S.~K. Biswas, ``Sentiment analysis: A survey on design framework, applications and future scopes,'' \emph{Artificial Intelligence Review}, pp. 1--56, 2023.

\bibitem[Mishev et~al.(2020)Mishev, Gjorgjevikj, Vodenska, Chitkushev, and Trajanov]{mishev2020evaluation}
K.~Mishev, A.~Gjorgjevikj, I.~Vodenska, L.~T. Chitkushev, and D.~Trajanov, ``Evaluation of sentiment analysis in finance: from lexicons to transformers,'' \emph{IEEE Access}, vol.~8, pp. 131\,662--131\,682, 2020.

\bibitem[Stone et~al.(1966)Stone, Dunphy, and Smith]{stone1966general}
P.~J. Stone, D.~C. Dunphy, and M.~S. Smith, ``The general inquirer: A computer approach to content analysis.'' 1966.

\bibitem[Pennebaker et~al.(2001)Pennebaker, Francis, and Booth]{pennebaker2001lexicon}
J.~W. Pennebaker, M.~E. Francis, and R.~J. Booth, ``Linguistic inquiry and word count: {LIWC} 2001,'' \emph{Mahway: Lawrence Erlbaum Associates}, vol.~71, 2001.

\bibitem[Taboada et~al.(2011)Taboada, Brooke, Tofiloski, Voll, and Stede]{taboada2011lexicon}
M.~Taboada, J.~Brooke, M.~Tofiloski, K.~Voll, and M.~Stede, ``Lexicon-based methods for sentiment analysis,'' \emph{Computational linguistics}, vol.~37, no.~2, pp. 267--307, 2011.

\bibitem[Loughran and McDonald(2011)]{loughran2011liability}
T.~Loughran and B.~McDonald, ``When is a liability not a liability? textual analysis, dictionaries, and {10-Ks},'' \emph{The Journal of finance}, vol.~66, no.~1, pp. 35--65, 2011.

\bibitem[Sohangir et~al.(2018{\natexlab{a}})Sohangir, Petty, and Wang]{sohangir2018financial}
S.~Sohangir, N.~Petty, and D.~Wang, ``Financial sentiment lexicon analysis,'' in \emph{2018 IEEE 12th International Conference on Semantic Computing (ICSC)}.\hskip 1em plus 0.5em minus 0.4em\relax IEEE, 2018, pp. 286--289.

\bibitem[Yekrangi and Abdolvand(2021)]{yekrangi2021financial}
M.~Yekrangi and N.~Abdolvand, ``Financial markets sentiment analysis: Developing a specialized lexicon,'' \emph{Journal of Intelligent Information Systems}, vol.~57, pp. 127--146, 2021.

\bibitem[Consoli et~al.(2022)Consoli, Barbaglia, and Manzan]{consoli2022fine}
S.~Consoli, L.~Barbaglia, and S.~Manzan, ``Fine-grained, aspect-based sentiment analysis on economic and financial lexicon,'' \emph{Knowledge-Based Systems}, vol. 247, p. 108781, 2022.

\bibitem[Chiong et~al.(2018)Chiong, Fan, Hu, Adam, Lutz, and Neumann]{chiong2018sentiment}
R.~Chiong, Z.~Fan, Z.~Hu, M.~T. Adam, B.~Lutz, and D.~Neumann, ``A sentiment analysis-based machine learning approach for financial market prediction via news disclosures,'' in \emph{Proceedings of the Genetic and Evolutionary Computation Conference Companion}, 2018, pp. 278--279.

\bibitem[Kalra and Prasad(2019)]{kalra2019efficacy}
S.~Kalra and J.~S. Prasad, ``Efficacy of news sentiment for stock market prediction,'' in \emph{2019 International Conference on Machine Learning, Big Data, Cloud and Parallel Computing (COMITCon)}.\hskip 1em plus 0.5em minus 0.4em\relax IEEE, 2019, pp. 491--496.

\bibitem[Kirange et~al.(2016)Kirange, Deshmukh, et~al.]{kirange2016sentiment}
D.~Kirange, R.~R. Deshmukh \emph{et~al.}, ``Sentiment analysis of news headlines for stock price prediction,'' \emph{Composoft, An International Journal of Advanced Computer Technology}, vol.~5, no.~3, pp. 2080--2084, 2016.

\bibitem[Dickinson et~al.(2015)Dickinson, Hu, et~al.]{dickinson2015sentiment}
B.~Dickinson, W.~Hu \emph{et~al.}, ``Sentiment analysis of investor opinions on {Twitter},'' \emph{Social Networking}, vol.~4, no.~03, p.~62, 2015.

\bibitem[Valencia et~al.(2019)Valencia, G{\'o}mez-Espinosa, and Vald{\'e}s-Aguirre]{valencia2019price}
F.~Valencia, A.~G{\'o}mez-Espinosa, and B.~Vald{\'e}s-Aguirre, ``Price movement prediction of cryptocurrencies using sentiment analysis and machine learning,'' \emph{Entropy}, vol.~21, no.~6, p. 589, 2019.

\bibitem[Yadav et~al.(2020)Yadav, Jha, Sharan, and Vaish]{yadav2020sentiment}
A.~Yadav, C.~Jha, A.~Sharan, and V.~Vaish, ``Sentiment analysis of financial news using unsupervised approach,'' \emph{Procedia Computer Science}, vol. 167, pp. 589--598, 2020.

\bibitem[Su et~al.(2022{\natexlab{a}})Su, Mulvey, and Poor]{su2022improving}
D.-J. Su, J.~M. Mulvey, and H.~V. Poor, ``Improving portfolio performance via natural language processing methods.'' \emph{Journal of Financial Data Science}, vol.~4, no.~2, 2022.

\bibitem[Mumtaz and Mumtaz(2023)]{mumtaz2023potential}
U.~Mumtaz and S.~Mumtaz, ``Potential of {ChatGPT} in predicting stock market trends based on {Twitter} sentiment analysis,'' \emph{arXiv preprint arXiv:2311.06273}, 2023.

\bibitem[Lopez-Lira and Tang(2023)]{lopez2023can}
A.~Lopez-Lira and Y.~Tang, ``Can {ChatGPT} forecast stock price movements? return predictability and large language models,'' \emph{arXiv preprint arXiv:2304.07619}, 2023.

\bibitem[Fatouros et~al.(2023)Fatouros, Soldatos, Kouroumali, Makridis, and Kyriazis]{fatouros2023transforming}
G.~Fatouros, J.~Soldatos, K.~Kouroumali, G.~Makridis, and D.~Kyriazis, ``Transforming sentiment analysis in the financial domain with {ChatGPT},'' \emph{Machine Learning with Applications}, vol.~14, p. 100508, 2023.

\bibitem[Luo and Gong(2024)]{luo2024pre}
W.~Luo and D.~Gong, ``Pre-trained large language models for financial sentiment analysis,'' \emph{arXiv preprint arXiv:2401.05215}, 2024.

\bibitem[Cook et~al.(2023)Cook, Kazinnik, Hansen, and McAdam]{cook2023evaluating}
T.~R. Cook, S.~Kazinnik, A.~L. Hansen, and P.~McAdam, ``Evaluating local language models: An application to financial earnings calls,'' \emph{Available at SSRN 4627143}, 2023.

\bibitem[Leippold(2023)]{leippold2023sentiment}
M.~Leippold, ``Sentiment spin: Attacking financial sentiment with {GPT-3},'' \emph{Finance Research Letters}, vol.~55, p. 103957, 2023.

\bibitem[Kim et~al.(2023{\natexlab{a}})Kim, Muhn, and Nikolaev]{kim2023bloated}
A.~G. Kim, M.~Muhn, and V.~V. Nikolaev, ``Bloated disclosures: can {ChatGPT} help investors process information?'' \emph{Chicago Booth Research Paper}, no. 23-07, 2023.

\bibitem[Aparicio et~al.(2024)Aparicio, Gordon, Huayamares, and Luo]{aparicio2024biofinbert}
V.~Aparicio, D.~Gordon, S.~G. Huayamares, and Y.~Luo, ``{BioFinBERT}: Finetuning large language models ({LLMs}) to analyze sentiment of press releases and financial text around inflection points of biotech stocks,'' \emph{arXiv preprint arXiv:2401.11011}, 2024.

\bibitem[Cao et~al.(2023{\natexlab{a}})Cao, Jiang, Yang, and Zhang]{cao2023talk}
S.~Cao, W.~Jiang, B.~Yang, and A.~L. Zhang, ``How to talk when a machine is listening: Corporate disclosure in the age of {AI},'' \emph{The Review of Financial Studies}, vol.~36, no.~9, pp. 3603--3642, 2023.

\bibitem[Kim et~al.(2023{\natexlab{b}})Kim, Kim, Kim, Park, Kim, Kim, Sung, Hong, and Lee]{kim2023llms}
S.~Kim, S.~Kim, Y.~Kim, J.~Park, S.~Kim, M.~Kim, C.~H. Sung, J.~Hong, and Y.~Lee, ``{LLMs} analyzing the analysts: Do {BERT} and {GPT} extract more value from financial analyst reports?'' in \emph{Proceedings of the Fourth ACM International Conference on AI in Finance}, 2023, pp. 383--391.

\bibitem[Shah et~al.(2023{\natexlab{a}})Shah, Paturi, and Chava]{shah2023trillion}
A.~Shah, S.~Paturi, and S.~Chava, ``Trillion dollar words: A new financial dataset, task \& market analysis,'' \emph{arXiv preprint arXiv:2305.07972}, 2023.

\bibitem[Kim et~al.(2023{\natexlab{c}})Kim, Sp{\"o}rer, and Handschuh]{kim2023analyzing}
W.~Kim, J.~F. Sp{\"o}rer, and S.~Handschuh, ``Analyzing {FOMC} minutes: Accuracy and constraints of language models,'' \emph{arXiv preprint arXiv:2304.10164}, 2023.

\bibitem[G{\"o}ssi et~al.(2023)G{\"o}ssi, Chen, Kim, Bermeitinger, and Handschuh]{gossi2023finbert}
S.~G{\"o}ssi, Z.~Chen, W.~Kim, B.~Bermeitinger, and S.~Handschuh, ``{FinBERT-FOMC}: Fine-tuned {FinBERT} model with sentiment focus method for enhancing sentiment analysis of {FOMC} minutes,'' in \emph{Proceedings of the Fourth ACM International Conference on AI in Finance}, 2023, pp. 357--364.

\bibitem[Fatouros et~al.(2024)Fatouros, Metaxas, Soldatos, and Kyriazis]{fatouros2024can}
G.~Fatouros, K.~Metaxas, J.~Soldatos, and D.~Kyriazis, ``Can large language models beat wall street? unveiling the potential of {AI} in stock selection,'' \emph{arXiv preprint arXiv:2401.03737}, 2024.

\bibitem[Kanelis and Siklos(2024)]{kanelis2024ecb}
D.~Kanelis and P.~L. Siklos, ``The {ECB} press conference statement: deriving a new sentiment indicator for the euro area,'' \emph{International Journal of Finance \& Economics}, 2024.

\bibitem[Renault(2020)]{renault2020sentiment}
T.~Renault, ``Sentiment analysis and machine learning in finance: a comparison of methods and models on one million messages,'' \emph{Digital Finance}, vol.~2, no.~1, pp. 1--13, 2020.

\bibitem[Mikolov et~al.(2013)Mikolov, Chen, Corrado, and Dean]{mikolov2013word2vec}
T.~Mikolov, K.~Chen, G.~Corrado, and J.~Dean, ``Efficient estimation of word representations in vector space,'' \emph{arXiv preprint arXiv:1301.3781}, 2013.

\bibitem[Pennington et~al.(2014)Pennington, Socher, and Manning]{pennington2014glove}
J.~Pennington, R.~Socher, and C.~D. Manning, ``{GloVe}: Global vectors for word representation,'' in \emph{Proceedings of the 2014 Conference on Empirical Methods in Natural Language Processing (EMNLP)}, 2014, pp. 1532--1543.

\bibitem[Bojanowski et~al.(2017)Bojanowski, Grave, Joulin, and Mikolov]{bojanowski2017fasttext}
P.~Bojanowski, E.~Grave, A.~Joulin, and T.~Mikolov, ``Enriching word vectors with subword information,'' \emph{Transactions of the Association for Computational Linguistics}, vol.~5, pp. 135--146, 2017.

\bibitem[Sarzynska-Wawer et~al.(2021)Sarzynska-Wawer, Wawer, Pawlak, Szymanowska, Stefaniak, Jarkiewicz, and Okruszek]{sarzynska2021elmo}
J.~Sarzynska-Wawer, A.~Wawer, A.~Pawlak, J.~Szymanowska, I.~Stefaniak, M.~Jarkiewicz, and L.~Okruszek, ``Detecting formal thought disorder by deep contextualized word representations,'' \emph{Psychiatry Research}, vol. 304, p. 114135, 2021.

\bibitem[Le and Mikolov(2014)]{le2014doc2vec}
Q.~Le and T.~Mikolov, ``Distributed representations of sentences and documents,'' in \emph{International conference on machine learning}.\hskip 1em plus 0.5em minus 0.4em\relax PMLR, 2014, pp. 1188--1196.

\bibitem[Sohangir et~al.(2018{\natexlab{b}})Sohangir, Wang, Pomeranets, and Khoshgoftaar]{sohangir2018bigdata}
S.~Sohangir, D.~Wang, A.~Pomeranets, and T.~M. Khoshgoftaar, ``Big data: Deep learning for financial sentiment analysis,'' \emph{Journal of Big Data}, vol.~5, no.~1, pp. 1--25, 2018.

\bibitem[Chen and Xing(2023)]{chen2023understanding}
S.~Chen and F.~Xing, ``Understanding emojis for financial sentiment analysis,'' 2023.

\bibitem[Vamossy and Skog(2023)]{vamossy2023emtract}
D.~F. Vamossy and R.~Skog, ``{EmTract}: Extracting emotions from social media,'' \emph{Available at SSRN 3975884}, 2023.

\bibitem[Jeong(2024)]{jeong2024fine}
C.~Jeong, ``Fine-tuning and utilization methods of domain-specific {LLMs},'' \emph{arXiv preprint arXiv:2401.02981}, 2024.

\bibitem[B{\u{a}}roiu and Tr{\u{a}}u{\c{s}}an-Matu(2023)]{buaroiu2023capable}
A.-C. B{\u{a}}roiu and {\c{S}}.~Tr{\u{a}}u{\c{s}}an-Matu, ``How capable are state-of-the-art language models to cope with sarcasm?'' in \emph{2023 24th International Conference on Control Systems and Computer Science (CSCS)}.\hskip 1em plus 0.5em minus 0.4em\relax IEEE, 2023, pp. 399--402.

\bibitem[Curti and Kazinnik(2023)]{curti2023let}
F.~Curti and S.~Kazinnik, ``Let's face it: Quantifying the impact of nonverbal communication in {FOMC} press conferences,'' \emph{Journal of Monetary Economics}, vol. 139, pp. 110--126, 2023.

\bibitem[Ebrahimi et~al.(2017)Ebrahimi, Yazdavar, and Sheth]{ebrahimi2017challenges}
M.~Ebrahimi, A.~H. Yazdavar, and A.~Sheth, ``Challenges of sentiment analysis for dynamic events,'' \emph{IEEE Intelligent Systems}, vol.~32, no.~5, pp. 70--75, 2017.

\bibitem[Arvanitis and Bassiliades(2017)]{arvanitis2017real}
K.~Arvanitis and N.~Bassiliades, ``Real-time investors’ sentiment analysis from newspaper articles,'' in \emph{Advances in Combining Intelligent Methods: Postproceedings of the 5th International Workshop CIMA-2015, Vietri sul Mare, Italy, November 2015 (at ICTAI 2015)}.\hskip 1em plus 0.5em minus 0.4em\relax Springer, 2017, pp. 1--23.

\bibitem[Chen et~al.(2022{\natexlab{a}})Chen, Kelly, and Xiu]{chen2022expected}
Y.~Chen, B.~T. Kelly, and D.~Xiu, ``Expected returns and large language models,'' \emph{Available at SSRN 4416687}, 2022.

\bibitem[Rosa(2013)]{rosa2013financial}
C.~Rosa, ``The financial market effect of {FOMC} minutes,'' \emph{Economic Policy Review}, vol.~19, no.~2, 2013.

\bibitem[Smales and Apergis(2017)]{smales2017does}
L.~A. Smales and N.~Apergis, ``Does more complex language in {FOMC} decisions impact financial markets?'' \emph{Journal of International Financial Markets, Institutions and Money}, vol.~51, pp. 171--189, 2017.

\bibitem[Klejdysz and Lumsdaine(2023)]{klejdysz2023shifts}
J.~Klejdysz and R.~L. Lumsdaine, ``Shifts in {ECB} communication: A textual analysis of the press conference,'' \emph{International Journal of Central Banking}, vol.~19, no.~2, pp. 473--542, 2023.

\bibitem[Anastasiou and Katsafados(2023)]{anastasiou2023bank}
D.~Anastasiou and A.~Katsafados, ``Bank deposits and textual sentiment: When an {European Central Bank} president's speech is not just a speech,'' \emph{The Manchester School}, vol.~91, no.~1, pp. 55--87, 2023.

\bibitem[Mody and Nedeljkovic(2024)]{mody2024central}
A.~Mody and M.~Nedeljkovic, ``Central bank policies and financial markets: Lessons from the euro crisis,'' \emph{Journal of Banking \& Finance}, vol. 158, p. 107033, 2024.

\bibitem[Nia et~al.(2022)Nia, Ahmadi, Bragazzi, Woldegerima, Mellado, Wu, Orbinski, Asgary, and Kong]{nia2022cross}
Z.~M. Nia, A.~Ahmadi, N.~L. Bragazzi, W.~A. Woldegerima, B.~Mellado, J.~Wu, J.~Orbinski, A.~Asgary, and J.~D. Kong, ``A cross-country analysis of macroeconomic responses to {COVID-19} pandemic using {Twitter} sentiments,'' \emph{Plos one}, vol.~17, no.~8, p. e0272208, 2022.

\bibitem[Shapiro et~al.(2022)Shapiro, Sudhof, and Wilson]{shapiro2022measuring}
A.~H. Shapiro, M.~Sudhof, and D.~J. Wilson, ``Measuring news sentiment,'' \emph{Journal of Econometrics}, vol. 228, no.~2, pp. 221--243, 2022.

\bibitem[Biswas et~al.(2020)Biswas, Ghosh, Chakraborty, Roy, and Bose]{biswas2020scope}
S.~Biswas, A.~Ghosh, S.~Chakraborty, S.~Roy, and R.~Bose, ``Scope of sentiment analysis on news articles regarding stock market and {GDP} in struggling economic condition,'' \emph{International Journal}, vol.~8, no.~7, pp. 3594--3609, 2020.

\bibitem[Lim and Zohren(2021)]{lim2021time}
B.~Lim and S.~Zohren, ``Time-series forecasting with deep learning: a survey,'' \emph{Philosophical Transactions of the Royal Society A}, vol. 379, no. 2194, p. 20200209, 2021.

\bibitem[Wang et~al.(2023{\natexlab{e}})Wang, Nie, Sun, Nguyen, Mulvey, and Poor]{wang2023st}
Z.~Wang, Y.~Nie, P.~Sun, N.~H. Nguyen, J.~Mulvey, and H.~V. Poor, ``{ST-MLP}: A cascaded spatio-temporal linear framework with channel-independence strategy for traffic forecasting,'' \emph{arXiv preprint arXiv:2308.07496}, 2023.

\bibitem[Wang et~al.(2023{\natexlab{f}})Wang, Zhuang, Li, Zhao, Sun, Wang, and Hu]{wang2023stgin}
Z.~Wang, D.~Zhuang, Y.~Li, J.~Zhao, P.~Sun, S.~Wang, and Y.~Hu, ``{ST-GIN}: An uncertainty quantification approach in traffic data imputation with spatio-temporal graph attention and bidirectional recurrent united neural networks,'' in \emph{2023 IEEE 26th International Conference on Intelligent Transportation Systems (ITSC)}.\hskip 1em plus 0.5em minus 0.4em\relax IEEE, 2023, pp. 1454--1459.

\bibitem[Pan et~al.(2024{\natexlab{b}})Pan, Jiang, Song, Garg, Rasul, Schneider, and Nevmyvaka]{pan2024structural}
Z.~Pan, Y.~Jiang, D.~Song, S.~Garg, K.~Rasul, A.~Schneider, and Y.~Nevmyvaka, ``Structural knowledge informed continual multivariate time series forecasting,'' \emph{arXiv preprint arXiv:2402.12722}, 2024.

\bibitem[Wang et~al.(2022{\natexlab{c}})Wang, Sun, Hu, and Boukerche]{wang2022novel}
Z.~Wang, P.~Sun, Y.~Hu, and A.~Boukerche, ``A novel mixed method of machine learning based models in vehicular traffic flow prediction,'' in \emph{Proceedings of the 25th International ACM Conference on Modeling Analysis and Simulation of Wireless and Mobile Systems}, 2022, pp. 95--101.

\bibitem[Chen et~al.(2022{\natexlab{b}})Chen, Gel, and Poor]{chen2022time}
Y.~Chen, Y.~Gel, and H.~V. Poor, ``Time-conditioned dances with simplicial complexes: Zigzag filtration curve based supra-hodge convolution networks for time-series forecasting,'' \emph{Advances in Neural Information Processing Systems}, vol.~35, pp. 8940--8953, 2022.

\bibitem[Jiang et~al.(2024)Jiang, Pan, Zhang, Garg, Schneider, Nevmyvaka, and Song]{jiang2024empowering}
Y.~Jiang, Z.~Pan, X.~Zhang, S.~Garg, A.~Schneider, Y.~Nevmyvaka, and D.~Song, ``Empowering time series analysis with large language models: A survey,'' \emph{arXiv preprint arXiv:2402.03182}, 2024.

\bibitem[Zhang et~al.(2024{\natexlab{a}})Zhang, Sun, Wang, Nie, Ma, Sun, and Li]{zhang2024large}
Z.~Zhang, Y.~Sun, Z.~Wang, Y.~Nie, X.~Ma, P.~Sun, and R.~Li, ``Large language models for mobility in transportation systems: A survey on forecasting tasks,'' \emph{arXiv preprint arXiv:2405.02357}, 2024.

\bibitem[Jin et~al.(2024)Jin, Zhang, Chen, Zhang, Liang, Yang, Wang, Pan, and Wen]{jin2024position}
M.~Jin, Y.~Zhang, W.~Chen, K.~Zhang, Y.~Liang, B.~Yang, J.~Wang, S.~Pan, and Q.~Wen, ``Position paper: What can large language models tell us about time series analysis,'' \emph{arXiv preprint arXiv:2402.02713}, 2024.

\bibitem[Pan et~al.(2024{\natexlab{c}})Pan, Jiang, Garg, Schneider, Nevmyvaka, and Song]{pan2024textbf}
Z.~Pan, Y.~Jiang, S.~Garg, A.~Schneider, Y.~Nevmyvaka, and D.~Song, ``{S$^2$IP-LLM}: Semantic space informed prompt learning with {LLM} for time series forecasting,'' \emph{arXiv preprint arXiv:2403.05798}, 2024.

\bibitem[Zhou et~al.(2022)Zhou, Ma, Wen, Wang, Sun, and Jin]{zhou2022fedformer}
T.~Zhou, Z.~Ma, Q.~Wen, X.~Wang, L.~Sun, and R.~Jin, ``{FEDformer}: Frequency enhanced decomposed transformer for long-term series forecasting,'' in \emph{International Conference on Machine Learning}.\hskip 1em plus 0.5em minus 0.4em\relax PMLR, 2022, pp. 27\,268--27\,286.

\bibitem[Nie et~al.(2022)Nie, Nguyen, Sinthong, and Kalagnanam]{nie2022time}
Y.~Nie, N.~H. Nguyen, P.~Sinthong, and J.~Kalagnanam, ``A time series is worth 64 words: Long-term forecasting with transformers,'' \emph{arXiv preprint arXiv:2211.14730}, 2022.

\bibitem[Wen et~al.(2022)Wen, Zhou, Zhang, Chen, Ma, Yan, and Sun]{wen2022transformers}
Q.~Wen, T.~Zhou, C.~Zhang, W.~Chen, Z.~Ma, J.~Yan, and L.~Sun, ``Transformers in time series: A survey,'' \emph{arXiv preprint arXiv:2202.07125}, 2022.

\bibitem[Zhu et~al.(2023)Zhu, Chen, Shen, Li, and Elhoseiny]{zhu2023minigpt}
D.~Zhu, J.~Chen, X.~Shen, X.~Li, and M.~Elhoseiny, ``{MiniGPT-4}: Enhancing vision-language understanding with advanced large language models,'' \emph{arXiv preprint arXiv:2304.10592}, 2023.

\bibitem[Zhang et~al.(2023{\natexlab{c}})Zhang, Gong, Zhang, Li, Qiao, Ouyang, and Yue]{zhang2023meta}
Y.~Zhang, K.~Gong, K.~Zhang, H.~Li, Y.~Qiao, W.~Ouyang, and X.~Yue, ``{Meta-Transformer}: A unified framework for multimodal learning,'' \emph{arXiv preprint arXiv:2307.10802}, 2023.

\bibitem[Zhou et~al.(2024{\natexlab{b}})Zhou, Niu, Sun, Jin, et~al.]{zhou2024one}
T.~Zhou, P.~Niu, L.~Sun, R.~Jin \emph{et~al.}, ``One fits all: Power general time series analysis by pretrained {LM},'' \emph{Advances in Neural Information Processing Systems}, vol.~36, 2024.

\bibitem[Gruver et~al.(2024)Gruver, Finzi, Qiu, and Wilson]{gruver2024large}
N.~Gruver, M.~Finzi, S.~Qiu, and A.~G. Wilson, ``Large language models are zero-shot time series forecasters,'' \emph{Advances in Neural Information Processing Systems}, vol.~36, 2024.

\bibitem[Jin et~al.(2023{\natexlab{a}})Jin, Wang, Ma, Chu, Zhang, Shi, Chen, Liang, Li, Pan, et~al.]{jin2023time}
M.~Jin, S.~Wang, L.~Ma, Z.~Chu, J.~Y. Zhang, X.~Shi, P.-Y. Chen, Y.~Liang, Y.-F. Li, S.~Pan \emph{et~al.}, ``{Time-LLM}: Time series forecasting by reprogramming large language models,'' \emph{arXiv preprint arXiv:2310.01728}, 2023.

\bibitem[Jin et~al.(2023{\natexlab{b}})Jin, Wen, Liang, Zhang, Xue, Wang, Zhang, Wang, Chen, Li, et~al.]{jin2023large}
M.~Jin, Q.~Wen, Y.~Liang, C.~Zhang, S.~Xue, X.~Wang, J.~Zhang, Y.~Wang, H.~Chen, X.~Li \emph{et~al.}, ``Large models for time series and spatio-temporal data: A survey and outlook,'' \emph{arXiv preprint arXiv:2310.10196}, 2023.

\bibitem[Liang et~al.(2024{\natexlab{b}})Liang, Wen, Nie, Jiang, Jin, Song, Pan, and Wen]{liang2024foundation}
Y.~Liang, H.~Wen, Y.~Nie, Y.~Jiang, M.~Jin, D.~Song, S.~Pan, and Q.~Wen, ``Foundation models for time series analysis: A tutorial and survey,'' \emph{arXiv preprint arXiv:2403.14735}, 2024.

\bibitem[Yu et~al.(2023)Yu, Chen, Ling, Dong, Liu, and Lu]{yu2023temporal}
X.~Yu, Z.~Chen, Y.~Ling, S.~Dong, Z.~Liu, and Y.~Lu, ``Temporal data meets {LLM}--explainable financial time series forecasting,'' \emph{arXiv preprint arXiv:2306.11025}, 2023.

\bibitem[Chen et~al.(2023{\natexlab{b}})Chen, Zheng, Lu, Yuan, and Zhu]{chen2023chatgpt}
Z.~Chen, L.~N. Zheng, C.~Lu, J.~Yuan, and D.~Zhu, ``{ChatGPT} informed graph neural network for stock movement prediction,'' \emph{arXiv preprint arXiv:2306.03763}, 2023.

\bibitem[Wimmer and Rekabsaz(2023)]{wimmer2023leveraging}
C.~Wimmer and N.~Rekabsaz, ``Leveraging vision-language models for granular market change prediction,'' \emph{arXiv preprint arXiv:2301.10166}, 2023.

\bibitem[Cao et~al.(2024)Cao, Chen, Pei, Dimino, Ausiello, Kumar, Subbalakshmi, and Ndiaye]{cao2024risklabs}
Y.~Cao, Z.~Chen, Q.~Pei, F.~Dimino, L.~Ausiello, P.~Kumar, K.~Subbalakshmi, and P.~M. Ndiaye, ``{RiskLabs}: Predicting financial risk using large language model based on multi-sources data,'' \emph{arXiv preprint arXiv:2404.07452}, 2024.

\bibitem[Xie et~al.(2023{\natexlab{b}})Xie, Han, Lai, Peng, and Huang]{xie2023wall}
Q.~Xie, W.~Han, Y.~Lai, M.~Peng, and J.~Huang, ``The wall street neophyte: A zero-shot analysis of {ChatGPT} over multimodal stock movement prediction challenges,'' \emph{arXiv preprint arXiv:2304.05351}, 2023.

\bibitem[Chandola et~al.(2009)Chandola, Banerjee, and Kumar]{chandola2009anomaly}
V.~Chandola, A.~Banerjee, and V.~Kumar, ``Anomaly detection: A survey,'' \emph{ACM Computing Surveys (CSUR)}, vol.~41, no.~3, pp. 1--58, 2009.

\bibitem[Zojaji et~al.(2016)Zojaji, Atani, Monadjemi, et~al.]{zojaji2016survey}
Z.~Zojaji, R.~E. Atani, A.~H. Monadjemi \emph{et~al.}, ``A survey of credit card fraud detection techniques: data and technique oriented perspective,'' \emph{arXiv preprint arXiv:1611.06439}, 2016.

\bibitem[Chen et~al.(2019)Chen, Xu, Zheng, Zhou, Yang, and Bian]{chen2019detecting}
W.~Chen, Y.~Xu, Z.~Zheng, Y.~Zhou, J.~E. Yang, and J.~Bian, ``Detecting {"Pump \& Dump Schemes"} on cryptocurrency market using an improved {Apriori Algorithm},'' in \emph{2019 IEEE International Conference on Service-Oriented System Engineering (SOSE)}.\hskip 1em plus 0.5em minus 0.4em\relax IEEE, 2019, pp. 293--2935.

\bibitem[Darban et~al.(2022)Darban, Webb, Pan, Aggarwal, and Salehi]{darban2022deep}
Z.~Z. Darban, G.~I. Webb, S.~Pan, C.~C. Aggarwal, and M.~Salehi, ``Deep learning for time series anomaly detection: A survey,'' \emph{arXiv preprint arXiv:2211.05244}, 2022.

\bibitem[Cr{\'e}pey et~al.(2022)Cr{\'e}pey, Lehdili, Madhar, and Thomas]{crepey2022anomaly}
S.~Cr{\'e}pey, N.~Lehdili, N.~Madhar, and M.~Thomas, ``Anomaly detection in financial time series by principal component analysis and neural networks,'' \emph{Algorithms}, vol.~15, no.~10, p. 385, 2022.

\bibitem[Zhu et~al.(2024{\natexlab{a}})Zhu, Cai, Deng, and Wu]{zhu2024llms}
J.~Zhu, S.~Cai, F.~Deng, and J.~Wu, ``Do {LLMs} understand visual anomalies? uncovering {LLM} capabilities in zero-shot anomaly detection,'' \emph{arXiv preprint arXiv:2404.09654}, 2024.

\bibitem[Park(2024)]{park2024enhancing}
T.~Park, ``Enhancing anomaly detection in financial markets with an {LLM}-based multi-agent framework,'' \emph{arXiv preprint arXiv:2403.19735}, 2024.

\bibitem[Bosancic et~al.(2024)Bosancic, Nie, and Mulvey]{bosancic2024regime}
T.~Bosancic, Y.~Nie, and J.~M. Mulvey, ``Regime-aware factor allocation with optimal feature selection,'' \emph{Available at SSRN 4825234}, 2024.

\bibitem[Nagy et~al.(2023)Nagy, Frey, Sapora, Li, Calinescu, Zohren, and Foerster]{nagy2023generative}
P.~Nagy, S.~Frey, S.~Sapora, K.~Li, A.~Calinescu, S.~Zohren, and J.~Foerster, ``Generative {AI} for end-to-end limit order book modelling: A token-level autoregressive generative model of message flow using a deep state space network,'' in \emph{Proceedings of the Fourth ACM International Conference on AI in Finance}, 2023, pp. 91--99.

\bibitem[Ding et~al.(2024{\natexlab{a}})Ding, Qin, Zhao, Luo, Li, Chen, Xia, Hu, Luu, and Joty]{ding2024data}
B.~Ding, C.~Qin, R.~Zhao, T.~Luo, X.~Li, G.~Chen, W.~Xia, J.~Hu, A.~T. Luu, and S.~Joty, ``Data augmentation using {LLMs}: Data perspectives, learning paradigms and challenges,'' \emph{arXiv preprint arXiv:2403.02990}, 2024.

\bibitem[Zhao et~al.(2023{\natexlab{a}})Zhao, Zhou, Li, Tang, Wang, Hou, Min, Zhang, Zhang, Dong, et~al.]{zhao2023survey}
W.~X. Zhao, K.~Zhou, J.~Li, T.~Tang, X.~Wang, Y.~Hou, Y.~Min, B.~Zhang, J.~Zhang, Z.~Dong \emph{et~al.}, ``A survey of large language models,'' \emph{arXiv preprint arXiv:2303.18223}, 2023.

\bibitem[Nguyen and Tulabandhula(2023)]{nguyen2023generative}
S.~T. Nguyen and T.~Tulabandhula, ``Generative {AI} for business strategy: Using foundation models to create business strategy tools,'' \emph{arXiv preprint arXiv:2308.14182}, 2023.

\bibitem[Ludwig and Bennetts(2023)]{ludwig2023streamlining}
E.~T. Ludwig and C.~R. Bennetts, ``Streamlining financial planning with {ChatGPT}: A collaborative approach between technology and human expertise: {AI} likely won't replace your job anytime soon, but those who successfully utilize it are bound to have an advantage over those who don't.'' \emph{Journal of Financial Planning}, vol.~36, no.~6, 2023.

\bibitem[Lakkaraju et~al.(2023{\natexlab{a}})Lakkaraju, Vuruma, Pallagani, Muppasani, and Srivastava]{lakkaraju2023can}
K.~Lakkaraju, S.~K.~R. Vuruma, V.~Pallagani, B.~Muppasani, and B.~Srivastava, ``Can {LLMs} be good financial advisors?: An initial study in personal decision making for optimized outcomes,'' \emph{arXiv preprint arXiv:2307.07422}, 2023.

\bibitem[de~Zarz{\`a} et~al.(2023)de~Zarz{\`a}, de~Curt{\`o}, Roig, and Calafate]{de2023optimized}
I.~de~Zarz{\`a}, J.~de~Curt{\`o}, G.~Roig, and C.~T. Calafate, ``Optimized financial planning: Integrating individual and cooperative budgeting models with {LLM} recommendations,'' \emph{AI}, vol.~5, no.~1, pp. 91--114, 2023.

\bibitem[Fava(2023)]{fava2023future}
D.~Fava, ``The future of tax planning: Leveraging generative {AI} in high-net-worth contexts: Artifical intelligence could help planers optimize their {HNW} clients' complex tax planning needs.'' \emph{Journal of Financial Planning}, no.~10, 2023.

\bibitem[Huang et~al.(2024)Huang, Che, Zheng, and Li]{huang2024research}
Z.~Huang, C.~Che, H.~Zheng, and C.~Li, ``Research on generative artificial intelligence for virtual financial robo-advisor,'' \emph{Academic Journal of Science and Technology}, vol.~10, no.~1, pp. 74--80, 2024.

\bibitem[Lu et~al.(2023{\natexlab{b}})Lu, Huang, and Li]{lu2023chatgpt}
F.~Lu, L.~Huang, and S.~Li, ``{ChatGPT, Generative AI, and Investment Advisory},'' \emph{Available at SSRN}, 2023.

\bibitem[Ramyadevi and Sasidharan(2024)]{ramyadevi2024cogniwealth}
R.~Ramyadevi and G.~Sasidharan, ``Cogniwealth: Revolutionizing finance, empowering investors, and shaping the future of wealth management,'' in \emph{2024 IEEE International Conference on Computing, Power and Communication Technologies (IC2PCT)}, vol.~5.\hskip 1em plus 0.5em minus 0.4em\relax IEEE, 2024, pp. 378--381.

\bibitem[Ko and Lee(2024)]{ko2024can}
H.~Ko and J.~Lee, ``Can {ChatGPT} improve investment decisions? from a portfolio management perspective,'' \emph{Finance Research Letters}, vol.~64, p. 105433, 2024.

\bibitem[Noguer~i Alonso and Dupouy(2024)]{noguer2024evaluating}
M.~Noguer~i Alonso and H.~Dupouy, ``Evaluating {LLMs} in financial tasks-code generation in trading strategies,'' \emph{Hanane, Evaluating LLMs in Financial Tasks-Code Generation in Trading Strategies (March 8, 2024)}, 2024.

\bibitem[Kim et~al.(2024)Kim, Muhn, and Nikolaev]{kim2024financial}
A.~Kim, M.~Muhn, and V.~V. Nikolaev, ``Financial statement analysis with large language models,'' \emph{Chicago Booth Research Paper Forthcoming, Fama-Miller Working Paper}, 2024.

\bibitem[Zhang et~al.(2024{\natexlab{b}})Zhang, Yoshie, and Huang]{zhang2024breakgpt}
K.~Zhang, O.~Yoshie, and W.~Huang, ``{BreakGPT}: A large language model with multi-stage structure for financial breakout detection,'' \emph{arXiv preprint arXiv:2402.07536}, 2024.

\bibitem[Chuang and Yang(2022)]{chuang2022buy}
C.~Chuang and Y.~Yang, ``{Buy Tesla, Sell Ford}: Assessing implicit stock market preference in pre-trained language models,'' in \emph{Proceedings of the 60th Annual Meeting of the Association for Computational Linguistics (Volume 2: Short Papers)}, 2022, pp. 100--105.

\bibitem[Caspi et~al.(2023)Caspi, Felber, and Gillis]{caspi2023generative}
I.~Caspi, S.~S. Felber, and T.~B. Gillis, ``Generative {AI} and the future of financial advice regulation,'' \emph{GenLaw Center}, 2023.

\bibitem[Niszczota and Abbas(2023)]{niszczota2023gpt}
P.~Niszczota and S.~Abbas, ``{GPT} has become financially literate: Insights from financial literacy tests of {GPT} and a preliminary test of how people use it as a source of advice,'' \emph{Finance Research Letters}, vol.~58, p. 104333, 2023.

\bibitem[Lakkaraju et~al.(2023{\natexlab{b}})Lakkaraju, Jones, Vuruma, Pallagani, Muppasani, and Srivastava]{lakkaraju2023llms}
K.~Lakkaraju, S.~E. Jones, S.~K.~R. Vuruma, V.~Pallagani, B.~C. Muppasani, and B.~Srivastava, ``{LLMs} for financial advisement: A fairness and efficacy study in personal decision making,'' in \emph{Proceedings of the Fourth ACM International Conference on AI in Finance}, 2023, pp. 100--107.

\bibitem[Berger et~al.(2023)Berger, Hillebrand, Leonhard, Deu{\ss}er, De~Oliveira, Dilmaghani, Khaled, Kliem, Loitz, Bauckhage, et~al.]{berger2023towards}
A.~Berger, L.~Hillebrand, D.~Leonhard, T.~Deu{\ss}er, T.~B.~F. De~Oliveira, T.~Dilmaghani, M.~Khaled, B.~Kliem, R.~Loitz, C.~Bauckhage \emph{et~al.}, ``Towards automated regulatory compliance verification in financial auditing with large language models,'' in \emph{2023 IEEE International Conference on Big Data (BigData)}.\hskip 1em plus 0.5em minus 0.4em\relax IEEE, 2023, pp. 4626--4635.

\bibitem[Hillebrand et~al.(2023)Hillebrand, Berger, Deu{\ss}er, Dilmaghani, Khaled, Kliem, Loitz, Pielka, Leonhard, Bauckhage, et~al.]{hillebrand2023improving}
L.~Hillebrand, A.~Berger, T.~Deu{\ss}er, T.~Dilmaghani, M.~Khaled, B.~Kliem, R.~Loitz, M.~Pielka, D.~Leonhard, C.~Bauckhage \emph{et~al.}, ``Improving zero-shot text matching for financial auditing with large language models,'' in \emph{Proceedings of the ACM Symposium on Document Engineering 2023}, 2023, pp. 1--4.

\bibitem[Cao and Feinstein(2024)]{cao2024large}
Z.~Cao and Z.~Feinstein, ``Large language model in financial regulatory interpretation,'' \emph{arXiv preprint arXiv:2405.06808}, 2024.

\bibitem[Choi and Kim(2024)]{Choi2024FirmLevelTax}
\BIBentryALTinterwordspacing
G.-Y. Choi and A.~G. Kim, ``Firm-level tax audits: A {Generative AI}-based measurement,'' Chicago Booth, Research Paper 23-23, 2024. [Online]. Available: \url{https://ssrn.com/abstract=4645865}
\BIBentrySTDinterwordspacing

\bibitem[Deu{\ss}er et~al.(2023)Deu{\ss}er, Leonhard, Hillebrand, Berger, Khaled, Heiden, Dilmaghani, Kliem, Loitz, Bauckhage, et~al.]{deusser2023uncovering}
T.~Deu{\ss}er, D.~Leonhard, L.~Hillebrand, A.~Berger, M.~Khaled, S.~Heiden, T.~Dilmaghani, B.~Kliem, R.~Loitz, C.~Bauckhage \emph{et~al.}, ``Uncovering inconsistencies and contradictions in financial reports using large language models,'' in \emph{2023 IEEE International Conference on Big Data (BigData)}.\hskip 1em plus 0.5em minus 0.4em\relax IEEE, 2023, pp. 2814--2822.

\bibitem[Feng et~al.(2023)Feng, Dai, Huang, Zhang, Xie, Han, Lopez-Lira, and Wang]{feng2023empowering}
D.~Feng, Y.~Dai, J.~Huang, Y.~Zhang, Q.~Xie, W.~Han, A.~Lopez-Lira, and H.~Wang, ``Empowering many, biasing a few: Generalist credit scoring through large language models,'' \emph{arXiv preprint arXiv:2310.00566}, 2023.

\bibitem[Zhao et~al.(2023{\natexlab{b}})Zhao, Zhu, Li, Wang, and Wang]{zhao2023generative}
Z.~Y. Zhao, Z.~Zhu, G.~Li, W.~Wang, and B.~Wang, ``Generative pretraining at scale: Transformer-based encoding of transactional behavior for fraud detection,'' \emph{arXiv preprint arXiv:2312.14406}, 2023.

\bibitem[Yang et~al.(2023{\natexlab{c}})Yang, Zhang, Sun, Pang, Jing, Wa, and Lv]{yang2023finchain}
X.~Yang, C.~Zhang, Y.~Sun, K.~Pang, L.~Jing, S.~Wa, and C.~Lv, ``{FinChain-BERT}: A high-accuracy automatic fraud detection model based on {NLP} methods for financial scenarios,'' \emph{Information}, vol.~14, no.~9, p. 499, 2023.

\bibitem[Bhattacharya and Mickovic(2024)]{bhattacharya2024accounting}
I.~Bhattacharya and A.~Mickovic, ``Accounting fraud detection using contextual language learning,'' \emph{International Journal of Accounting Information Systems}, vol.~53, p. 100682, 2024.

\bibitem[Aggarwal et~al.(2023)Aggarwal, Mehra, and Mitra]{aggarwal2023multi}
S.~Aggarwal, S.~Mehra, and P.~Mitra, ``Multi-purpose {NLP} chatbot: Design, methodology \& conclusion,'' \emph{arXiv preprint arXiv:2310.08977}, 2023.

\bibitem[Yue and Au(2023)]{yue2023gptquant}
T.~Yue and D.~Au, ``{GPTQuant's Conversational AI}: Simplifying investment research for all,'' \emph{Available at SSRN 4380516}, 2023.

\bibitem[Yadav et~al.(2024)Yadav, Zhang, Jin, Krishnan, and Clarke]{yadav2024generative}
D.~Yadav, S.~Zhang, T.~Jin, P.~Krishnan, and D.~Clarke, ``Generative {AI} based virtual assistant for reconciliation research,'' 2024.

\bibitem[Arun et~al.(2023)Arun, Dhiman, Soni, and Hu]{arun2023numerical}
A.~Arun, A.~Dhiman, M.~Soni, and Y.~Hu, ``Numerical reasoning for financial reports,'' \emph{arXiv preprint arXiv:2312.14870}, 2023.

\bibitem[Phogat et~al.(2023)Phogat, Harsha, Dasaratha, Ramakrishna, and Puranam]{phogat2023zero}
K.~S. Phogat, C.~Harsha, S.~Dasaratha, S.~Ramakrishna, and S.~A. Puranam, ``Zero-shot question answering over financial documents using large language models,'' \emph{arXiv preprint arXiv:2311.14722}, 2023.

\bibitem[Srivastava et~al.(2024)Srivastava, Malik, and Ganu]{srivastava2024assessing}
P.~Srivastava, M.~Malik, and T.~Ganu, ``Assessing {LLMs}' mathematical reasoning in financial document question answering,'' \emph{arXiv preprint arXiv:2402.11194}, 2024.

\bibitem[Guo et~al.(2024)Guo, Chen, Wang, Chang, Pei, Chawla, Wiest, and Zhang]{guo2024large}
T.~Guo, X.~Chen, Y.~Wang, R.~Chang, S.~Pei, N.~V. Chawla, O.~Wiest, and X.~Zhang, ``Large language model based multi-agents: A survey of progress and challenges,'' \emph{arXiv preprint arXiv:2402.01680}, 2024.

\bibitem[Xi et~al.(2023)Xi, Chen, Guo, He, Ding, Hong, Zhang, Wang, Jin, Zhou, et~al.]{xi2023rise}
Z.~Xi, W.~Chen, X.~Guo, W.~He, Y.~Ding, B.~Hong, M.~Zhang, J.~Wang, S.~Jin, E.~Zhou \emph{et~al.}, ``The rise and potential of large language model based agents: A survey,'' \emph{arXiv preprint arXiv:2309.07864}, 2023.

\bibitem[Ma et~al.(2024{\natexlab{b}})Ma, Xue, Zhou, Yu, Liu, Zhang, Zhao, Shen, Ji, Li, et~al.]{ma2024computational}
Q.~Ma, X.~Xue, D.~Zhou, X.~Yu, D.~Liu, X.~Zhang, Z.~Zhao, Y.~Shen, P.~Ji, J.~Li \emph{et~al.}, ``Computational experiments meet large language model based agents: A survey and perspective,'' \emph{arXiv preprint arXiv:2402.00262}, 2024.

\bibitem[Zhang et~al.(2024{\natexlab{c}})Zhang, Mao, Ge, Wang, de~Wynter, Xia, Wu, Song, Lan, and Wei]{zhang2024llm}
Y.~Zhang, S.~Mao, T.~Ge, X.~Wang, A.~de~Wynter, Y.~Xia, W.~Wu, T.~Song, M.~Lan, and F.~Wei, ``{LLM} as a mastermind: A survey of strategic reasoning with large language models,'' \emph{arXiv preprint arXiv:2404.01230}, 2024.

\bibitem[Epstein(1999)]{epstein1999agent}
J.~M. Epstein, ``Agent-based computational models and generative social science,'' \emph{Complexity}, vol.~4, no.~5, pp. 41--60, 1999.

\bibitem[Liu et~al.()Liu, Zhang, Jin, Zhang, Wang, Shu, Zhu, Li, Du, and Zhang]{liuai}
X.~Liu, C.~Zhang, M.~Jin, Z.~Zhang, Z.~Wang, D.~Shu, S.~Zhu, S.~Li, M.~Du, and Y.~Zhang, ``{When AI Meets Finance (StockAgent)}: Large language model-based stock trading in simulated real-world environments.''

\bibitem[Zhang et~al.(2024{\natexlab{d}})Zhang, Zhao, Xia, Sun, Sun, Qin, Li, Zhao, Zhao, Cai, et~al.]{zhang2024finagent}
W.~Zhang, L.~Zhao, H.~Xia, S.~Sun, J.~Sun, M.~Qin, X.~Li, Y.~Zhao, Y.~Zhao, X.~Cai \emph{et~al.}, ``{FinAgent}: A multimodal foundation agent for financial trading: Tool-augmented, diversified, and generalist,'' \emph{arXiv preprint arXiv:2402.18485}, 2024.

\bibitem[Yu et~al.(2024{\natexlab{b}})Yu, Li, Chen, Jiang, Li, Zhang, Liu, Suchow, and Khashanah]{yu2024finmem}
Y.~Yu, H.~Li, Z.~Chen, Y.~Jiang, Y.~Li, D.~Zhang, R.~Liu, J.~W. Suchow, and K.~Khashanah, ``{FinMem}: A performance-enhanced {LLM} trading agent with layered memory and character design,'' in \emph{Proceedings of the AAAI Symposium Series}, vol.~3, no.~1, 2024, pp. 595--597.

\bibitem[Wang et~al.(2024{\natexlab{c}})Wang, Yuan, Ni, and Guo]{wang2024quantagent}
S.~Wang, H.~Yuan, L.~M. Ni, and J.~Guo, ``{QuantAgent}: Seeking holy grail in trading by self-improving large language model,'' \emph{arXiv preprint arXiv:2402.03755}, 2024.

\bibitem[Wang et~al.(2023{\natexlab{g}})Wang, Yuan, Zhou, Ni, Shum, and Guo]{wang2023alpha}
S.~Wang, H.~Yuan, L.~Zhou, L.~M. Ni, H.-Y. Shum, and J.~Guo, ``{Alpha-GPT}: {Human-AI} interactive alpha mining for quantitative investment,'' \emph{arXiv preprint arXiv:2308.00016}, 2023.

\bibitem[Yuan et~al.(2024{\natexlab{a}})Yuan, Wang, and Guo]{yuan2024alpha}
H.~Yuan, S.~Wang, and J.~Guo, ``{Alpha-GPT 2.0}: {Human-in-the-Loop AI} for quantitative investment,'' \emph{arXiv preprint arXiv:2402.09746}, 2024.

\bibitem[Sims(1980)]{sims1980macroeconomics}
C.~A. Sims, ``Macroeconomics and reality,'' \emph{Econometrica: journal of the Econometric Society}, pp. 1--48, 1980.

\bibitem[Smets and Wouters(2003)]{smets2003estimated}
F.~Smets and R.~Wouters, ``An estimated dynamic stochastic general equilibrium model of the euro area,'' \emph{Journal of the European Economic Association}, vol.~1, no.~5, pp. 1123--1175, 2003.

\bibitem[Tesfatsion(2006)]{tesfatsion2006agent}
L.~Tesfatsion, ``Agent-based computational economics: A constructive approach to economic theory,'' \emph{Handbook of Computational Economics}, vol.~2, pp. 831--880, 2006.

\bibitem[Farmer and Foley(2009)]{farmer2009economy}
J.~D. Farmer and D.~Foley, ``The economy needs agent-based modelling,'' \emph{Nature}, vol. 460, no. 7256, pp. 685--686, 2009.

\bibitem[Li et~al.(2023{\natexlab{d}})Li, Gao, Li, and Liao]{li2023large2}
N.~Li, C.~Gao, Y.~Li, and Q.~Liao, ``Large language model-empowered agents for simulating macroeconomic activities,'' \emph{arXiv preprint arXiv:2310.10436}, 2023.

\bibitem[Horton(2023)]{horton2023large}
J.~J. Horton, ``Large language models as simulated economic agents: What can we learn from homo silicus?'' National Bureau of Economic Research, Tech. Rep., 2023.

\bibitem[Zhao et~al.(2023{\natexlab{c}})Zhao, Wang, Zhang, Jin, Zhu, Chen, and Xie]{zhao2023competeai}
Q.~Zhao, J.~Wang, Y.~Zhang, Y.~Jin, K.~Zhu, H.~Chen, and X.~Xie, ``{CompeteAI}: Understanding the competition behaviors in large language model-based agents,'' \emph{arXiv preprint arXiv:2310.17512}, 2023.

\bibitem[Zeng et~al.(2023)Zeng, Watson, Cho, Rahimi, Reynolds, Balch, and Veloso]{zeng2023flowmind}
Z.~Zeng, W.~Watson, N.~Cho, S.~Rahimi, S.~Reynolds, T.~Balch, and M.~Veloso, ``{FlowMind}: Automatic workflow generation with {LLMs},'' in \emph{Proceedings of the Fourth ACM International Conference on AI in Finance}, 2023, pp. 73--81.

\bibitem[Chen et~al.(2023{\natexlab{c}})Chen, Yuan, Ye, Majumder, and Richardson]{chen2023put}
J.~Chen, S.~Yuan, R.~Ye, B.~P. Majumder, and K.~Richardson, ``Put your money where your mouth is: Evaluating strategic planning and execution of {LLM} agents in an auction arena,'' \emph{arXiv preprint arXiv:2310.05746}, 2023.

\bibitem[Li et~al.(2023{\natexlab{e}})Li, Yu, Li, Chen, and Khashanah]{li2023tradinggpt}
Y.~Li, Y.~Yu, H.~Li, Z.~Chen, and K.~Khashanah, ``{TradingGPT}: Multi-agent system with layered memory and distinct characters for enhanced financial trading performance,'' \emph{arXiv preprint arXiv:2309.03736}, 2023.

\bibitem[Tsao and AILAB(2023)]{tsao2023multi}
W.-K. Tsao and T.~AILAB, ``Multi-agent reasoning with large language models for effective corporate planning,'' in \emph{The 10th International Conf. on Computational Science and Computational Intelligence}, 2023.

\bibitem[Xing(2024)]{xing2024designing}
F.~Xing, ``Designing heterogeneous {LLM} agents for financial sentiment analysis,'' \emph{arXiv preprint arXiv:2401.05799}, 2024.

\bibitem[Wan et~al.(2024)Wan, Deng, Zou, and Xu]{wan2024enhancing}
X.~Wan, H.~Deng, K.~Zou, and S.~Xu, ``Enhancing the efficiency and accuracy of underlying asset reviews in structured finance: The application of multi-agent framework,'' \emph{arXiv preprint arXiv:2405.04294}, 2024.

\bibitem[Su et~al.(2022{\natexlab{b}})Su, Lee, Mulvey, and Poor]{su2022competitive}
D.~Su, J.~D. Lee, J.~M. Mulvey, and H.~V. Poor, ``Competitive multi-agent reinforcement learning with self-supervised representation,'' in \emph{ICASSP 2022-2022 IEEE International Conference on Acoustics, Speech and Signal Processing (ICASSP)}.\hskip 1em plus 0.5em minus 0.4em\relax IEEE, 2022, pp. 4098--4102.

\bibitem[Koa et~al.(2024)Koa, Ma, Ng, and Chua]{koa2024learning}
K.~J. Koa, Y.~Ma, R.~Ng, and T.-S. Chua, ``Learning to generate explainable stock predictions using self-reflective large language models,'' \emph{arXiv preprint arXiv:2402.03659}, 2024.

\bibitem[Sharma et~al.(2024)Sharma, Rao, Brockett, Malhotra, Jojic, and Dolan]{sharma2024investigating}
A.~Sharma, S.~Rao, C.~Brockett, A.~Malhotra, N.~Jojic, and W.~B. Dolan, ``Investigating agency of {LLMs} in {Human-AI} collaboration tasks,'' in \emph{Proceedings of the 18th Conference of the European Chapter of the Association for Computational Linguistics (Volume 1: Long Papers)}, 2024, pp. 1968--1987.

\bibitem[Zhang et~al.(2024{\natexlab{e}})Zhang, Bo, Ma, Li, Chen, Dai, Zhu, Dong, and Wen]{zhang2024survey}
Z.~Zhang, X.~Bo, C.~Ma, R.~Li, X.~Chen, Q.~Dai, J.~Zhu, Z.~Dong, and J.-R. Wen, ``A survey on the memory mechanism of large language model based agents,'' \emph{arXiv preprint arXiv:2404.13501}, 2024.

\bibitem[Hua et~al.(2024)Hua, Yang, Li, Wei, and Zhang]{hua2024trustagent}
W.~Hua, X.~Yang, Z.~Li, C.~Wei, and Y.~Zhang, ``{TrustAgent}: Towards safe and trustworthy {LLM-based} agents through agent constitution,'' \emph{arXiv preprint arXiv:2402.01586}, 2024.

\bibitem[Kathiriya et~al.()Kathiriya, Challla, and Devineni]{kathiriyaserverless}
S.~Kathiriya, N.~Challla, and S.~K. Devineni, ``Serverless architecture in {LLMs}: Transforming the financial industry's {AI} landscape.''

\bibitem[George()]{georgetransforming}
J.~G. George, ``Transforming banking in the digital age: The strategic integration of large language models and multi-cloud environments.''

\bibitem[Xie et~al.(2024)Xie, Han, Chen, Xiang, Zhang, He, Xiao, Li, Dai, Feng, et~al.]{xie2024finben}
Q.~Xie, W.~Han, Z.~Chen, R.~Xiang, X.~Zhang, Y.~He, M.~Xiao, D.~Li, Y.~Dai, D.~Feng \emph{et~al.}, ``The {FinBen}: An holistic financial benchmark for large language models,'' \emph{arXiv preprint arXiv:2402.12659}, 2024.

\bibitem[Li et~al.(2024)Li, Li, Shi, Xu, Du, Tan, Huang, and Lin]{li2024alphafin}
X.~Li, Z.~Li, C.~Shi, Y.~Xu, Q.~Du, M.~Tan, J.~Huang, and W.~Lin, ``Alphafin: Benchmarking financial analysis with retrieval-augmented stock-chain framework,'' \emph{arXiv preprint arXiv:2403.12582}, 2024.

\bibitem[Li et~al.(2023{\natexlab{f}})Li, Chan, Zhu, Pei, Ma, Liu, and Shah]{li2023chatgpt}
X.~Li, S.~Chan, X.~Zhu, Y.~Pei, Z.~Ma, X.~Liu, and S.~Shah, ``Are {ChatGPT} and {GPT-4} general-purpose solvers for financial text analytics? a study on several typical tasks,'' in \emph{Proceedings of the 2023 Conference on Empirical Methods in Natural Language Processing: Industry Track}, 2023, pp. 408--422.

\bibitem[Koncel-Kedziorski et~al.(2023)Koncel-Kedziorski, Krumdick, Lai, Reddy, Lovering, and Tanner]{koncel2023bizbench}
R.~Koncel-Kedziorski, M.~Krumdick, V.~Lai, V.~Reddy, C.~Lovering, and C.~Tanner, ``{BizBench}: A quantitative reasoning benchmark for business and finance,'' \emph{arXiv preprint arXiv:2311.06602}, 2023.

\bibitem[Zhao et~al.(2023{\natexlab{d}})Zhao, Long, Liu, Nan, Chen, Kamoi, Liu, Tang, Zhang, and Cohan]{zhao2023docmath}
Y.~Zhao, Y.~Long, H.~Liu, L.~Nan, L.~Chen, R.~Kamoi, Y.~Liu, X.~Tang, R.~Zhang, and A.~Cohan, ``{DocMath-Eval}: Evaluating numerical reasoning capabilities of {LLMs} in understanding long documents with tabular data,'' \emph{arXiv preprint arXiv:2311.09805}, 2023.

\bibitem[Quan and Liu(2024)]{quan2024econlogicqa}
Y.~Quan and Z.~Liu, ``{EconLogicQA}: A question-answering benchmark for evaluating large language models in economic sequential reasoning,'' \emph{arXiv preprint arXiv:2405.07938}, 2024.

\bibitem[Islam et~al.(2023)Islam, Kannappan, Kiela, Qian, Scherrer, and Vidgen]{islam2023financebench}
P.~Islam, A.~Kannappan, D.~Kiela, R.~Qian, N.~Scherrer, and B.~Vidgen, ``{FinanceBench}: A new benchmark for financial question answering,'' \emph{arXiv preprint arXiv:2311.11944}, 2023.

\bibitem[El-Haj(2019)]{el2019multiling}
M.~El-Haj, ``{MultiLing} 2019: Financial narrative summarisation,'' in \emph{Proceedings of the Workshop MultiLing 2019: Summarization Across Languages, Genres and Sources}, 2019, pp. 6--10.

\bibitem[Yuan et~al.(2024{\natexlab{b}})Yuan, He, Dong, Wang, Zhao, Xia, Xu, Zhou, Li, Zhang, et~al.]{yuan2024r}
T.~Yuan, Z.~He, L.~Dong, Y.~Wang, R.~Zhao, T.~Xia, L.~Xu, B.~Zhou, F.~Li, Z.~Zhang \emph{et~al.}, ``{R-Judge}: Benchmarking safety risk awareness for {LLM} agents,'' \emph{arXiv preprint arXiv:2401.10019}, 2024.

\bibitem[Lei et~al.(2023)Lei, Li, Jiang, Hu, Cheng, Ding, and Jiang]{lei2023cfbenchmark}
Y.~Lei, J.~Li, M.~Jiang, J.~Hu, D.~Cheng, Z.~Ding, and C.~Jiang, ``{CFBenchmark}: {Chinese} financial assistant benchmark for large language model,'' \emph{arXiv preprint arXiv:2311.05812}, 2023.

\bibitem[Hirano(2024)]{hirano2024construction}
M.~Hirano, ``Construction of a {Japanese} financial benchmark for large language models,'' \emph{arXiv preprint arXiv:2403.15062}, 2024.

\bibitem[Zhang et~al.(2024{\natexlab{f}})Zhang, Xiang, Yuan, Feng, Han, Lopez-Lira, Liu, Ananiadou, Peng, Huang, et~al.]{zhang2024d}
X.~Zhang, R.~Xiang, C.~Yuan, D.~Feng, W.~Han, A.~Lopez-Lira, X.-Y. Liu, S.~Ananiadou, M.~Peng, J.~Huang \emph{et~al.}, ``{D$\backslash$'olares or Dollars}? unraveling the bilingual prowess of financial {LLMs} between {Spanish} and {English},'' \emph{arXiv preprint arXiv:2402.07405}, 2024.

\bibitem[Zhang et~al.(2023{\natexlab{d}})Zhang, Cai, Liu, Yang, Dai, Liao, Qin, Li, Liu, Liu, et~al.]{zhang2023fineval}
L.~Zhang, W.~Cai, Z.~Liu, Z.~Yang, W.~Dai, Y.~Liao, Q.~Qin, Y.~Li, X.~Liu, Z.~Liu \emph{et~al.}, ``{FinEval}: A {Chinese} financial domain knowledge evaluation benchmark for large language models,'' \emph{arXiv preprint arXiv:2308.09975}, 2023.

\bibitem[Hu et~al.(2024)Hu, Qin, Yuan, Peng, Lopez-Lira, Wang, Ananiadou, Yu, Huang, and Xie]{hu2024no}
G.~Hu, K.~Qin, C.~Yuan, M.~Peng, A.~Lopez-Lira, B.~Wang, S.~Ananiadou, W.~Yu, J.~Huang, and Q.~Xie, ``No language is an island: Unifying {Chinese} and {English} in financial large language models, instruction data, and benchmarks,'' \emph{arXiv preprint arXiv:2403.06249}, 2024.

\bibitem[Xu et~al.(2024)Xu, Zhu, Wu, and Xue]{xu2024superclue}
L.~Xu, L.~Zhu, Y.~Wu, and H.~Xue, ``{SuperCLUE-Fin}: Graded fine-grained analysis of {Chinese} {LLMs} on diverse financial tasks and applications,'' \emph{arXiv preprint arXiv:2404.19063}, 2024.

\bibitem[Malo et~al.(2014)Malo, Sinha, Korhonen, Wallenius, and Takala]{malo2014good}
P.~Malo, A.~Sinha, P.~Korhonen, J.~Wallenius, and P.~Takala, ``Good debt or bad debt: Detecting semantic orientations in economic texts,'' \emph{Journal of the Association for Information Science and Technology}, vol.~65, no.~4, pp. 782--796, 2014.

\bibitem[Maia et~al.(2018)Maia, Handschuh, Freitas, Davis, McDermott, Zarrouk, and Balahur]{maia201818}
M.~Maia, S.~Handschuh, A.~Freitas, B.~Davis, R.~McDermott, M.~Zarrouk, and A.~Balahur, ``{WWW'18 Open Challenge}: financial opinion mining and question answering,'' in \emph{Companion Proceedings of the Web Conference 2018}, 2018, pp. 1941--1942.

\bibitem[Chen et~al.(2021)Chen, Chen, Smiley, Shah, Borova, Langdon, Moussa, Beane, Huang, Routledge, et~al.]{chen2021finqa}
Z.~Chen, W.~Chen, C.~Smiley, S.~Shah, I.~Borova, D.~Langdon, R.~Moussa, M.~Beane, T.-H. Huang, B.~Routledge \emph{et~al.}, ``{FinQA}: A dataset of numerical reasoning over financial data,'' \emph{arXiv preprint arXiv:2109.00122}, 2021.

\bibitem[Mukherjee et~al.(2022)Mukherjee, Bohra, Banerjee, Sharma, Hegde, Shaikh, Shrivastava, Dasgupta, Ganguly, Ghosh, et~al.]{mukherjee2022ectsum}
R.~Mukherjee, A.~Bohra, A.~Banerjee, S.~Sharma, M.~Hegde, A.~Shaikh, S.~Shrivastava, K.~Dasgupta, N.~Ganguly, S.~Ghosh \emph{et~al.}, ``{ECTSum}: A new benchmark dataset for bullet point summarization of long earnings call transcripts,'' \emph{arXiv preprint arXiv:2210.12467}, 2022.

\bibitem[Shah et~al.(2023{\natexlab{b}})Shah, Vithani, Gullapalli, and Chava]{shah2023finer}
A.~Shah, R.~Vithani, A.~Gullapalli, and S.~Chava, ``{FiNER}: Financial named entity recognition dataset and weak-supervision model,'' \emph{arXiv preprint arXiv:2302.11157}, 2023.

\bibitem[Sharma et~al.(2022)Sharma, Nayak, Bose, Meena, Dasgupta, Ganguly, and Goyal]{sharma2022finred}
S.~Sharma, T.~Nayak, A.~Bose, A.~K. Meena, K.~Dasgupta, N.~Ganguly, and P.~Goyal, ``{FinRED}: A dataset for relation extraction in financial domain,'' in \emph{Companion Proceedings of the Web Conference 2022}, 2022, pp. 595--597.

\bibitem[Au et~al.(2021)Au, Ait-Azzi, and Kang]{au2021finsbd}
W.~Au, A.~Ait-Azzi, and J.~Kang, ``{FinSBD-2021}: the 3rd shared task on structure boundary detection in unstructured text in the financial domain,'' in \emph{Companion Proceedings of the Web Conference 2021}, 2021, pp. 276--279.

\bibitem[Zhu et~al.(2024{\natexlab{b}})Zhu, Li, Wen, and Guo]{zhu2024benchmarking}
J.~Zhu, J.~Li, Y.~Wen, and L.~Guo, ``Benchmarking large language models on {CFLUE}--a {Chinese} financial language understanding evaluation dataset,'' \emph{arXiv preprint arXiv:2405.10542}, 2024.

\bibitem[Ding et~al.(2024{\natexlab{b}})Ding, Mallick, Wang, Sim, Mukherjee, Ruhle, Lakshmanan, and Awadallah]{ding2024hybrid}
D.~Ding, A.~Mallick, C.~Wang, R.~Sim, S.~Mukherjee, V.~Ruhle, L.~V. Lakshmanan, and A.~H. Awadallah, ``{Hybrid LLM}: Cost-efficient and quality-aware query routing,'' \emph{arXiv preprint arXiv:2404.14618}, 2024.

\bibitem[Sarkar and Vafa(2024)]{sarkar2024lookahead}
S.~K. Sarkar and K.~Vafa, ``Lookahead bias in pretrained language models,'' \emph{Available at SSRN}, 2024.

\bibitem[Drinkall et~al.(2024)Drinkall, Rahimikia, Pierrehumbert, and Zohren]{drinkall2024time}
F.~Drinkall, E.~Rahimikia, J.~B. Pierrehumbert, and S.~Zohren, ``Time machine {GPT},'' \emph{arXiv preprint arXiv:2404.18543}, 2024.

\bibitem[Krishna et~al.(2024)Krishna, Ramprasad, Gupta, Wallace, Lipton, and Bigham]{krishna2024genaudit}
K.~Krishna, S.~Ramprasad, P.~Gupta, B.~C. Wallace, Z.~C. Lipton, and J.~P. Bigham, ``{GenAudit}: Fixing factual errors in language model outputs with evidence,'' \emph{arXiv preprint arXiv:2402.12566}, 2024.

\bibitem[Yao et~al.(2024)Yao, Duan, Xu, Cai, Sun, and Zhang]{yao2024survey}
Y.~Yao, J.~Duan, K.~Xu, Y.~Cai, Z.~Sun, and Y.~Zhang, ``A survey on large language model ({LLM}) security and privacy: The good, the bad, and the ugly,'' \emph{High-Confidence Computing}, p. 100211, 2024.

\bibitem[Cao et~al.(2023{\natexlab{b}})Cao, Cao, Lin, and Chen]{cao2023defending}
B.~Cao, Y.~Cao, L.~Lin, and J.~Chen, ``Defending against alignment-breaking attacks via robustly aligned {LLM},'' \emph{arXiv preprint arXiv:2309.14348}, 2023.

\bibitem[He et~al.(2024)He, Xia, and Henderson]{he2024s}
L.~He, M.~Xia, and P.~Henderson, ``{What's in Your "Safe" Data?}: Identifying benign data that breaks safety,'' \emph{arXiv preprint arXiv:2404.01099}, 2024.

\bibitem[{Association for Computing Machinery}(2024)]{acm_code_of_ethics}
\BIBentryALTinterwordspacing
{Association for Computing Machinery}, ``{ACM Code of Ethics and Professional Conduct},'' 2024. [Online]. Available: \url{https://www.acm.org/code-of-ethics}
\BIBentrySTDinterwordspacing

\bibitem[Commission(2024)]{eu_ai_regulation}
\BIBentryALTinterwordspacing
E.~Commission, ``Regulatory framework for {AI},'' 2024. [Online]. Available: \url{https://digital-strategy.ec.europa.eu/en/policies/regulatory-framework-ai}
\BIBentrySTDinterwordspacing

\end{thebibliography}

\end{document}